\documentclass[12pt]{cernart}

\usepackage{graphicx}
\usepackage{amsmath}
\usepackage{amssymb}
\usepackage{times}
\usepackage{cite}
\usepackage{multirow}

\begin{document}
\setcounter{topnumber}{3}
\renewcommand{\topfraction}{0.999}
\renewcommand{\bottomfraction}{0.99}
\renewcommand{\textfraction}{0.0}
\setcounter{totalnumber}{6}

\begin{titlepage}
\docnum{CERN-PH-EP/2009-006}
\date{5 March 2009}
%
\title{\large{Inclusive production of protons, anti-protons and neutrons \\ 
in p+p collisions at 158 GeV/c beam momentum}}
\begin{Authlist}
\vspace{2mm}
\noindent
T.~Anticic$^{15}$,
B.~Baatar$^{6}$, J.~Bartke$^{4}$, L.~Betev$^{8}$, H.~Bia{\l}\-kowska$^{14}$, C.~Blume$^{7}$, 
B.~Boimska$^{14}$, J.~Bracinik$^{1,a}$,
V.~Cerny$^{1}$,  O.~Chvala$^{11,b}$,
J.~Dolejsi$^{11}$,
V.~Eckardt$^{10}$,
H.G.~Fischer$^{8}$, Z.~Fodor$^{3}$, P.~Foka$^{5}$, V.~Friese$^{5}$,
M.~Ga\'zdzicki$^{7}$,
C.~H\"{o}hne$^{5}$,
K.~Kadija$^{15}$, A.~Karev$^{8}$, V.~Kolesnikov$^{6}$, M.~Kowalski$^{4}$, M.~Kreps$^{1,c}$,
M.~Makariev$^{13}$, A.~Malakhov$^{6}$, M.~Mateev$^{12}$,
G.~Melkumov$^{6}$, M.~Mitrovski$^{7}$, S.~Mr\'owczy\'nski$^{9}$,
R.~Renfordt$^{7}$, M. Rybczy\'nski$^{9}$, A.~Rybicki$^{4}$,
A.~Sandoval$^{5}$, N.~Schmitz$^{10}$, P.~Seyboth$^{10}$, G.~Stefanek$^{9}$, R.~Stock$^{7}$, 
H.~Str\"{o}bele$^{7}$, T.~Susa$^{15}$, P.~Szymanski$^{14}$,
V.~Trubnikov$^{14}$,
D.~Varga$^{2}$, G.~Vesztergombi$^{3}$, D.~Vrani\'{c}$^{5}$,
S.~Wenig$^{8}$,
Z.~W{\l}odarczyk$^{9}$, A.~Wojtaszek$^{9}$
\vspace*{2mm} 

\noindent
{\it (The NA49 Collaboration)}  \\
\vspace*{2mm}
\noindent
$^{1}$Comenius University, Bratislava, Slovakia\\
$^{2}$ E\"otv\"os Lor\'ant University, Budapest, Hungary \\
$^{3}$KFKI Research Institute for Particle and Nuclear Physics, Budapest, Hungary\\
$^{4}$H. Niewodnicza\'nski Institute of Nuclear Physics,
             Polish Academy of Sciences, Cracow, Poland \\
$^{5}$Gesellschaft f\"{u}r Schwerionenforschung (GSI), Darmstadt, Germany.\\
$^{6}$Joint Institute for Nuclear Research, Dubna, Russia.\\
$^{7}$Fachbereich Physik der Universit\"{a}t, Frankfurt, Germany.\\
$^{8}$CERN, Geneva, Switzerland\\
$^{9}$Institute of Physics \'Swi\c{e}tokrzyska Academy, Kielce, Poland.\\
$^{10}$Max-Planck-Institut f\"{u}r Physik, Munich, Germany.\\
$^{11}$Charles University, Faculty of Mathematics and Physics, Institute of
             Particle and Nuclear Physics, Prague, Czech Republic \\
$^{12}$Atomic Physics Department, Sofia University St. Kliment Ohridski, Sofia, Bulgaria\\
$^{13}$Institute for Nuclear Research and Nuclear Energy, BAS, Sofia, Bulgaria\\
$^{14}$Institute for Nuclear Studies, Warsaw, Poland\\
$^{15}$Rudjer Boskovic Institute, Zagreb, Croatia\\ 
$^{a}$now at School of Physics and Astronomy, University of Birmingham, Birmingham, UK \\
$^{b}$now at UC Riverside, Riverside, CA, USA\\
$^{c}$now at Institut fur Experimentelle Kernphysik, Karlsruhe, DE\\
\end{Authlist}

\begin{center}
{\small{\it to be published in EPJC }}
\end{center}
\vspace*{2mm} 
\clearpage

\begin{abstract}
\vspace{-3mm}
New data on the production of protons, anti-protons and neutrons
in p+p interactions are presented. The data come from a sample
of 4.8 million inelastic events obtained with the NA49 detector
at the CERN SPS at 158~GeV/c beam momentum. The charged baryons
are identified by energy loss measurement in a large TPC tracking
system. Neutrons are detected in a forward hadronic calorimeter.
Inclusive invariant cross sections are obtained in intervals
from 0 to 1.9~GeV/c (0 to 1.5~GeV/c) in transverse momentum
and from -0.05 to 0.95 (-0.05 to 0.4) in Feynman x for protons
(anti-protons), respectively. $p_T$ integrated neutron cross sections
are given in the interval from 0.1 to 0.9 in Feynman x. The data 
are compared to a wide sample of existing results in the SPS and ISR 
energy ranges as well as to proton and neutron measurements from HERA and RHIC. 
\end{abstract}
 
\clearpage
\end{titlepage}

%
%
\section{Introduction} 
\vspace{3mm}
\label{sec:intro}

In the framework of its extensive experimental programme concerning
soft hadronic interactions at SPS energies, the NA49
collaboration has recently published detailed data on the inclusive
production of charged pions in p+p collisions \cite{bib:pp_paper}. The present paper 
extends this study to the baryonic sector by providing inclusive
cross sections for protons, anti-protons and neutrons. The aim is
again to obtain precise sets of data covering the available phase
space  as densely and completely as possible in accordance with
the available event statistics and the limitations set by the NA49
detector layout.

As in the case of pions, the experimental situation in the SPS energy
range is far from being satisfactory also for baryons. The presently
available data sets suffer from insufficient coverage and at least 
partially large systematic and statistical error margins. It is 
therefore one of the main aims of this study to provide a concise
overview and evaluation of the experimental situation on a quantitative
basis.

This paper is arranged as follows. The present experimental situation
is discussed in Sect.~\ref{sec:exp_sit}. Section~\ref{sec:na49} concentrates on those 
aspects of the NA49 experiment which are special to baryon detection, as for
instance high momentum tracking and neutron calorimetry. 
The acceptance coverage and the binning scheme are presented in 
Sect.~\ref{sec:bin_scheme}, followed by the description of charged particle
identification in Sect.~\ref{sec:pid}.
The evaluation of invariant cross sections and of the
applied corrections is given in Sect.~\ref{sec:corr}. Results concerning double
differential cross sections for protons and anti-protons are 
presented in Sect.~\ref{sec:data}, followed by a detailed comparison to existing
data in Sects.~\ref{sec:comp} and \ref{sec:comp_isr}. Sections~\ref{sec:ptint} and \ref{sec:neut}
show $p_T$ integrated results for protons and neutrons including a comparison
to other experiments. Finally in Sect.~\ref{sec:hera} the NA49 results on proton 
and neutron production are compared to baryon production in deep inelastic lepton scattering
from HERA.

%
%
\section{The Experimental Situation}
\vspace{3mm}
\label{sec:exp_sit}

Concerning the present publication we are interested in the 
available measurements of the double differential cross section of
identified baryons,

\begin{equation}
  \frac{d^2\sigma}{dx_Fdp_T^2}   ,
\end{equation}
as a function of the phase space variables defined in this paper as
transverse momentum $p_T$ and reduced longitudinal momentum

\begin{equation}
x_F = \frac{p_L}{\sqrt{s}/2}
\end{equation}
where $p_L$ denotes the longitudinal momentum component in the cms.

\begin{figure}[b]
  \begin{center}
  	\includegraphics[width=15.4cm]{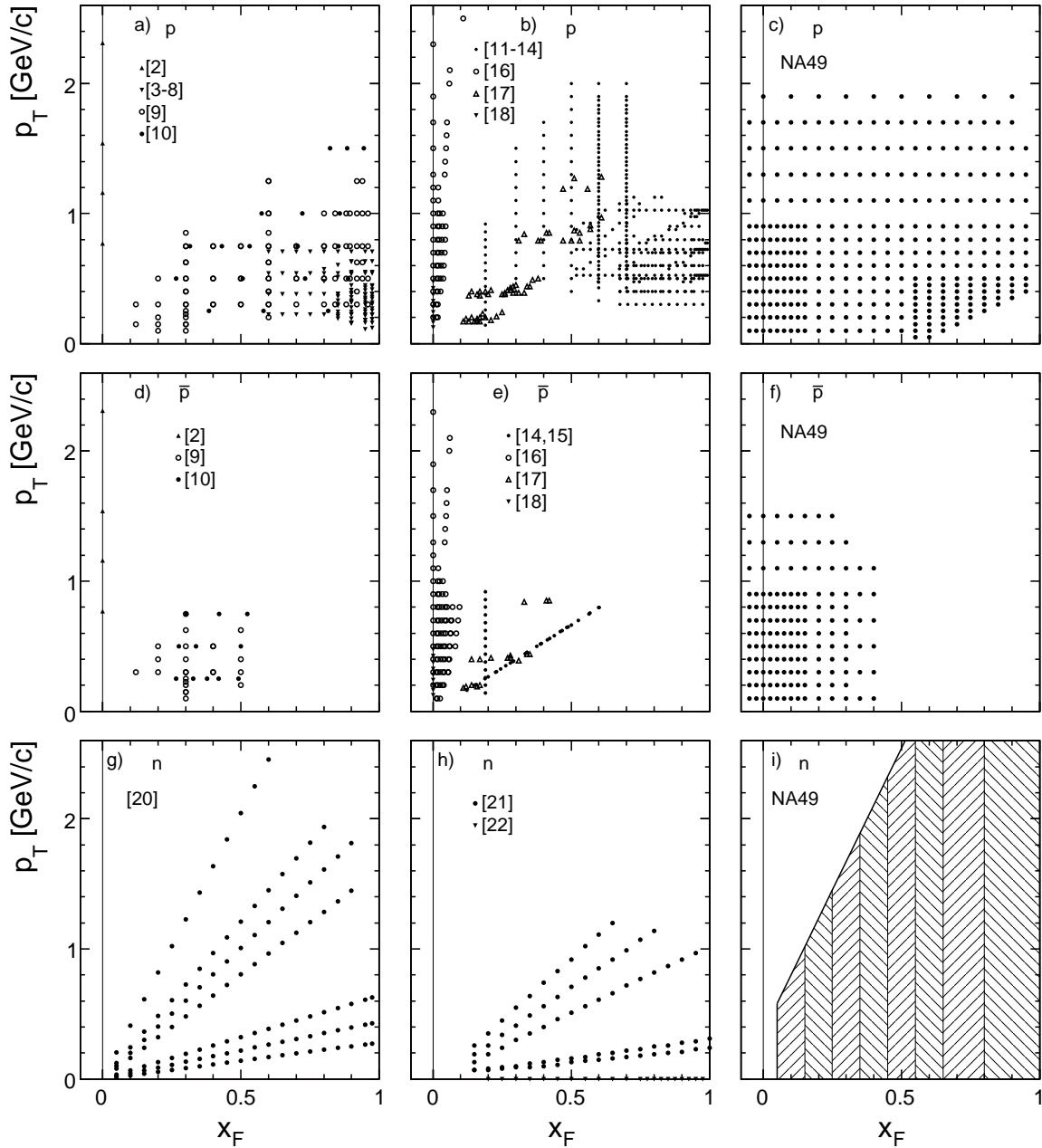}
  	\caption{Phase space coverage of existing data}
  	\label{fig:cov}
  \end{center}
\end{figure}

Defining a range of beam momenta from 100 to about 400~GeV/c as
SPS/Fermilab energy range, quite a few experiments have published
inclusive particle yields 
\cite{bib:cronin,bib:sannes,bib:childress,bib:scham,bib:akimov,bib:chapman,bib:whitmore,bib:brenner,bib:john}.
The corresponding data coverage
of the $p_T$/$x_F$ plane is shown in Fig.~\ref{fig:cov}a for protons and in 
Fig.~\ref{fig:cov}d for anti-protons. It is apparent from these plots that data are
scarce in the regions of $p_T$ below 0.3~GeV/c and above 1~GeV/c as
well as $x_F$ below 0.3. At large $x_F$ there is abundant coverage only for
protons in a $p_T$ interval from about 0.2 to 0.6~GeV/c from experiments
concentrating on single diffraction.
It is therefore mandatory to also regard data from the ISR 
\cite{bib:alb_prot1,bib:alb_prot2,bib:alb_prot3,bib:alb_prot4,bib:alb_aprot1,bib:alb_aprot2,bib:capi,bib:alper,bib:guettler}
at least in the overlapping region of $\sqrt{s}$ up to 30~GeV for 
this comparison. The corresponding phase space regions are presented 
in Figs.~\ref{fig:cov}b and \ref{fig:cov}e for protons and anti-protons, respectively. 
Except for a rather complete coverage at $x_F$ close to zero 
a lack of data in the intermediate region 0.1~$< x_F <$~0.4, at $p_T$
below 0.3~GeV/c and above about 1.5~GeV/c is evident.

The NA49 phase space coverage, Figs.~\ref{fig:cov}c and \ref{fig:cov}f, is essentially only
limited by counting statistics at large $p_T$, and at large $x_F$ for the
anti-protons. In addition there is a small phase space gap not
accessible due to the interaction trigger, in a $p_T$ range below
0.05~GeV/c at $x_F$~=~0.6 to 0.4~GeV/c at $x_F$~=~0.95 which only concerns
protons.

For neutrons, the situation is less favourable. There is only one
measurement from Fermilab \cite{bib:fermilab_neut} and one ISR experiment
\cite{bib:engl_neut,bib:flau_neut}, with coverages
shown in Fig.~\ref{fig:cov}g and \ref{fig:cov}h. Due to lack of transversal granularity,
the NA49 calorimeter only allows for the measurement of $p_T$ integrated
neutron yields. The corresponding $p_T$/$x_F$ coverage, limited by the
fiducial dimension of the calorimeter, is shown in Fig.~\ref{fig:cov}i. 

It is useful to repeat here that the main aim of the present paper
is to contribute precise new data covering the accessible phase space
as densely and continuously as possible in a single experiment in
order to clarify the unsatisfactory experimental situation and to
provide a sound base for the comparative study of the more complex
nuclear interactions.

%
%
\section{The NA49 Experiment}
\vspace{3mm}
\label{sec:na49}

\begin{figure}[h]
  \begin{center}
  	\includegraphics[width=12cm]{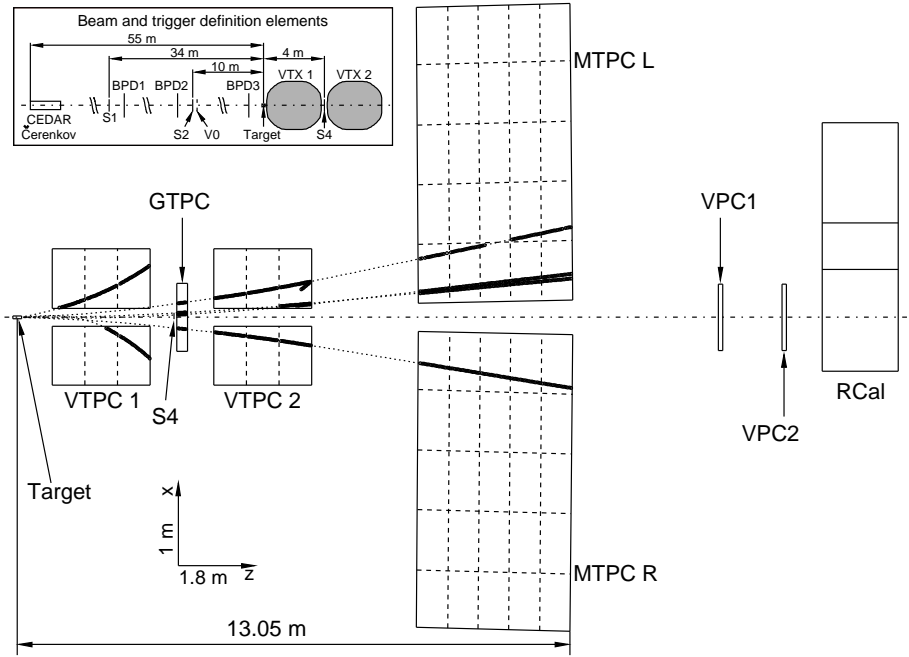}
  	\caption{NA49 detector layout and real tracks of a typical mean multiplicity
                  p+p event. The thick lines give the points registered in the TPC's, the dotted 
                  lines are the interpolation trajectories between the track segments and the 
                  extrapolations to the event vertex in the LH$_2$ target. The beam and trigger
                  definition counters are presented in the inset}
  	\label{fig:exp}
  \end{center}
\end{figure}

The basic features of the NA49 detector have been described
in detail in references \cite{bib:pp_paper,bib:nim}. The top view shown in
Fig.~\ref{fig:exp} recalls the main components.

The beam is defined by 
a CEDAR Cerenkov counter, several scintillation counters
(S1, S2, V0) and a set of high precision proportional chambers
(BPD1-3). The hydrogen target is placed in front of two
superconducting Magnets (VTX1 and VTX2). Four large volume
Time Projection Chambers (VTPC1 and VTPC2 inside the magnetic
fields, MTPCL and MTPCR downstream of the magnets) provide
for charged particle tracking and identification. A smaller
Time Projection Chamber (GTPC) placed between the two magnets
together with two Multiwire Proportional Chambers (VPC1 and VPC2) 
in forward direction allows tracking in the high momentum region 
through the gaps between the principal track detectors. A Ring 
Calorimeter (RCal) closes the detector setup 18~m downstream 
of the target.

As details of the beam and target setup, the trigger definition
as well as the event and track selection have been given in \cite{bib:pp_paper} 
only those parts of the detector which are of special interest for
the present paper will be described here. This concerns in particular
the extension of the acceptance into the large $x_F$ region and the
neutron calorimetry.

%
%
\subsection{Tracking at high momenta using the GTPC and VPC's}
\vspace{3mm}
\label{sec:high_mom}

\begin{figure}[h]
  \begin{center}
  	\rotatebox{270}{\includegraphics[width=5.5cm]{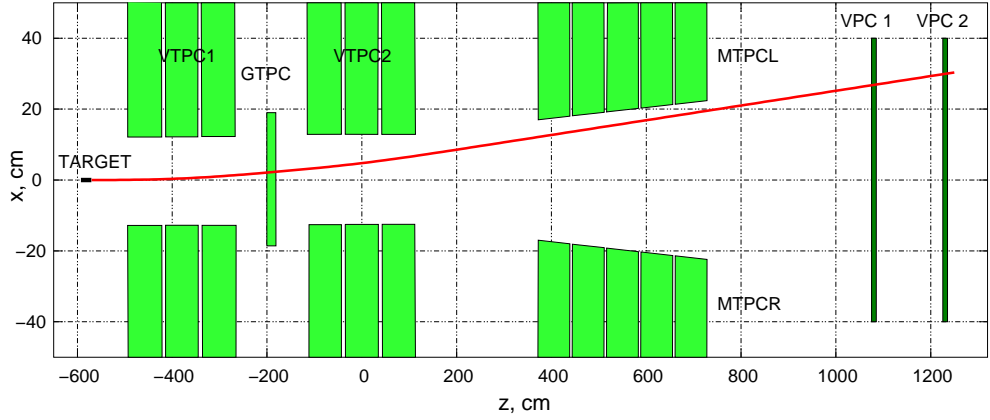}}
  	\caption{Forward proton detection with the GTPC and the VPC-s: the
                   trajectory of a 110~GeV particle is shown. The scale is stretched in the
                   $x$ direction}
  	\label{fig:gtpc_vpc}
  \end{center}
\end{figure}

The particles originating from the primary interaction vertex and
missing, at high momentum, the main TPC arrangement, are detected
in the GTPC and VPC's. These three sets of points are sufficiently
far from each other to provide a reasonable lever arm for momentum
measurement. A sketch of this detector part is shown in Fig.~\ref{fig:gtpc_vpc}.
For experimental details see \cite{bib:dezso}.

The VPC proportional chambers feature a single sense wire plane with 
strip readout ($\pm$30 degree inclination) on both cathode surfaces 
resulting in a space resolution of 2~mm. This results, together 
with the GTPC resolution of less than 150~$\mu$m, in a longitudinal 
momentum resolution of 

\begin{figure}[b]
  \begin{center}
  	\rotatebox{270}{\includegraphics[width=6cm]{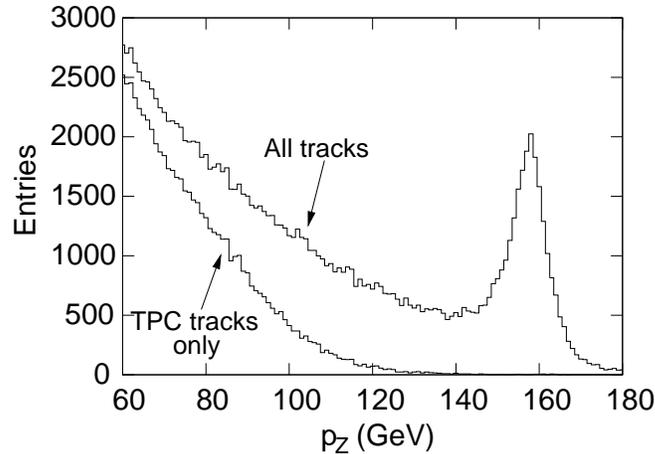}}
  	\caption{Raw $p_z$ distribution in p+p interaction. The diffractive
                   peak is clearly visible. Lower histogram VTPC+MTPC tracking,
                   upper histogram GTPC+VPC tracking added}
  	\label{fig:rawpz}
  \end{center}
\end{figure}

\begin{equation}
    \frac{\Delta p_z}{p_z} \approx 0.013\% \times  p_z(\textrm{GeV/c})
\end{equation}
where the error is dominated by the VPC position resolution.
The momentum resolution at maximum momentum was controlled using a 
trigger on beam particles. For inelastic events, it is also established 
by the width of the diffractive peak as shown in the raw $p_z$ 
distribution in Fig.~\ref{fig:rawpz}.

The general improvement of charged particle acceptance at $x_F >$~0.4
due to this detector combination, as compared to the tracks
visible in the main TPC detector only, is also apparent from
Fig.~\ref{fig:rawpz}.

The corresponding transverse momentum resolution is given by

\begin{equation}
       \Delta p_T \approx 2 \times 10^{-4} p_z.
\end{equation}

It is dominated by both the GTPC and the transverse vertex resolutions.
The resulting uncertainty of 30~MeV/c at beam momentum is small enough
to allow the extraction of transverse momentum distributions up to
the kinematic limit.
   
%
%
\subsection{Neutron detection}
\vspace{3mm}
\label{sec:neut_det}

Forward neutrons as well as fast forward charged particles are detected
in the Ring Calorimeter (RCal). This device, originally designed for
the study of jet production in deep inelastic interactions by the
CERN NA5 experiment \cite{bib:marzo1,bib:marzo2,bib:eckardt}, is placed 18~m 
downstream of the target. It is a cylindrical structure with azimuthal and radial subdivision
into 240 cells, each with an electromagnetic and hadronic
compartment. 

For the present purpose, it was off-centered with respect to the
beam axis such that a fully sensitive fiducial area of 80$\times$160 cm$^2$
corresponding to the size of the VPC chambers could be established, see Fig.~\ref{fig:exp}.
This corresponds to a $p_T$ cut-off of 1.25~GeV/c at $x_F$~=~0.2, increasing
to more than 2~GeV/c at $x_F >$~0.4, for neutral particles.

Each RCal cell is built up from 2 parts: an electromagnetic part
(20 radiation lengths of Pb/scintillator sandwich) and a hadronic
part (4 interaction lengths of Fe/scintillator sandwich) \cite{bib:marzo2}.
Energy deposits in the two parts are recorded separately. As the
position resolution of the RCal is rather limited in the transverse
plane due to the substantial cell size, only $p_T$ integrated $x_F$
distributions are presented in this paper. For experimental details see \cite{bib:dezso}.

%
%
\subsubsection{Veto against charged particles}
\vspace{3mm}
\label{sec:veto}

\begin{figure}[h]
  \begin{center}
  	\rotatebox{0}{\includegraphics[width=7cm]{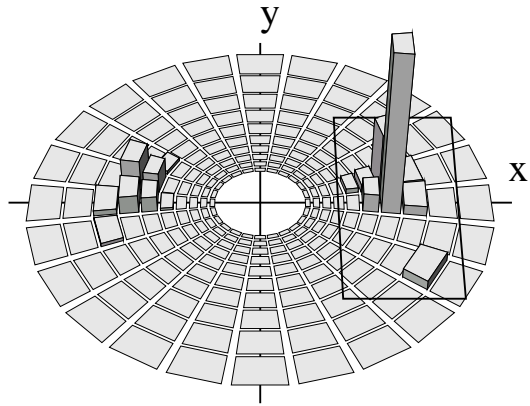}}
  	\caption{Example of an event in which the RCal energy deposit is not
                   associated with a VPC hit. The VPC fiducial area is projected on the RCal}
  	\label{fig:rcal_neut}
  \end{center}
\end{figure}

The VPC detectors are essential for the discrimination between charged 
and neutral hadrons impinging on the RCal. The geometrical situation is
shown in Fig.~\ref{fig:rcal_neut}, where the VPC acceptance is superposed
to the $r/\phi$ structure of the calorimeter. 

As the efficiency of the VPC detectors for charged particles has been
measured to be higher than 99\%, the presence of a calorimeter cluster
together with the absence of a corresponding hit in the VPC surface yields
a clean selection of neutral particles. In the case of multiple-hit patterns the 
equality of the signal amplitudes induced on the cathode surface by
a traversing particle was used for pattern recognition by matching
equal-amplitude strip combinations. 

Two reliable tests of the VPC-RCal performance using external
constraints were developed. The first one uses the fact that the
vast majority of fast forward tracks is of positive charge.
In case of VPC inefficiency this would lead, due to the bending
of charged tracks in the magnetic field, to a noticeable left-right 
asymmetry of neutron detection. The second test uses the
GTPC information as additional constraint on charged trajectories.
In both cases a reliable assessment of the systematic errors is
obtained.

%
%
\subsubsection{Calorimeter calibration and performance}
\vspace{3mm}
\label{sec:calo}

The RCal calibration was performed with beam particles of 40 and 
158~GeV/c momentum. The resulting hadronic energy resolution can be 
parametrized by the following expression:

\begin{equation}
       \frac{\sigma(E)}{E} = \sqrt{\frac{(0.9 \pm 0.1)}{E} + ( 0.02 \pm 0.005 )  }.
\end{equation}

This is well compatible with earlier detailed studies \cite{bib:marzo2}. The
constant term in addition to the square-root behaviour is mainly
due to the non-uniformity of the response over the calorimeter
surface. The energy response was found to be non-Gaussian which
was taken into account in the unfolding procedure.
  
\begin{figure}[h]
  \begin{center}
  	\rotatebox{270}{\includegraphics[width=7cm]{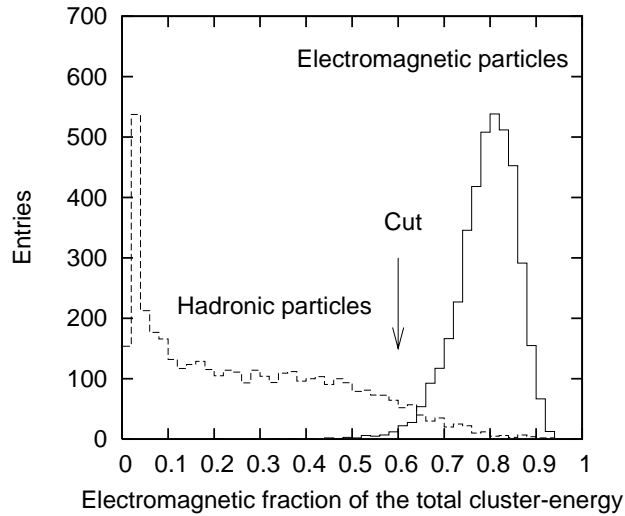}}
  	\caption{Electromagnetic and hadronic particle response (40 GeV pion and
                   electron beam)}
  	\label{fig:encut}
  \end{center}
\end{figure}

Using beams of identified electrons and pions, a precise separation
of the RCal response to hadronic and electromagnetic
particles has been obtained. This separation is quantified by a cut
in the electromagnetic fraction of the cluster-energy which was
placed at 0.6, as shown in Fig.~\ref{fig:encut}.

With this cut, the contamination from electromagnetic particles
(mainly photons from $\pi^0$ decay) is negligible at all energies. 
The loss of hadrons due to the cut has been determined using identified
beam particles at different momenta and also by matching identified
tracks in the TPC system and the corresponding calorimeter clusters
in the region of common acceptance. It can be parametrized as:

\begin{equation}
\delta(\%) = \frac{51 \pm 5}{\sqrt{E}} .
\end{equation}

The analysis of the calorimeter response required the development
of an optimized cluster finding algorithm which fully exploits the
analog response of the RCal cells. As the magnetic field suppresses
low energy hadrons on the RCal fiducial surface, in most cases
only single high energy protons or neutrons which may be accompanied
by lower-energy K$^0_L$ or anti-neutrons, have to be accounted for.
The cluster-finding algorithm thus first tries to find the largest
cluster and verifies its shape-compatibility with the cluster
model as obtained from calibration data. If needed, clusters are
split further. Monte Carlo methods were used to estimate the effects
of cluster overlap, demonstrating that this causes only small
and well controllable systematic errors on the 2\% level.

%
%
\subsubsection{Energy resolution unfolding}
\vspace{3mm}
\label{sec:unfold}

\begin{figure}[b]
  \begin{center}
  	\rotatebox{270}{\includegraphics[width=7cm]{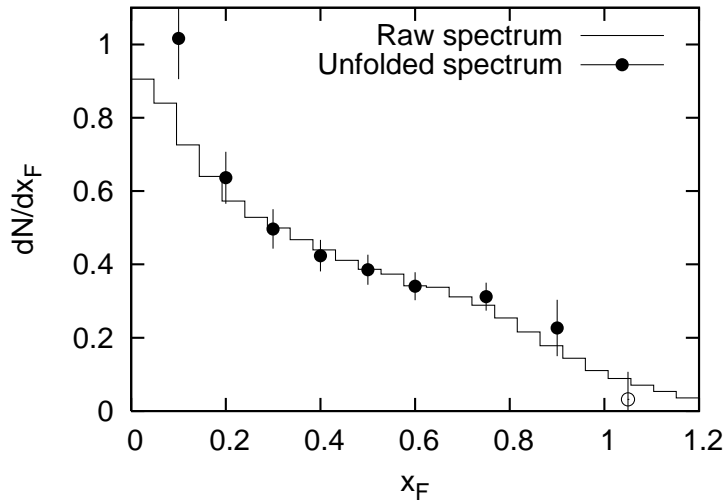}}
  	\caption{Raw measured energy distribution of neutrons compared with the
                   unfolded neutron distribution. The increase of the latter at $x_F <$~0.1 is
                  due to reduction in transverse acceptance. The open circle indicates 
                  the consistency of the unfolded spectra with zero beyond the kinematic limit}
  	\label{fig:neut_raw}
  \end{center}
\end{figure}

A critical step in the analysis of the neutron data is the unfolding
of the calorimeter resolution from the measured momentum distribution. 
With a starting estimate of the real neutron distribution as an input,
a Monte Carlo simulation is used to predict the distribution
modified by the calorimeter resolution. The difference between the
real measurement and the Monte Carlo output is fed back to correct
the input estimation. In a few steps, this iterative process results
in a precise description of the raw neutral particle energy 
distribution. Due to the approximately linear behaviour of the
measured spectrum as a function of $x_F$, the raw and the unfolded 
distributions are consistent with each other over most of the
$x_F$ range with the exception of the regions around $x_F$~=~1 and $x_F$~=~0.1. As the
real neutron distribution is constrained to the physical region
$x_F <$~1, the unphysical tail beyond the kinematic limit is removed. 
This is demonstrated in Fig.~\ref{fig:neut_raw}. 

Indeed the unfolded spectrum is consistent with zero at $x_F$ beyond
1 and the corrected distribution is increased over the measured
one above $x_F \sim$~0.7 in accordance with the width of the RCal energy
resolution.  

%
%
\subsubsection{Transformation to $x_F$ and acceptance correction}
\vspace{3mm}
\label{sec:neut_xf}

In the absence of resolution in transverse momentum the transformation
from the neutron energy as measured in the SPS lab system to the
cms variable $x_F$ introduces a spread in $x_F$ which depends on
the range in $p_T$ and on the energy. This spread diverges with 
decreasing energy assuming a fixed $p_T$ window. Taking however
account of the transverse momentum cut-off at low energy
shown in Fig.~\ref{fig:cov}i, and limiting the $p_T$ range to 2~GeV/c
in the high energy region, this divergence is regularized 
such that the actual spread in $x_F$ varies between 0.012 and
0.024 with the maximum value at $x_F$~=~0.5. This spread is small 
compared to the bin width of 0.1 in $x_F$. The actual transformation 
was performed using Monte Carlo methods under the assumption that the 
$p_T$ distribution of the neutrons would be equal to the one for 
protons. As shown in the later Sects.~\ref{sec:neut} and \ref{sec:hera}
of this paper this assumption has been verified experimentally. 
The resulting systematic errors are negligible. 

The same assumption concerning the neutron $p_T$ distribution has been
made concerning the correction for the $p_T$ cut-off at low $x_F$.
Here the correction decreases rapidly from 20\% at $x_F$~=~0.1 to less than 1\%
at $x_F$~=~0.3. Allowing for a 10\% variation in surface of the
assumed neutron $p_T$ distribution beyond the experimental cut-off,
this leads to the systematic error estimate of less than 2\% given in Table~\ref{tab:syst}.

%
%
\section{Acceptance Coverage and Binning}
\vspace{3mm}
\label{sec:bin_scheme}

The NA49 detector acceptance allows for the extraction of baryon
yields over most of the forward cms hemisphere, with a welcome 
extension to negative $x_F$ which may be used for a test of the 
experimental forward-backward symmetry.

The available event statistics limits the transverse momentum
range to $p_T <$~1.9~GeV/c for protons and $p_T <$~1.7~GeV/c for anti-
protons. The strong decrease of the anti-proton yield with
increasing $x_F$ defines a further limit at $x_F <$~0.4. For protons
there is an acceptance gap at $x_F >$~0.6 and $p_T <$~0.4~GeV/c. This is
a result of the interaction trigger: a small scintillation
counter, S4 (see Fig.~\ref{fig:exp}), vetoes non-interacting beam particles
and, unavoidably, also events with charged secondaries 
in this region.
 
As described in Sect.~\ref{sec:na49} the granularity of the hadron 
calorimeter used for neutron detection does not allow for binning 
in transverse momentum. In addition the size of the fiducial
region in the transverse plane progressively cuts off $p_T$ values 
at below 2~GeV/c with decreasing $x_F$. This effect, together with
the uncertainties of estimating the inseparable anti-neutron and
K$^0_L$ yields at low $x_F$, leads to a cutoff at $x_F$~=~0.05 for neutrons.

The accessible kinematical regions for baryons described above
were subdivided into bins in the $x_F$/$p_T$ plane which vary according
to the available particle yields. Effects of finite bin width
are corrected for in the enumeration of the inclusive cross 
sections, see Sect.~\ref{sec:corr}.
 
\begin{figure}[h]
  	\begin{center}
  		\includegraphics[width=15.5cm]{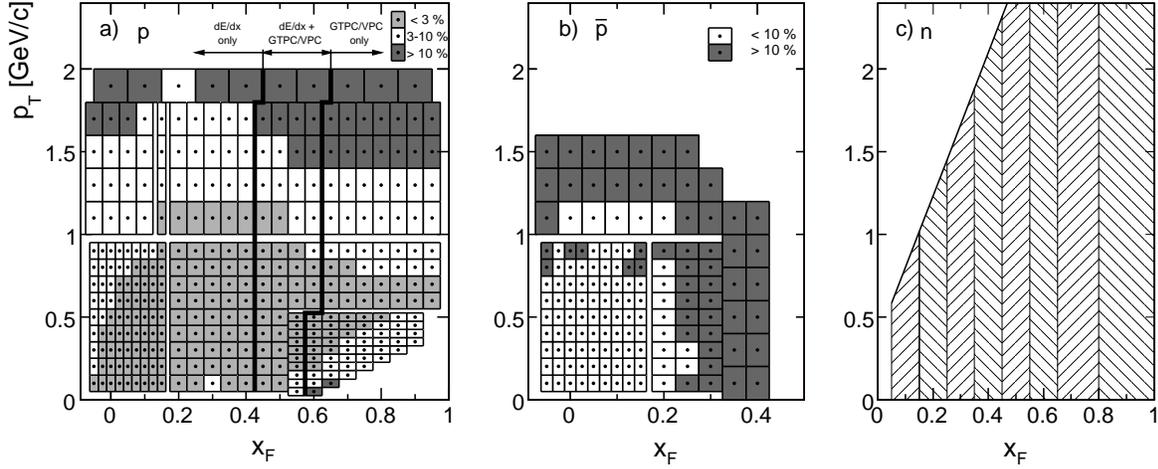}
 		 \caption{Binning scheme for a) protons, b) anti-protons and c) neutrons.
                        In panels a) and b) the different regions of
                        statistical uncertainty are indicated by different shades }
  	  \label{fig:accept}
   \end{center}
\end{figure}

The resulting binning schemes are shown in Fig.~\ref{fig:accept}.

For protons in the forward direction, the extended acceptance 
region using the tracking combination of GTPC and VPC is indicated
by the thick line in Fig.~\ref{fig:accept}a at $x_F \sim$~0.6. This procedure is cross-checked in the region 
of overlap with the main TPC tracking down to the second thick line at $x_F \sim$~0.4.
As particle identification via energy loss measurement ($dE/dx$)
does not operate in the region beyond $x_F$~=~0.6, $\pi$/p and K$^+$/p ratios
from other experiments have been used to extract the proton cross
sections, see Sect.~\ref{sec:forw_rat}.

%
%
\section{Particle Identification}
\vspace{3mm}
\label{sec:pid}

As compared to the preceding publication of pion production \cite{bib:pp_paper}
the identification of charged particles using energy loss 
measurement in the TPC detector system has been further improved.
In fact the extraction of pion yields by a four-parameter fit to the
truncated $dE/dx$ distribution of a track sample in a given bin,
see  \cite{bib:pp_paper}, is insensitive to small imperfections of the 
analog response of the detectors. In addition it has been shown 
that the method used does not introduce additional fluctuations over 
and above the purely statistical error of the extracted pion sample.

This is not quite the case for the other particle species, especially
for kaons and anti-protons which have generally small yields in
relation to pions. Here the fit procedure introduces non-negligible
additional fluctuations which are to be described by an error matrix
with terms that create effective errors beyond the ones related
to the particle yields proper. In this context it is mandatory to
reduce the possible variation of the absolute position of the energy
loss for the different particle species to a minimum in order to
constrain the possible variations of the fit parameters.

%
%
\subsection{Scaling of the truncated mean distributions}
\vspace{3mm}
\label{sec:truncated}

\begin{figure}[h]
  \begin{center}
  	\rotatebox{270}{\includegraphics[width=7cm]{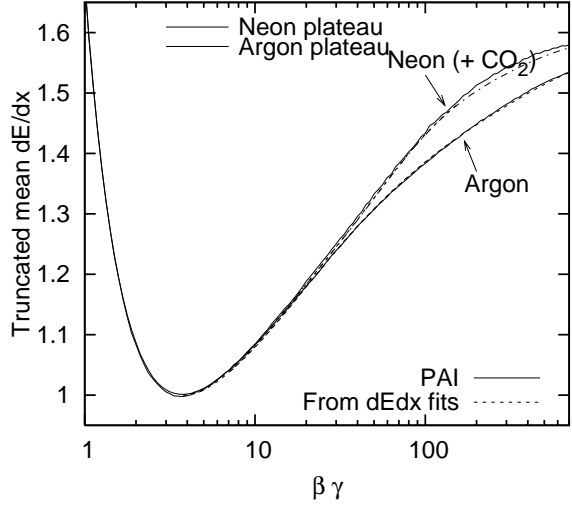}}
  	\caption{Truncated mean Bethe-Bloch functions for Ar+CH$_4$+CO$_2$ (90:5:5)
  					and Ne+CO$_2$ (91:9) from the Photon Absorption Ionization (PAI) model and from
  					direct $dE/dx$ fits}
  	\label{fig:dedxBB}
  \end{center}
\end{figure}

The distribution of truncated means as a function of $p/m = \beta\gamma$
shows non-linear deviations from the Bethe-Bloch parametrization
which is formulated for the total ionization energy loss.
It may be calculated using elementary photon absorption data \cite{bib:photo}
taking account of the effects of truncation using Monte Carlo
methods. For the two gas mixtures used in the NA49 experiment
(Ne+CO$_2$ 91/9 and Ar+CH$_4$+CO$_2$ 90/5/5) it has also been extracted
experimentally by a careful re-analysis of all data. The resulting 
distributions show agreement on the sub-percent level as presented 
in Fig.~\ref{fig:dedxBB}.

\begin{figure}[t]
  \begin{center}
  	\rotatebox{270}{\includegraphics[width=7cm]{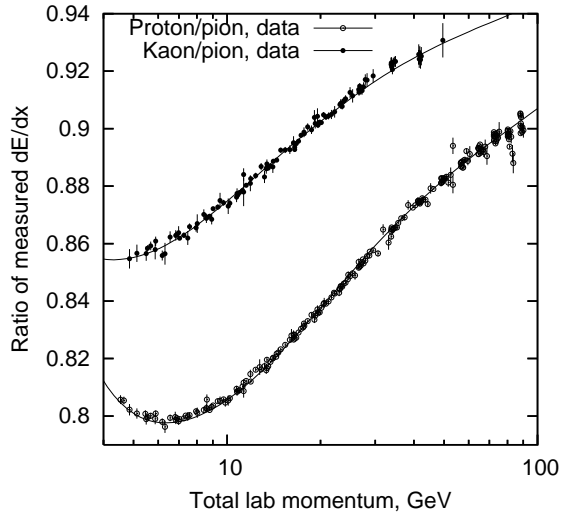}}
  	\caption{Ratios of the measured truncated mean $dE/dx$, protons and kaons
                  relative to pions. The lines correspond to the calibrated Argon
                  Bethe-Bloch curve, the data points are individual fits on the p+p data}
  	\label{fig:dedx_rel}
  \end{center}
\end{figure}

The precision of the predictivity of the absolute energy deposit
is exemplified in Fig.~\ref{fig:dedx_rel} on an extended scale by the ratio of the
truncated means of protons and kaons to pions as a function of
the lab momentum. The calibrated Bethe-Bloch references are
superimposed as full lines.

%
%
\subsection{Control of the analog detector response}
\vspace{3mm}
\label{sec:shifts}

\begin{figure}[h]
  \begin{center}
  	\includegraphics[width=11cm]{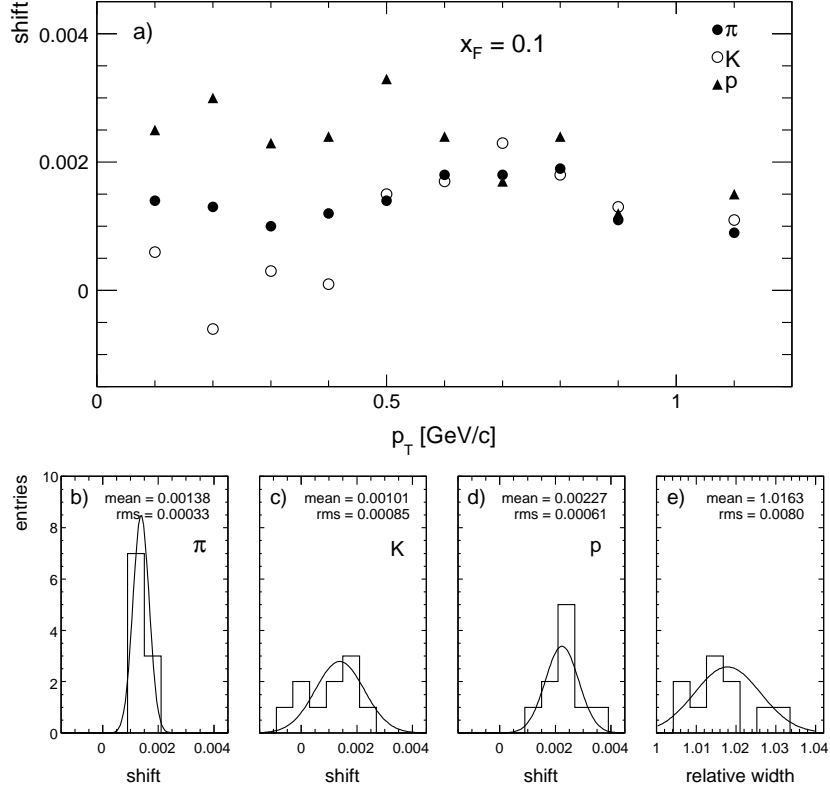}
  	\caption{a) Mean $dE/dx$ position relative to  the Bethe-Bloch parametrization (shift) as a
  	               function of $p_T$ at fixed $x_F$ and distribution of b) $\pi$ shift, c) K shift, d) p shift
  	               and e) relative width }
  	\label{fig:shifts}
  \end{center}
\end{figure}

A thorough re-analysis of the particle identification methods compared
to the earlier work on pion extraction \cite{bib:pp_paper} has been performed.
This concerns a re-calibration  of time dependences, detector edge effects 
and the various corrections due to track length variations at the pad plane 
including the influence of $E \times B$ effects in the inhomogeneous magnetic 
fields. It results in an improvement of the predictivity of the mean $dE/dx$ 
position relative to the Bethe-Bloch parametrization, in particular for
kaons and baryons with respect to pions.  An example is
shown in Fig.~\ref{fig:shifts} for the $dE/dx$ shifts of pions, kaons and protons
in a bin at $x_F$~=~0.1, as a function of transverse momentum together
with the variation of the relative width of the fitted $dE/dx$ distribution.

It is evident that the local variation as well as the difference in energy deposit
for the different particle species stay at the permille level of absolute energy
loss, and in the percent level for the width relative to the absolute prediction.

%
%
\subsection{Error estimation}
\vspace{3mm}
\label{sec:error}

Particle identification proceeds, in each chosen bin of phase
space, via a $\chi^2$ optimization procedure between the measured
truncated energy loss distribution and the sum of four single
particle $dE/dx$ distributions of known shape but a priori
unknown positions and widths for electrons, pions, kaons and
protons, respectively. Due to the small fraction of electrons
and their position on the density plateau of the energy loss
function, and due to the known dependence of the $dE/dx$
resolution on the $dE/dx$ value of each particle species \cite{bib:pp_paper},
the problem reduces in practice to the determination of eight
quantities: three positions, one width parameter and four
yield parameters which correspond to the predicted number of
particles. The statistical error of the four particle yields
thus obtained may be determined from the dependence of $\chi^2$
on all parameters (covariance matrix). It is to be noted that
the inverse square root of the predicted numbers for each
particle species is only a first approximation to the relative
statistical error of the yields. The fluctuations of the fitted
particle positions, Fig.~\ref{fig:shifts}, and their contributions to the
error of the yield parameters are intercorrelated with the
particle ratios and with the relative distances of the energy
deposits in the $dE/dx$ variable. The proper evaluation of the
covariance matrix thus gives the effective statistical fluctuation
of the yield parameters to be quoted as the experimental statistical
error. 

The method may be cross-checked using Monte Carlo methods by 
creating, in a given bin, statistically independent samples
using the yield parameters fitted to the experimental $dE/dx$
distribution as input and allowing for their proper statistical 
fluctuation.  Thus the assumption of a Gaussian parameter 
distribution used in the covariance matrix approach has been justified 
in particular also for phase space bins of small statistics or strongly 
correlated fit parameters.

It is interesting to compare the predicted relative statistical
error of the yield parameters to the inverse square root of the
fitted particle numbers. In case of "perfect" identification the
two figures should be equal; the square of their ratio 
determines how much more statistics the real detector should
collect in order to achieve the same precision as a "perfect" one.
As an example in Table~\ref{tab:example} the fitted yields, N, of pions, kaons
and protons in one bin at $x_F$~=~0.1 and $p_T$~=~0.5~GeV/c are given
together with the effective statistical error and the $1/\sqrt{N}$
value. The ratio of these two numbers is very close to one for
the prevailing pion samples. In contrast it amounts to 1.44 and
1.23 for kaons and protons, respectively. For negative particles
and in accordance with the inverted particle ratios, it is
larger for anti-protons (1.4) than for negative kaons (1.25).
Concerning the present work on proton and anti-proton cross sections
the mean factors are, averaged over all phase space bins, about
1.1 for protons and 1.3 for anti-protons. The statistical errors
given in the data tables, Sect.~\ref{sec:dtables}, correspond to the error
evaluation described above.

\begin{table}[h]
	\begin{center}
		\begin{tabular}{|c|ccc|ccc|}
		\hline
		                                                     & $\pi^+$ &   p       & K$^+$ & $\pi^-$ & $\overline{\textrm{p}}$ & K$^-$ \\ 
		 \hline
		 number of entries $N$                 &  28 388  &  6786  &  3088    & 20 851 &             1019                &  1917  \\
		 $1/\sqrt{N}$  [\%]                       &    0.594  &   1.21  &   1.80    &  0.693  &             3.13                 &  2.28   \\
		 $\sigma_{\textrm{stat}}$  [\%] &     0.605 &   1.50  &   2.60   &  0.701   &            4.38                 &  2.80   \\
		 \hline    
		\end{tabular}
	\end{center}
	\caption{Yields and statistical errors for protons, kaons and pions at $x_F$~=~0.1 and $p_T$~=~0.5~GeV/c }
	\label{tab:example}
\end{table}

Another, independent cross check of the validity of the evaluation
of the statistical errors is given by the two dimensional 
interpolation of the final cross sections described in Sect.~\ref {sec:data}. 
As this interpolation reduces the local statistical uncertainty
by a factor of between 3 and 4, the deviations of the data points from
the interpolated value in each bin should measure the real point
by point statistical fluctuation. In fact the compatibility of 
the distribution of the relative deviations shown in Fig.~\ref{fig:diff}
with an rms of unity confirms the correctness of the error estimate given above.   

%
%
\subsection{Estimation of K$^+$ and $\pi^+$ contributions in the extreme forward direction}
\vspace{3mm}
\label{sec:forw_rat}

\begin{figure}[h]
  \begin{center}
  	\includegraphics[width=15.6cm]{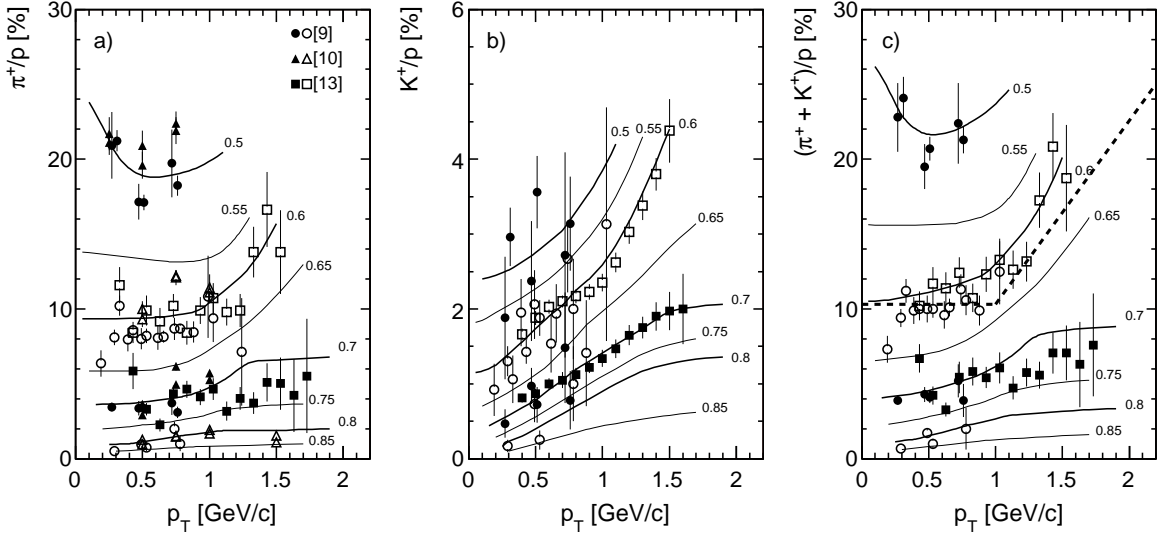}
  	\caption{Ratios a) $\pi^+$/p, b) K$^+$/p and c) ($\pi^+$+K$^+$)/p as a function of $p_T$ 
  	               in the forward direction. The $x_F$ values are indicated in the figure. Below the thick dashed line in panel c) 
  	               the tracking combination of GTPC and VPC was used}
  	\label{fig:ratios}
  \end{center}
\end{figure}

As the GTPC and VPC combination does not allow for particle 
identification via energy loss measurement, the proton extraction
in the region $x_F >$~0.6, see in Fig.~\ref{fig:accept}, has to rely on the
measurement of $\pi^+$/p and K$^+$/p ratios from other experiments.
In fact there are sufficient and mutually consistent data sets 
available to establish a reliable data base. The problem is
alleviated by the fact that particle ratios are relatively
stable against systematic errors of the different experiments
and that their absolute values decrease rapidly to a few percent
margin in the phase space region in question.
The situation is shown in Fig.~\ref{fig:ratios} in detail for $\pi^+$/p,
K$^+$/p and  ($\pi^+$+K$^+$)/p for different $x_F$ values as a function of
transverse momentum. In both cases the ratios obtained by NA49
\cite{bib:pp_paper,bib:ka_paper} overlap consistently with the other data sets.

The interpolated lines shown in Fig.~\ref{fig:ratios} have been used for the
determination of proton cross sections from the total positive
particle yields. The uncertainties connected with
this procedure have been taken into account by an increase of
the given statistical errors for the bins in question.

%
%
\section{Evaluation of Invariant Cross Sections and Corrections}
\vspace{3mm}
\label{sec:corr}

The experimental evaluation of the invariant cross section

\begin{equation}
  f(x_F,p_T) = E(x_F,p_T) \cdot \frac{d^3\sigma}{dp^3} (x_F,p_T)
\end{equation}
follows the methods described in \cite{bib:pp_paper}. The normalization and
the corrections are discussed below, concentrating on those
issues specific for baryon measurements.

%
%
\subsection{Empty target correction}
\vspace{3mm}
\label{sec:empty}

\begin{figure}[b]
  \begin{center}
  	\rotatebox{270}{\includegraphics[width=5.5cm]{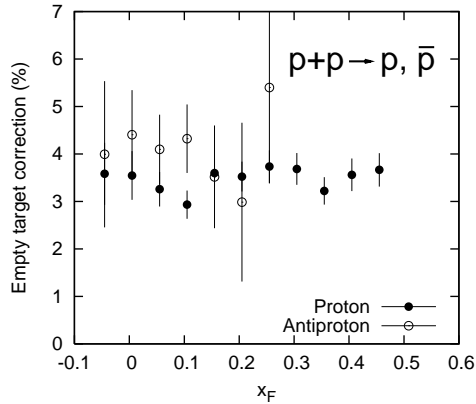}}
  	\caption{Empty target correction for protons and anti-protons (averaged over all $p_T$)}
  	\label{fig:empty}
  \end{center}
\end{figure}

The empty target background is treated as a correction factor
as described in \cite{bib:pp_paper} by determining the baryon yields in the
full and empty target samples and establishing their normalized
difference relative to the full target sample. The resulting 
correction factor is shown in Fig.~\ref{fig:empty}. It is, within the statistical
errors, $p_T$ and $x_F$ independent and is compatible with the one
given for pions \cite{bib:pp_paper}. The correction for neutrons is equal to
the one for protons.

%
%
\subsection{Trigger bias correction}
\vspace{3mm}
\label{sec:s4}

The interaction trigger uses a circular scintillator of 2~cm
diameter placed at a distance of 4~m from the target in
anti-coincidence (S4 counter in Fig.~\ref{fig:exp}). 
It accepts 89\% of the total inelastic cross section. The
majority of the vetoed events contain one fast proton in the small
S4 acceptance. As explained in detail in \cite{bib:pp_paper} this event loss creates
an $x_F$ and eventually $p_T$ dependent bias for the extracted data which
has to be carefully examined as it depends on short range and long
range correlations in the hadronic final state.

This trigger bias is determined by an off-line increase of the S4 radius.
With this method the limiting value of each measured cross section at zero 
radius may be obtained. The S4
radius increase is possible as all tracks in the corresponding
momentum region are detected via the GTPC+VPC+RCal combination
(Sect.~\ref{sec:na49}).

\begin{figure}[h]
  \begin{center}
  	\rotatebox{270}{\includegraphics[width=6.5cm]{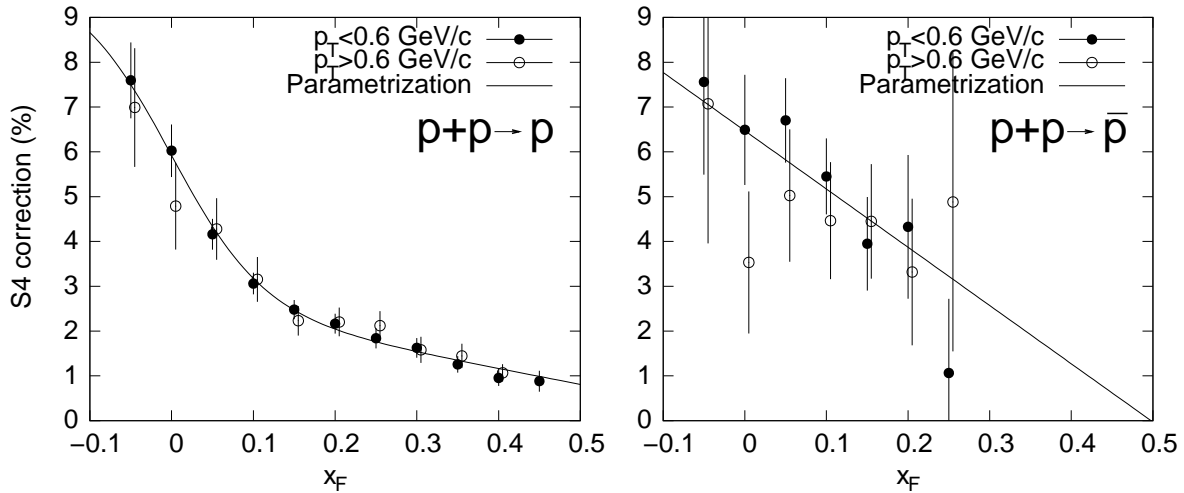}}
  	\caption{Trigger bias correction for protons (left) and anti-protons (right)}
  	\label{fig:s4}
  \end{center}
\end{figure}

Figure~\ref {fig:s4} shows the size of the correction as a function of $x_F$ for
protons and anti-protons. The results for two $p_T$ regions
demonstrate that the correction is within errors independent of
$p_T$. Anti-protons exhibit a different $x_F$ dependence, again $p_T$
independent within the (larger) statistical errors in this case.
The correction varies from the one for pions
in the forward hemisphere due to the different correlation between
leading protons and secondary baryons in the projectile fragmentation.
For neutrons, the trigger bias correction is equal to the one for protons. 

%
%
\subsection{Re-interaction and absorption}
\vspace{3mm}
\label{sec:reint}

\begin{figure}[h]
  \begin{center}
  	\includegraphics[width=8.5cm]{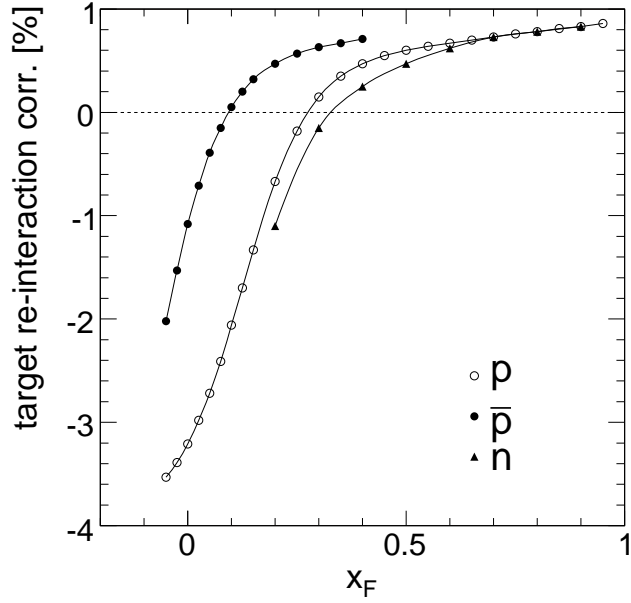}
  	\caption{Target re-interaction correction}
  	\label{fig:treint}
  \end{center}
\end{figure}

The re-interaction of baryons in the hydrogen target has been 
evaluated, as in the case of pions \cite{bib:pp_paper}, using the PYTHIA event 
generator. The corresponding corrections are shown in Fig.~\ref{fig:treint}.

The absorption of baryons by interaction with the detector
material has been elaborated based on the results for pions,
modifying the absorption length in accordance with the higher
baryonic interaction cross section.

%
%
\subsection{Feed-down from weak decays}
\vspace{3mm}
\label{sec:feed}

The determination of the contribution from weakly decaying baryons
($\Lambda$, $\Sigma$ and their anti-particles) is based on the methods
discussed in \cite{bib:pp_paper}. The parent particle input distributions are 
taken from published data and a subsequent Monte Carlo simulation
is used to estimate the on-vertex reconstruction efficiency for
baryonic daughters. 

\begin{figure}[h]
  \begin{center}
  	\includegraphics[width=15.5cm]{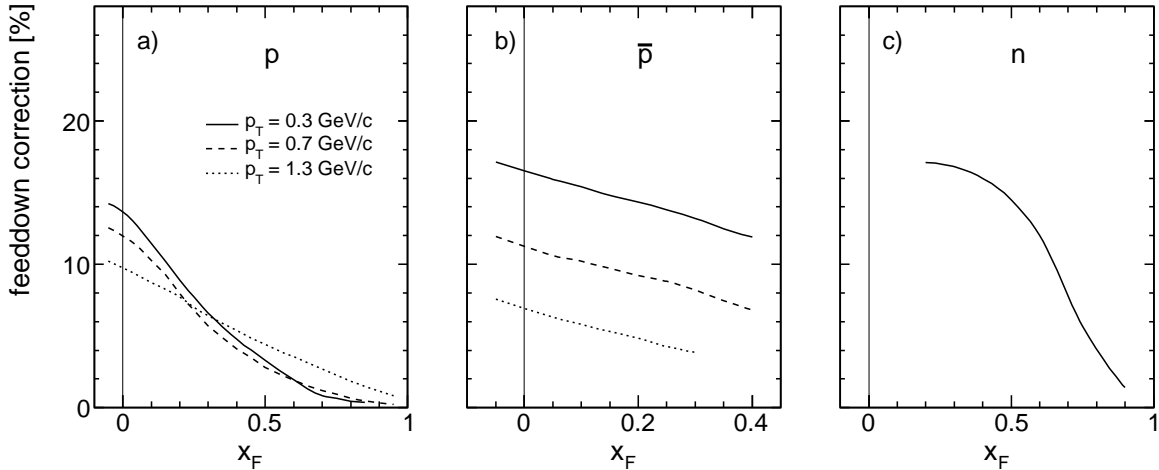}
  	\caption{Relative size of the feed-down correction for a) protons, b) anti-protons and c) neutrons}
  	\label{fig:fd}
  \end{center}
\end{figure}

As the decay baryons are close in mass to the parent hyperons,
they take up most of the parent momentum. Their distribution
over the measured phase space is therefore much wider than the
one for decay pions and extends over the complete $x_F$ and $p_T$
ranges. As shown in Fig.~\ref{fig:fd} this correction amounts to up to
15\% for protons and 20\% for anti-protons with $p_T$ dependences which 
are different for protons and anti-protons. For protons
at large $x_F$, where the $\Sigma^+$ contribution dominates the
feed-down, it even increases at large $p_T$.

The main systematic error source is the uncertainty of earlier
measurements, especially for $\Sigma^+$. For anti-protons, besides
the $\overline{\Lambda}$ also $\overline{\Sigma}^-$ contributes for which there are
no existing measurements. The yield of this particle was
estimated from general anti-baryon/ baryon ratio and isospin
arguments. To first order it was assumed that the $x_F$ and $p_T$
shapes are the same as for $\overline{\Lambda}$, and that the $\overline{\Sigma}^-$
to $\Sigma^+$ ratio is 80\% of the $\overline{\Lambda}/\Lambda$ ratio.
For neutrons the feed-down correction corresponds to the full relative yield from $\Lambda$
and $\Sigma$ decays, as shown in Fig.~\ref{fig:fd}c.
%
%
\subsection{Binning correction}
\vspace{3mm}
\label{sec:bin}

\begin{figure}[h]
  \begin{center}
  	\includegraphics[width=14cm]{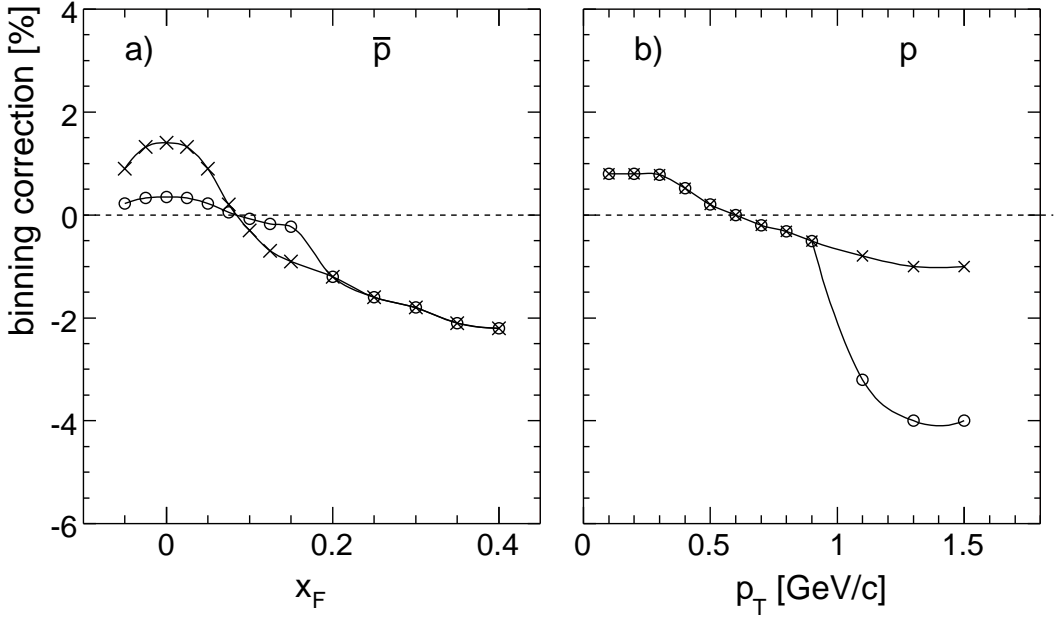}
  	\caption{Correction due to the binning in a) $x_F$ for $\overline{\textrm{p}}$ and b) $p_T$ for p. 
  	               The crosses represent the correction at fixed bin widths of 
                   $\Delta x_F$~= 0.05 and $\Delta p_T$~= 0.1~GeV/c, respectively, and the open
           circles describe the correction for the bins actually used}
  	\label{fig:bincor}
  \end{center}
\end{figure}

The effect of finite bin sizes on the extracted inclusive cross
sections was discussed in detail in \cite{bib:pp_paper} and shown to depend
on the second derivative of the $x_F$ or $p_T$ distributions. Due to
the approximately linear rather than exponential $x_F$ distribution
of protons, the binning effects can in fact be neglected in
longitudinal direction for the modest bin widths chosen.
Also in transverse direction, due to the larger mean transverse
momentum of baryons, the effect is smaller than for pions. As shown
in Fig. \ref{fig:bincor} it reaches values in excess of 1\% only at large $p_T$ due to
the bin width of 0.2~GeV/c in this region.

%
%
\subsection{Systematic errors}
\vspace{3mm}
\label{sec:sys}

\begin{figure}[h]
  \begin{center}
  	\rotatebox{270}{\includegraphics[width=8cm]{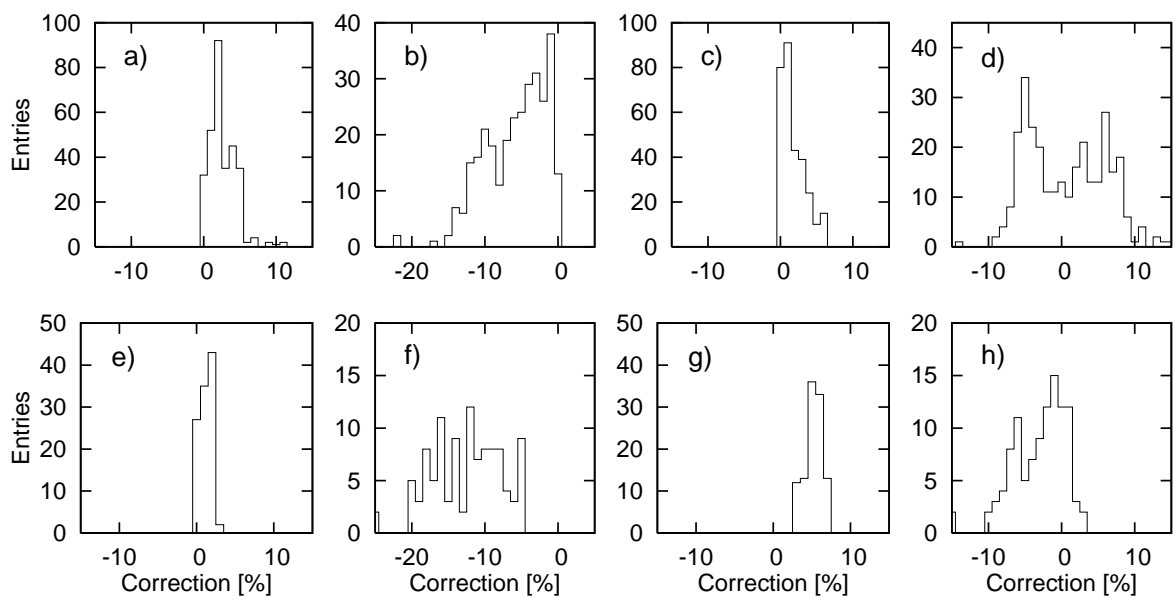}}
  	\caption{Distribution of corrections for protons (upper four panels) and anti-protons
  	               (lower four panels); a) and e) detector absorption, b) and f) feed-down, c) and g) trigger bias
  	               and d) and h) total}
  	\label{fig:syst}
  \end{center}
\end{figure}

The systematic errors of the extracted cross sections are given
by the normalization procedure and the uncertainties of the
applied corrections. These contributions are estimated in
Table~\ref{tab:syst}. They are governed by the fluctuation of the detector
absorption, feed-down and trigger bias corrections which are
shown in Fig.~\ref{fig:syst} over all phase space bins, for protons and
anti-protons. 

\begin{table}[h] 
\footnotesize
\begin{center}
\begin{tabular}{|lccc|lcc|}
\hline
                                           &        &  p        & $\overline{\textrm{p}}$  &                                               &  &   n      \\ \hline
Normalization                      &      &  1.5\%   & 1.5\%                                & Normalization                        &  &  1.5\%        \\ 
Tracking efficiency               &        & 0.5\%  & 0.5\%                                &                                               &   &                 \\
Trigger bias                          &        &  0.5\%  & 1.0\%                                & Trigger bias                           &  &  1\%         \\ 
Feed-down                           &       & 1.5\%    & 2.5\%                                & Feed-down                            &  &  3\%          \\ 
Detector absorption             & \multirow{3}{0.mm}{\large $\biggr\}$}&  &
                                                                    & Detector absorption             & \multirow{3}{0.mm}{\large $\biggr\}$}&    \\
Target re-interaction             &        &   0.5 -- 1.5\%  &     1.0\%            & Target re-interaction             &    &   0.5 -- 1.5\%   \\
Binning correction                 &        &             &                                          & Binning correction                 &    &      \\  
                                              &        &             &                                          & Acceptance                           &   &   0 -- 2\%   \\
                                              &        &             &                                          & Energy scale error                &   &    4 -- 8\%   \\
                                              &        &             &                                          & Energy resolution unfolding  &   &   3 -- 8\%   \\
                                              &        &             &                                          & Charged veto efficiency       &   &    2 -- 3\%  \\
                                              &        &             &                                          & Cluster overlap                     &   &    2\%     \\
                                              &        &             &                                          & Hadron identification            &   &     2 -- 5\%   \\
                                              &        &             &                                          & K$_L^0$ contribution           &   &     0 -- 3\%    \\ 
                                              \hline
Total (upper limit)                   &    & 5.0\%      & 6.5\%                                & Total (upper limit)                  &    &   28\%   \\ 
Total (quadratic sum)             &     & 2.5\%     & 3.3\%                                & Total (quadratic sum)            &    &    10\%  \\
 \hline
\end{tabular}
\end{center}
\caption{Summary of systematic errors}
\label{tab:syst}
\end{table}

With a linear sum of 5.0\% and 6.5\%, respectively, for protons and
anti-protons, and quadratic sums of 2.5\% and 3.3\% they are only
slightly larger than the ones estimated for pion production \cite{bib:pp_paper}.
The larger systematic uncertainty of the neutron yields reflects the 
difficulties inherent in hadronic calorimetry as compared to charged track detection.

%
%
\section{Results on double differential cross sections}
\vspace{3mm}
\label{sec:data}
%
%
\subsection{Data tables}
\vspace{3mm}
\label{sec:dtables}

The binning scheme presented in Sect.~\ref{sec:bin_scheme} results in 333 and 143
data values for protons and anti-protons, respectively. These
are presented in Tables~\ref{tab:prot_cs} and \ref{tab:aprot_cs}.

%
%
\begin{table}[b]
\renewcommand{\tabcolsep}{0.14pc} 
\renewcommand{\arraystretch}{1.0}
\scriptsize
\begin{center}
\begin{tabular}{|c|cr|cr|cr|cr|cr|cr|cr|cr|cr|}
\hline
\multicolumn{19}{|c|}{$f(x_F,p_T), \Delta f$} \\ 
\hline
$p_T \backslash x_F$ & \multicolumn{2}{|c|}{-0.05} & \multicolumn{2}{|c|}{-0.025} & \multicolumn{2}{|c|}{0.0} & \multicolumn{2}{|c|}{0.025} & \multicolumn{2}{|c|}{0.05} & \multicolumn{2}{|c|}{0.075} & \multicolumn{2}{|c|}{0.1} & \multicolumn{2}{|c|}{0.125} & \multicolumn{2}{|c|}{0.15}\\ \hline
0.1 & 2.349&2.76 & 2.161&1.90 & 2.121&1.78 & 2.128&1.75 & 2.224&1.69 & 2.340&1.76 & 2.620&1.83 & 3.039&1.55 & 3.479&1.76\\
0.2 & 2.158&3.53 & 1.921&1.60 & 1.837&1.46 & 1.854&1.42 & 1.901&1.44 & 2.095&1.49 & 2.349&1.50 & 2.666&1.36 & 2.998&1.49\\
0.3 & 1.664&3.72 & 1.570&2.63 & 1.631&1.81 & 1.575&1.41 & 1.664&1.26 & 1.750&1.31 & 1.935&1.33 & 2.179&1.34 & 2.549&1.28\\
0.4 & 1.297&4.24 & 1.286&2.74 & 1.287&2.26 & 1.238&1.71 & 1.334&1.37 & 1.379&1.43 & 1.486&1.52 & 1.742&1.49 & 1.880&1.52\\
0.5 & 1.094&3.96 & 0.951&3.25 & 0.977&2.45 & 0.929&1.98 & 0.973&1.47 & 1.057&1.45 & 1.161&1.52 & 1.224&1.56 & 1.353&1.72\\
0.6 & 0.707&4.38 & 0.689&3.62 & 0.645&3.24 & 0.688&2.46 & 0.712&1.98 & 0.709&1.78 & 0.807&1.67 & 0.884&1.74 & 0.948&1.91\\
0.7 & 0.493&5.38 & 0.451&4.31 & 0.449&4.07 & 0.494&2.76 & 0.482&2.47 & 0.533&2.15 & 0.526&1.93 & 0.601&1.78 & 0.654&2.11\\
0.8 & 0.378&6.13 & 0.357&5.10 & 0.314&4.45 & 0.322&3.52 & 0.329&3.19 & 0.3386&2.95 & 0.3686&2.42 & 0.3984&2.42 & 0.430&2.48\\
0.9 & 0.244&6.33 & 0.203&6.38 & 0.233&5.44 & 0.2196&4.49 & 0.2274&3.90 & 0.2399&3.46 & 0.2572&3.19 & 0.2503&3.33 & 0.2694&2.71\\
1.1 & 0.0956&4.75 && & 0.0899&4.28 && & 0.0927&3.14 && & 0.0988&2.76 && & 0.1078&2.42\\
1.3 & 0.0445&7.71 && & 0.0366&6.10 && & 0.0384&4.67 && & 0.0410&4.15 && & 0.0398&3.64\\
1.5 & 0.0199&9.72 && & 0.0173&8.74 && & 0.0188&6.90 && & 0.0161&6.48 && & 0.0163&6.46\\
1.7 & 0.0087&14.2 && & 0.00640&13.4 && & 0.00617&11.3 && & 0.00705&9.45 && & 0.00607&9.60\\
1.9 && && & 0.00284&14.3 && && && & 0.00296&10.3 && &&\\
\hline
$p_T \backslash x_F$ & \multicolumn{2}{|c|}{0.2} & \multicolumn{2}{|c|}{0.25} & \multicolumn{2}{|c|}{0.3} & \multicolumn{2}{|c|}{0.35} & \multicolumn{2}{|c|}{0.4} & \multicolumn{2}{|c|}{0.45} & \multicolumn{2}{|c|}{0.5} & \multicolumn{2}{|c|}{0.55}& \multicolumn{2}{|c|}{0.6}\\ \hline
0.05 && && && && && && && &17.1&7.46 &16.8&14.6\\
0.1 & 4.654&1.53 & 5.71&1.83 & 6.64&3.44 & 7.749&1.28 & 8.92&1.58 &10.02&1.57 &11.68&1.65 &14.38&4.31 &15.07&6.53\\
0.15 && && && && && && && &13.64&3.21 &13.67&4.24\\
0.2 & 4.086&1.24 & 5.132&1.44 & 6.10&2.54 & 6.958&0.96 & 7.500&1.21 & 8.52&1.22 & 9.59&1.24 &11.92&2.96 &12.80&2.94\\
0.25 && && && && && && && & 9.95&2.37 &11.55&2.76\\
0.3 & 3.265&1.18 & 4.231&1.63 & 5.016&1.70 & 5.876&0.83 & 6.261&1.08 & 6.766&1.14 & 7.232&1.16 & 8.78&2.30 & 9.61&2.25\\
0.35 && && && && && && && & 7.30&2.34 & 7.67&2.33\\
0.4 & 2.518&1.29 & 3.145&1.74 & 3.904&1.87 & 4.561&0.89 & 4.863&1.06 & 5.098&1.10 & 5.350&1.14 & 6.09&2.06 & 6.21&2.11\\
0.45 && && && && && && && & 4.84&2.19 & 5.03&2.21\\
0.5 & 1.815&1.28 & 2.292&1.58 & 2.783&1.57 & 3.341&0.93 & 3.609&1.08 & 3.772&1.17 & 3.781&1.21 & 3.908&2.31 & 3.938&2.36\\
0.6 & 1.247&1.43 & 1.516&1.96 & 1.957&2.27 & 2.224&1.10 & 2.544&1.18 & 2.628&1.31 & 2.649&1.34 & 2.709&1.80 & 2.738&1.84\\
0.7 & 0.798&1.65 & 0.989&2.09 & 1.252&2.01 & 1.489&1.32 & 1.652&1.35 & 1.771&1.40 & 1.753&1.53 & 1.741&2.08 & 1.692&2.17\\
0.8 & 0.5092&1.71 & 0.620&1.94 & 0.756&2.35 & 0.894&1.59 & 1.021&1.61 & 1.140&1.61 & 1.127&1.71 & 1.128&2.43 & 1.028&2.61\\
0.9 & 0.3201&2.08 & 0.3672&2.38 & 0.458&2.63 & 0.548&1.92 & 0.605&1.94 & 0.676&1.97 & 0.689&2.11 & 0.672&2.91 & 0.614&3.12\\
1.1 & 0.1197&2.26 & 0.1388&2.42 & 0.1535&2.83 & 0.1657&2.22 & 0.1970&2.14 & 0.2071&2.25 & 0.2172&2.29 & 0.2124&3.43 & 0.1944&3.67\\
1.3 & 0.0439&3.72 & 0.0485&3.71 & 0.0527&4.54 & 0.0584&3.48 & 0.0587&3.63 & 0.0602&3.78 & 0.0629&3.94 & 0.0590&6.28 & 0.0567&6.57\\
1.5 & 0.0176&5.77 & 0.0156&6.43 & 0.0177&8.02 & 0.0207&5.42 & 0.0187&6.07 & 0.0182&6.38 & 0.0165&7.05 & 0.0153&11.4 & 0.0142&12.1\\
1.7 & 0.00578&8.75 & 0.00614&8.95 & 0.00628&9.37 & 0.00653&9.44 & 0.00709&9.45 & 0.00592&10.3 & 0.00549&17.8 & 0.0059&17.5 & 0.00235&27.6\\
1.9 & 0.00313&7.98 && & 0.00245&10.6 && & 0.00242&10.8 && & 0.00161&21.2 && & 0.00082&29.4\\
\hline
$p_T \backslash x_F$ & \multicolumn{2}{|c|}{0.65} & \multicolumn{2}{|c|}{0.7} & \multicolumn{2}{|c|}{0.75} & \multicolumn{2}{|c|}{0.8} & \multicolumn{2}{|c|}{0.85} & \multicolumn{2}{|c|}{0.9} & \multicolumn{2}{|c|}{0.95}\\ \cline{1-15}
0.05  && && && && && && &&\\
0.1 &18.5&10.1 && && && && && &&\\
0.15 &14.09&5.64 &14.0&9.72 && && && && &&\\
0.2 &12.68&3.88 &14.74&4.94 &15.0&8.35 && && && &&\\
0.25 &11.26&3.20 &11.75&3.70 &14.9&7.49 &10.61&9.23 && && &&\\
0.3 & 9.74&2.83 &10.86&3.06 & 9.76&5.11 &10.00&8.62 &12.5&8.11 && &&\\
0.35 & 8.09&2.34 & 7.34&3.09 & 7.93&3.42 & 8.48&4.52 &10.18&4.94 &12.39&7.76 &&\\
0.4 & 6.55&2.42 & 6.19&2.58 & 6.47&3.53 & 6.46&3.67 & 7.71&4.64 & 9.00&5.07 &&\\
0.45 & 5.13&2.24 & 5.03&2.70 & 4.93&2.82 & 5.08&3.20 & 5.43&3.91 & 6.78&4.80 &&\\
0.5 & 4.028&2.41 & 3.930&2.52 & 3.752&2.65 & 4.05&2.63 & 3.55&3.36 & 4.98&3.21 &&\\
0.6 & 2.592&1.95 & 2.507&2.04 & 2.322&2.18 & 2.247&2.28 & 2.238&2.35 & 2.464&2.28 & 4.954&1.81\\
0.7 & 1.619&2.28 & 1.581&2.38 & 1.436&2.57 & 1.288&2.78 & 1.278&2.86 & 1.324&2.87 & 2.339&2.20\\
0.8 & 0.984&2.75 & 0.973&2.84 & 0.886&3.06 & 0.768&3.35 & 0.686&3.66 & 0.711&3.68 & 1.130&2.97\\
0.9 & 0.609&3.22 & 0.557&3.45 & 0.506&3.72 & 0.455&4.05 & 0.396&4.44 & 0.368&4.74 & 0.544&3.97\\
1.1 & 0.1942&3.77 & 0.1757&4.07 & 0.1618&4.36 & 0.1334&4.92 & 0.1338&5.05 & 0.1185&5.50 & 0.1292&5.42\\
1.3 & 0.0442&7.55 & 0.0392&8.24 & 0.0463&7.87 & 0.0350&9.24 & 0.0364&9.32 & 0.0404&9.08 & 0.0447&8.89\\
1.5 & 0.0147&12.3 & 0.0136&13.0 & 0.0109&14.9 & 0.0084&17.4 & 0.0108&15.9 & 0.0131&14.8 & 0.0147&14.3\\
1.7 & 0.00437&20.9 & 0.00421&22.1 & 0.00295&26.1 & 0.0045&22.4 & 0.00293&28.0 & 0.00379&25.9 & 0.0051&23.0\\
1.9 && & 0.00076&32.3 && & 0.00119&28.4 && & 0.00083&36.1 &&\\
\cline{1-15}
\end{tabular}
\end{center}
\caption{Invariant cross section, $f(x_F,p_T)$, in mb/(GeV$^2$/c$^3$) for protons in p+p collisions at 158~GeV/c 
               beam momentum. The relative statistical errors, $\Delta f$, are given in \%} 
\label{tab:prot_cs}
\end{table}
\clearpage

\begin{table}
\renewcommand{\tabcolsep}{0.4pc} 
\renewcommand{\arraystretch}{1.0}
\scriptsize
\begin{center}
\begin{tabular}{|c|cr|cr|cr|cr|cr|cr|cr|}
\hline
\multicolumn{15}{|c|}{$f(x_F,p_T), \Delta f$} \\ 
\hline
$p_T \backslash x_F$ & \multicolumn{2}{|c|}{-0.05} & \multicolumn{2}{|c|}{-0.025} & \multicolumn{2}{|c|}{0.0} & \multicolumn{2}{|c|}{0.025} & \multicolumn{2}{|c|}{0.05} & \multicolumn{2}{|c|}{0.075} & \multicolumn{2}{|c|}{0.1}\\ \hline
0.1 & 0.563&6.71 & 0.590&4.28 & 0.581&3.66 & 0.545&3.67 & 0.499&3.87 & 0.489&4.24 & 0.391&4.52\\
0.2 & 0.508&6.15 & 0.505&3.55 & 0.509&2.98 & 0.518&2.92 & 0.473&3.12 & 0.405&3.69 & 0.379&4.04\\
0.3 & 0.391&6.75 & 0.450&5.52 & 0.450&3.88 & 0.404&2.99 & 0.401&2.79 & 0.353&3.12 & 0.320&3.56\\
0.4 & 0.287&7.61 & 0.338&6.23 & 0.324&4.90 & 0.311&3.61 & 0.2848&3.17 & 0.2918&3.38 & 0.237&4.38\\
0.5 & 0.222&8.16 & 0.260&6.77 & 0.230&5.48 & 0.262&4.31 & 0.2324&3.22 & 0.2031&3.52 & 0.1761&4.38\\
0.6 & 0.163&8.52 & 0.178&8.99 & 0.178&6.67 & 0.1587&5.37 & 0.1511&4.82 & 0.1211&4.82 & 0.1255&4.72\\
0.7 & 0.117&9.82 & 0.103&10.4 & 0.1102&8.17 & 0.1168&6.29 & 0.1141&5.46 & 0.0929&5.56 & 0.0830&4.71\\
0.8 & 0.0534&15.7 & 0.0862&9.83 & 0.0845&8.84 & 0.0773&7.93 & 0.0637&7.22 & 0.0614&7.24 & 0.0600&6.58\\
0.9 & 0.0432&15.5 & 0.0427&13.9 & 0.0481&12.2 & 0.0409&10.7 & 0.0385&9.07 & 0.0441&8.11 & 0.0363&8.59\\
1.1 & 0.0155&12.3 && & 0.0194&8.12 && & 0.0153&7.39 && & 0.0142&7.32\\
1.3 & 0.0095&14.7 && & 0.00532&16.9 && & 0.00515&13.3 && & 0.00521&12.6\\
1.5 & 0.00272&27.1 && & 0.00258&23.8 && & 0.00241&20.2 && & 0.00179&21.4\\
\hline
\hline
$p_T \backslash x_F$ & \multicolumn{2}{|c|}{0.125} & \multicolumn{2}{|c|}{0.15} & \multicolumn{2}{|c|}{0.2} & \multicolumn{2}{|c|}{0.25} & \multicolumn{2}{|c|}{0.3} & \multicolumn{2}{|c|}{0.35} & \multicolumn{2}{|c|}{0.4}\\ \hline
0.1 & 0.355&4.98 & 0.260&7.36 & 0.194&7.09 & 0.115&11.5 & 0.074&18.1 & 0.0418&18.9 & 0.0173&55.1\\
0.2 & 0.300&4.49 & 0.275&5.57 & 0.183&7.15 & 0.0900&9.63 & 0.0613&14.4 && &&\\
0.3 & 0.255&4.53 & 0.199&5.84 & 0.1654&4.69 & 0.0941&8.96 & 0.0429&14.7 & 0.0270&13.8 & 0.0137&29.5\\
0.4 & 0.203&4.98 & 0.165&6.14 & 0.1154&7.52 & 0.0645&10.5 & 0.0425&12.5 && &&\\
0.5 & 0.1350&5.55 & 0.1231&5.73 & 0.0860&5.76 & 0.0463&10.3 & 0.0266&13.9 & 0.0161&13.6 & 0.0082&25.9\\
0.6 & 0.1005&6.69 & 0.0787&6.64 & 0.0578&8.62 & 0.0427&8.40 & 0.0190&14.9 && &&\\
0.7 & 0.0705&5.72 & 0.0596&8.66 & 0.0443&9.05 & 0.0191&14.2 & 0.0162&14.7 & 0.0092&16.4 & 0.0062&21.8\\
0.8 & 0.0382&9.29 & 0.0346&10.8 & 0.0235&8.27 & 0.0151&14.8 & 0.0132&15.0 && &&\\
0.9 & 0.0275&10.9 & 0.0242&10.0 & 0.0190&11.5 & 0.0141&11.7 & 0.0061&21.7 & 0.00435&20.6 & 0.00189&38.0\\
1.1 && & 0.01008&9.15 & 0.00796&8.63 & 0.00447&16.1 & 0.00230&22.1 & 0.00182&27.0 & 0.00073&54.0\\
1.3 && & 0.00403&15.5 & 0.00255&15.2 & 0.00152&21.7 & 0.00128&27.1 && &&\\
1.5 && & 0.00127&22.8 & 0.00068&30.4 & 0.00057&34.5 && && &&\\
\hline
\end{tabular}
\end{center}
\caption{Invariant cross section, $f(x_F,p_T)$, in mb/(GeV$^2$/c$^3$) for anti-protons in p+p collisions at 
              158~GeV/c beam momentum. The relative statistical errors, $\Delta f$, are given in \%}
\label{tab:aprot_cs}
\end{table}%
%
\subsection{Extension of the data to high $\bf x_F$ and low $\bf p_T$}
\vspace{3mm}
\label{sec:high_xf}

As shown in Sect.~\ref{sec:na49} the NA49 detector acceptance is limited 
at large $x_F$ and low $p_T$ by the necessity of using an interaction
trigger, vetoing through-going beam tracks. The corresponding
acceptance gap extends from $p_T <$~0.05 at $x_F$~=~0.65 to $p_T <$~0.6 at
$x_F$~=~0.95, see Fig.~\ref{fig:cov}. In order to maintain the possibility of
precise $p_T$ integration in this phase space region it is mandatory
to use data from other experiments to supplement the NA49 results.
Fortunately there are data from seven different experiments,
all conducted at Fermilab in the years 1973 to 1982
\cite{bib:sannes,bib:childress,bib:scham,bib:akimov,bib:chapman,bib:whitmore,bib:brenner}
in exactly this region which also partially overlap with the NA49
data. These data come from internal target \cite{bib:sannes,bib:childress,bib:scham,bib:akimov}
and bubble chamber experiments \cite{bib:chapman,bib:whitmore}, all performed in the target region 
at low proton lab momenta, and from a spectrometer experiment
\cite{bib:brenner} in the forward hemisphere. If applicable the data have been
transformed from the coordinate pair momentum transfer $t$ and missing mass into the
$p_T$ and $x_F$ coordinates, interpolated to the $x_F$ values defined
by the NA49 binning scheme and corrected for $s$-dependence. This
latter correction will be quantified in section 10 below. In 
total 123 data points are thus available as given in Table~\ref{tab:extension}.

\begin{table}
\renewcommand{\tabcolsep}{0.2pc} 
\renewcommand{\arraystretch}{1.0}
\scriptsize
\begin{center}
\begin{tabular}{|cr@{}lr@{}lc@{\hspace{4.5mm}}|cr@{}lr@{}lc@{\hspace{4.5mm}}|
                cr@{}lr@{}lc@{\hspace{4.5mm}}|cr@{}lr@{}lc@{\hspace{4.5mm}}|cr@{}lr@{}lc|}
\hline
$p_T$ & \multicolumn{2}{c}{$f$} & \multicolumn{2}{c}{$\Delta f$} & ref &
$p_T$ & \multicolumn{2}{c}{$f$} & \multicolumn{2}{c}{$\Delta f$} & ref &
$p_T$ & \multicolumn{2}{c}{$f$} & \multicolumn{2}{c}{$\Delta f$} & ref &
$p_T$ & \multicolumn{2}{c}{$f$} & \multicolumn{2}{c}{$\Delta f$} & ref &
$p_T$ & \multicolumn{2}{c}{$f$} & \multicolumn{2}{c}{$\Delta f$} & ref \\ \hline
\multicolumn{6}{|c|}{$x_F$ = 0.6} & \multicolumn{6}{c|}{$x_F$ = 0.65} &
\multicolumn{6}{c|}{$x_F$ = 0.7} & \multicolumn{6}{c|}{$x_F$ = 0.75} &
\multicolumn{6}{c|}{$x_F$ = 0.8} \\ \hline 
0.224 & 14&.10 & 15&.0 & 7 & 0.224 & 15&.30 & 15&.0 & 7 & 0.224 & 16&.60 & 15&.0 & 7 &
0.224 & 16&.90 & 15&.0 & 7 & 0.478 &  4&.70 &  2&.0 & 3 \\ 
0.381 &  5&.89 & 15&.0 & 7 & 0.381 &  6&.46 & 15&.0 & 7 & 0.381 &  6&.56 & 15&.0 & 7 &
0.381 &  6&.94 & 15&.0 & 7 & 0.570 &  2&.70 &  2&.0 & 3 \\ 
0.540 &  3&.61 & 15&.0 & 7 & 0.540 &  3&.80 & 15&.0 & 7 & 0.540 &  3&.89 & 15&.0 & 7 &
0.540 &  3&.71 & 15&.0 & 7 &       &   &    &   &   &   \\ 
0.707 &  1&.35 & 15&.0 & 7 & 0.707 &  1&.23 & 15&.0 & 7 & 0.707 &  1&.25 & 15&.0 & 7 &
0.707 &  1&.24 & 15&.0 & 7 & 0.224 & 16&.20 & 15&.0 & 7 \\ 
      &   &    &   &   &   &       &   &    &   &   &   &       &   &    &   &   &   &
      &   &    &   &   &   & 0.381 &  7&.41 & 15&.0 & 7 \\ 
0.200 & 13&.34 &  3&.3 & 9 &       &   &    &   &   &   & 0.300 & 10&.45 &  2&.0 & 9 &     
      &   &    &   &   &   & 0.540 &  3&.52 & 15&.0 & 7 \\ 
0.300 &  9&.43 &  2&.0 & 9 &       &   &    &   &   &   & 0.500 &  4&.04 &  2&.4 & 9 &
      &   &    &   &   &   & 0.707 &  1&.14 & 15&.0 & 7 \\
0.400 &  6&.31 &  2&.6 & 9 &       &   &    &   &   &   & 0.750 &  1&.32 &  4&.4 & 9 &
      &   &    &   &   &   &       &   &    &   &   &   \\ 
0.500 &  4&.14 &  2&.3 & 9 &       &   &    &   &   &   & 0.500 &  4&.06 &  1&.0 & 9 &
      &   &    &   &   &   & 0.300 & 11&.67 &  1&.4 & 9 \\
0.625 &  2&.42 &  2&.7 & 9 &       &   &    &   &   &   & 0.750 &  1&.26 &  1&.6 & 9 &
      &   &    &   &   &   & 0.500 &  4&.10 &  1&.5 & 9 \\
0.750 &  1&.41 &  3&.3 & 9 &       &   &    &   &   &   &       &   &    &   &   &   &
      &   &    &   &   &   & 0.750 &  1&.00 &  2&.9 & 9 \\
0.300 &  8&.82 &  5&.4 & 9 &       &   &    &   &   &   &       &   &    &   &   &   &
      &   &    &   &   &   & 0.500 &  3&.90 &  1&.3 & 9 \\
0.400 &  5&.93 &  1&.0 & 9 &       &   &    &   &   &   &       &   &    &   &   &   &
      &   &    &   &   &   & 0.750 &  1&.00 &  2&.0 & 9 \\
0.500 &  4&.03 &  1&.6 & 9 &       &   &    &   &   &   &       &   &    &   &   &   &
      &   &    &   &   &   &       &   &    &   &   &   \\ 
0.625 &  2&.40 &  1&.2 & 9 &       &   &    &   &   &   &       &   &    &   &   &   &
      &   &    &   &   &   &       &   &    &   &   &   \\ 
0.750 &  1&.41 &  1&.0 & 9 &       &   &    &   &   &   &       &   &    &   &   &   &
      &   &    &   &   &   &       &   &    &   &   &   \\ \hline
\multicolumn{6}{|c|}{$x_F$ = 0.85} & \multicolumn{6}{c|}{$x_F$ = 0.9} &
\multicolumn{6}{c|}{$x_F$ = 0.95} & \multicolumn{6}{c|}{$x_F$ = 0.975} & 
\multicolumn{6}{c|}{}\\ \hline 
0.511 &  4&.18 &  2&.0 & 3 & 0.537 &  3&.91 &  2&.0 & 3 & 0.182 & 34&.10 &  5&.0 & 4 &
0.188 & 44&.51 &  5&.0 & 4 &       &   &    &   &   &  \\
0.602 &  2&.33 &  2&.0 & 3 & 0.629 &  2&.12 &  2&.0 & 3 & 0.246 & 23&.35 &  5&.0 & 4 &
0.253 & 31&.24 &  5&.0 & 4 &       &   &    &   &   &  \\
      &   &    &   &   &   &       &   &    &   &   &   & 0.299 & 21&.15 &  5&.0 & 4 &
0.302 & 30&.31 &  5&.0 & 4 &       &   &    &   &   &  \\      
0.190 & 20&.83 &  5&.0 & 4 & 0.157 & 28&.41 &  5&.0 & 4 & 0.337 & 18&.94 &  5&.0 & 4 &
0.344 & 28&.09 &  5&.0 & 4 &       &   &    &   &   &  \\  
0.245 & 16&.73 &  5&.0 & 4 & 0.225 & 20&.83 &  5&.0 & 4 & 0.375 & 14&.52 &  5&.0 & 4 &
0.384 & 21&.79 &  5&.0 & 4 &       &   &    &   &   &  \\
0.290 & 15&.47 &  5&.0 & 4 & 0.275 & 17&.68 &  5&.0 & 4 & 0.409 & 12&.15 &  5&.0 & 4 &
0.416 & 20&.20 &  5&.0 & 4 &       &   &    &   &   &  \\
0.328 & 12&.00 &  5&.0 & 4 & 0.318 & 16&.10 &  5&.0 & 4 &       &   &    &   &   &   &
      &   &    &   &   &   &       &   &    &   &   &  \\
0.363 &  9&.63 &  5&.0 & 4 & 0.355 & 12&.31 &  5&.0 & 4 & 0.224 & 33&.30 & 15&.0 & 7 &
0.224 & 53&.20 & 15&.0 & 7 &       &   &    &   &   &  \\
      &   &    &   &   &   & 0.389 & 10&.10 &  5&.0 & 4 & 0.381 & 14&.44 & 15&.0 & 7 &
0.381 & 25&.18 & 15&.0 & 7 &       &   &    &   &   &  \\      
0.224 & 17&.20 & 15&.0 & 7 &       &   &    &   &   &   & 0.540 &  6&.84 & 15&.0 & 7 &
0.540 & 12&.16 & 15&.0 & 7 &       &   &    &   &   &  \\
0.381 &  7&.80 & 15&.0 & 7 & 0.224 & 21&.40 & 15&.0 & 7 & 0.707 &  1&.60 & 15&.0 & 7 &
0.707 &  3&.14 & 15&.0 & 7 &       &   &    &   &   &  \\
0.540 &  3&.32 & 15&.0 & 7 & 0.381 &  9&.12 & 15&.0 & 7 &       &   &    &   &   &   &
      &   &    &   &   &   &       &   &    &   &   &  \\
0.707 &  1&.05 & 15&.0 & 7 & 0.540 &  3&.52 & 15&.0 & 7 & 0.179 & 43&.10 &  2&.0 & 5 &
0.186 & 69&.80 &  2&.0 & 5 &       &   &    &   &   &  \\
      &   &    &   &   &   & 0.707 &  1&.06 & 15&.0 & 7 & 0.263 & 24&.20 &  2&.0 & 5 &
0.269 & 41&.30 &  2&.0 & 5 &       &   &    &   &   &  \\      
0.210 & 16&.70 &  5&.0 & 5 &       &   &    &   &   &   & 0.350 & 17&.90 &  2&.0 & 5 &
0.357 & 30&.20 &  2&.0 & 5 &       &   &    &   &   &  \\ 
0.302 & 11&.80 &  5&.0 & 5 & 0.154 & 29&.50 &  6&.0 & 5 & 0.430 & 11&.00 &  2&.0 & 5 &
0.438 & 19&.70 &  2&.0 & 5 &       &   &    &   &   &  \\
0.384 &  7&.47 &  5&.0 & 5 & 0.244 & 17&.70 &  6&.0 & 5 &       &   &    &   &   &   &
      &   &    &   &   &   &       &   &    &   &   &  \\
      &   &    &   &   &   & 0.330 & 12&.80 &  4&.0 & 5 & 0.110 & 41&.30 &  5&.9 & 6 &
0.119 & 58&.20 &  5&.0 & 6 &       &   &    &   &   &  \\      
      &   &    &   &   &   & 0.410 &  7&.87 &  4&.0 & 5 & 0.212 & 30&.20 &  5&.8 & 6 &
0.220 & 49&.10 &  4&.0 & 6 &       &   &    &   &   &  \\      
      &   &    &   &   &   &       &   &    &   &   &   &       &   &    &   &   &   &
      &   &    &   &   &   &       &   &    &   &   &  \\      
      &   &    &   &   &   & 0.190 & 25&.70 &  4&.0 & 6 & 0.160 & 46&.23 & 15&.0 & 8 & 
0.160 & 67&.60 & 15&.0 & 8 &       &   &    &   &   &  \\           
      &   &    &   &   &   &       &   &    &   &   &   & 0.316 & 19&.50 & 15&.0 & 8 &
0.316 & 42&.65 & 15&.0 & 8 &       &   &    &   &   &  \\            
      &   &    &   &   &   & 0.160 & 29&.17 & 15&.0 & 8 & 0.447 & 10&.07 & 15&.0 & 8 &
0.447 & 14&.79 & 15&.0 & 8 &       &   &    &   &   &  \\            
      &   &    &   &   &   & 0.316 & 14&.96 & 15&.0 & 8 & 0.548 &  7&.24 & 15&.0 & 8 &
0.548 & 10&.47 & 15&.0 & 8 &       &   &    &   &   &  \\           
      &   &    &   &   &   & 0.447 &  7&.32 & 15&.0 & 8 & 0.632 &  4&.07 & 15&.0 & 8 &
0.632 &  6&.03 & 15&.0 & 8 &       &   &    &   &   &  \\            
      &   &    &   &   &   & 0.548 &  5&.01 & 15&.0 & 8 &       &   &    &   &   &   &
      &   &    &   &   &   &       &   &    &   &   &  \\            
      &   &    &   &   &   & 0.632 &  2&.72 & 15&.0 & 8 &       &   &    &   &   &   &
      &   &    &   &   &   &       &   &    &   &   &  \\      
      &   &    &   &   &   &       &   &    &   &   &   &       &   &    &   &   &   & 
      &   &    &   &   &   &       &   &    &   &   &  \\           
      &   &    &   &   &   & 0.500 &  4&.55 &  1&.1 & 9 &       &   &    &   &   &   &
      &   &    &   &   &   &       &   &    &   &   &  \\            
      &   &    &   &   &   & 0.750 &  1&.01 &  1&.0 & 9 &       &   &    &   &   &   &
      &   &    &   &   &   &       &   &    &   &   &  \\  \hline                                     
\end{tabular}
\end{center}
\caption{Invariant cross section in mb/(GeV$^2$/c$^3$) for protons at very forward region
($x_F \geq$~0.6) in p+p collisions measured by
\cite{bib:sannes,bib:childress,bib:scham,bib:akimov,bib:chapman,bib:whitmore,bib:brenner} }
\label{tab:extension}
\end{table}

The data are well consistent within their statistical errors, both
between the different experiments and with the NA49 results in
the overlap region. The only exception is given by the bubble
chamber experiment \cite{bib:whitmore} where at $x_F$ below 0.9 the cross sections
deviate from all other experiments by +20\% to +30\% independent
of $p_T$. This difference cannot be understood by eventual 
mis-identification nor by binning effects. Data from \cite{bib:whitmore} are therefore
only used at $x_F \geq$~0.9.

%
%
\subsection{Interpolation scheme}
\vspace{3mm}
\label{sec:interp}

\begin{figure}[h]
  \begin{center}
  	\includegraphics[width=14cm]{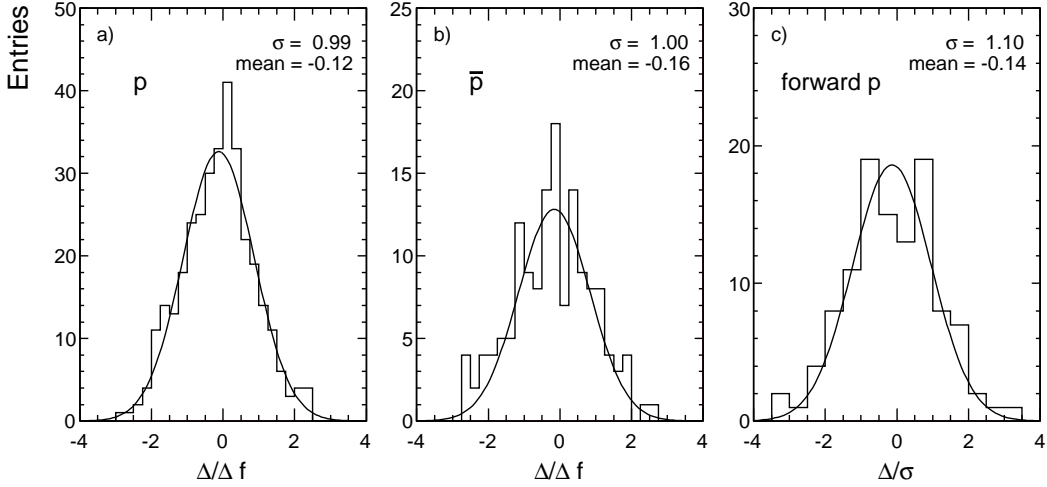}
  	\caption{Normalized difference plots between data and interpolation for a) protons, 
  	              b) anti-protons and c) protons at high $x_F$}
  	\label{fig:diff}
  \end{center}
\end{figure}

As in the preceding publications concerning pions \cite{bib:pp_paper,bib:pc_paper}
a two-dimensional interpolation is applied to the data which 
reduces the local statistical fluctuations given by the 
errors of the data points by a factor of 3-4. As there is
no possibility to describe the detailed $x_F$ and $p_T$ distributions
by simple functions and as any algebraic approximation risks
to dilute the data quality by introducing systematic biases,
the interpolation scheme relies on a multi-step recursive
method using eyeball fits. The quality of this procedure may
be controlled by plotting the differences between data points
and interpolation, normalized to the statistical errors. The
resulting distribution should be a Gaussian centered at zero with 
variance unity. This is demonstrated in Fig.~\ref{fig:diff} for protons and 
anti-protons as far as the NA49 data points are concerned, and 
separately for the extension to higher $x_F$ at low $p_T$ described above.

\begin{figure}[h]
  \begin{center}
  	\includegraphics[width=6cm]{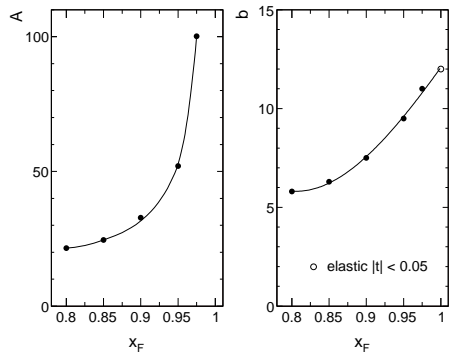}
  	\caption{Parameters $A$ and $b$ as a function of $x_F$}
  	\label{fig:tpar}
  \end{center}
\end{figure}

As, in this latter region, there are practically no measurements below 
$p_T$~=~0.2-0.3~GeV/c the extrapolation to $p_T$~=~0 has to be independently
quantified. In this limited range of transverse momentum and at
$x_F >$~0.6 a parametrization of the form

 \begin{equation}
f = Ae^{-b|t|},
\label{eq:t}
\end{equation} 
with  $p_T^2 \sim |t|x_F $ has been applied. The parameters $A$ and $b$ 
are shown in Fig.~\ref{fig:tpar} as a function of $x_F$.

The slope $b$ extrapolates well to the value for low-$t$ elastic
scattering at SPS energy also shown in Fig.~\ref{fig:tpar}

\begin{figure}[t]
  \begin{center}
  	\includegraphics[width=13.4cm]{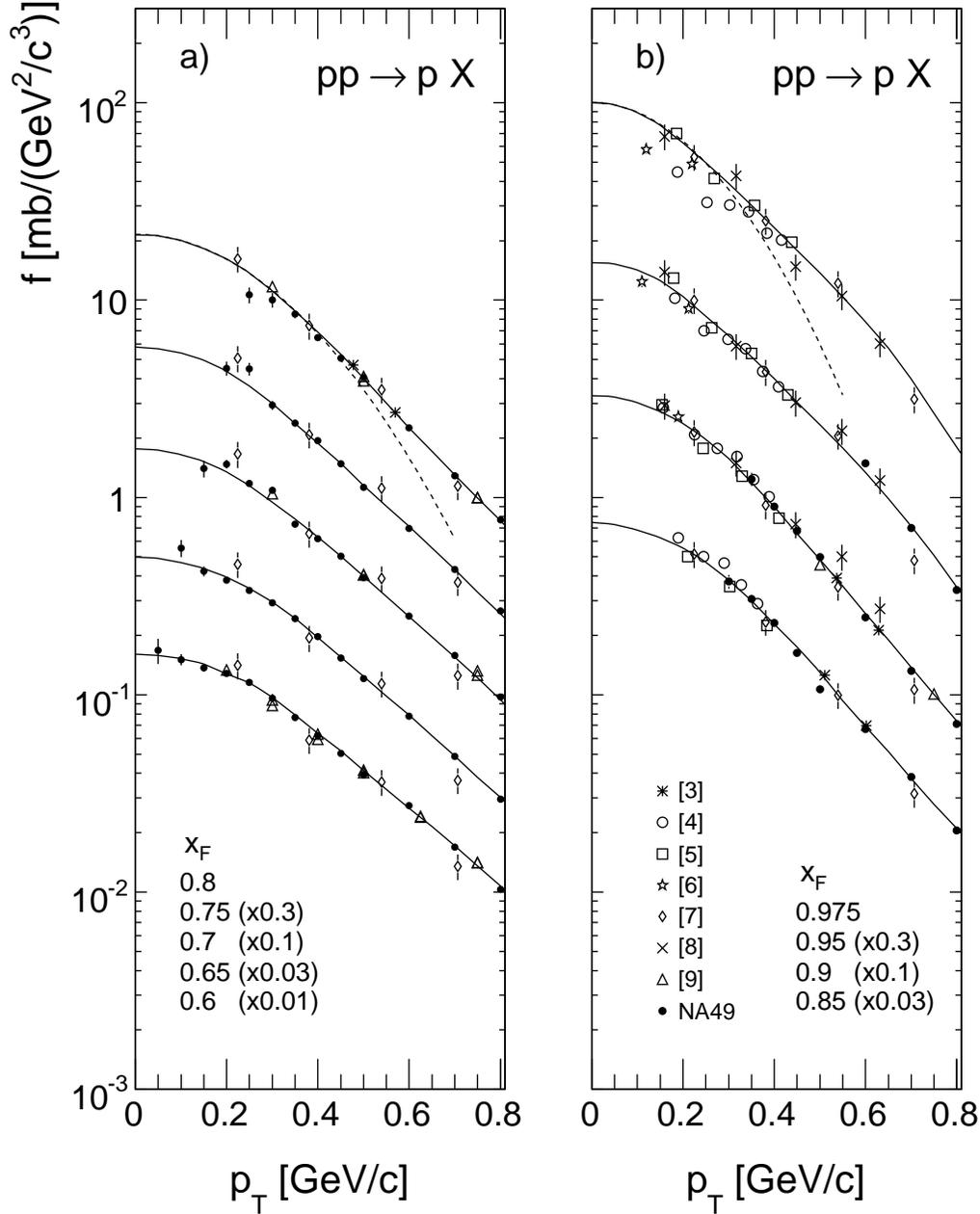}
  	\caption{Invariant cross section as function of $p_T$ at fixed $x_F$ taken from
  	\cite{bib:sannes,bib:childress,bib:scham,bib:akimov,bib:chapman,bib:whitmore,bib:brenner}  and NA49.
  	            The full lines represent the data interpolation, the dashed lines the
                 exponential parametrization \cite{bib:whitmore}}
  	\label{fig:cs_ext}
  \end{center}
\end{figure}

The internal consistency of the data sets used and their compatibility
with the interpolation scheme as well as with the extrapolation to very
low $p_T$ is presented in Fig.~\ref{fig:cs_ext}.

It should be noted that the measured cross sections deviate rapidly
from the low-$t$ parametrization, Eq.~\ref{eq:t}, already at $p_T$ values of $\sim$0.4~GeV/c.
This is exemplified by the dashed lines in Fig.~\ref{fig:cs_ext} for two $x_F$ values. 
Fits over larger regions of $p_T^2$ therefore result systematically in 
smaller values of $b$ \cite{bib:whitmore}, see also the discussion in Sect.~\ref{sec:hera}. 

%
%
\subsection{Dependence of the invariant cross sections on $\bf p_T$ and $\bf x_F$}
\vspace{3mm}
\label{sec:dist}

\begin{figure}[t!]
  \begin{center}
  	\includegraphics[width=15.4cm]{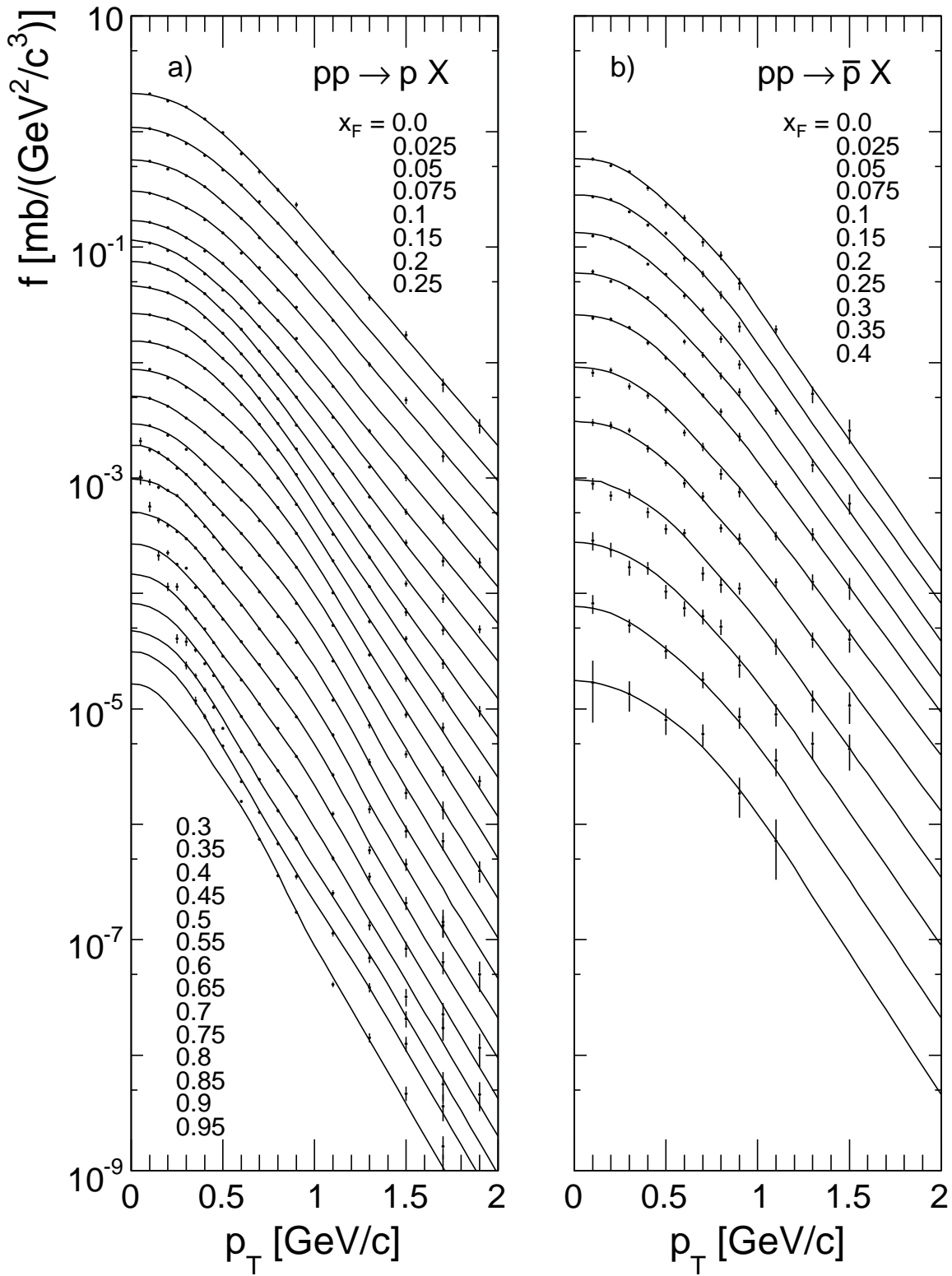}
  	\caption{Double differential invariant cross section $f(x_F,p_T)$ [mb/(GeV$^2$/c$^3$)] as a function 
  	              of $p_T$ at fixed $x_F$ for a) protons
                 and b) anti-protons produced in p+p collisions at 158~GeV/c beam momentum. 
                The distributions for different $x_F$ values are successively scaled down by 0.5 for better separation}
  	\label{fig:ptdist}
  \end{center}
\end{figure}

\begin{figure}[t!]
  \begin{center}
  	\includegraphics[width=15.8cm]{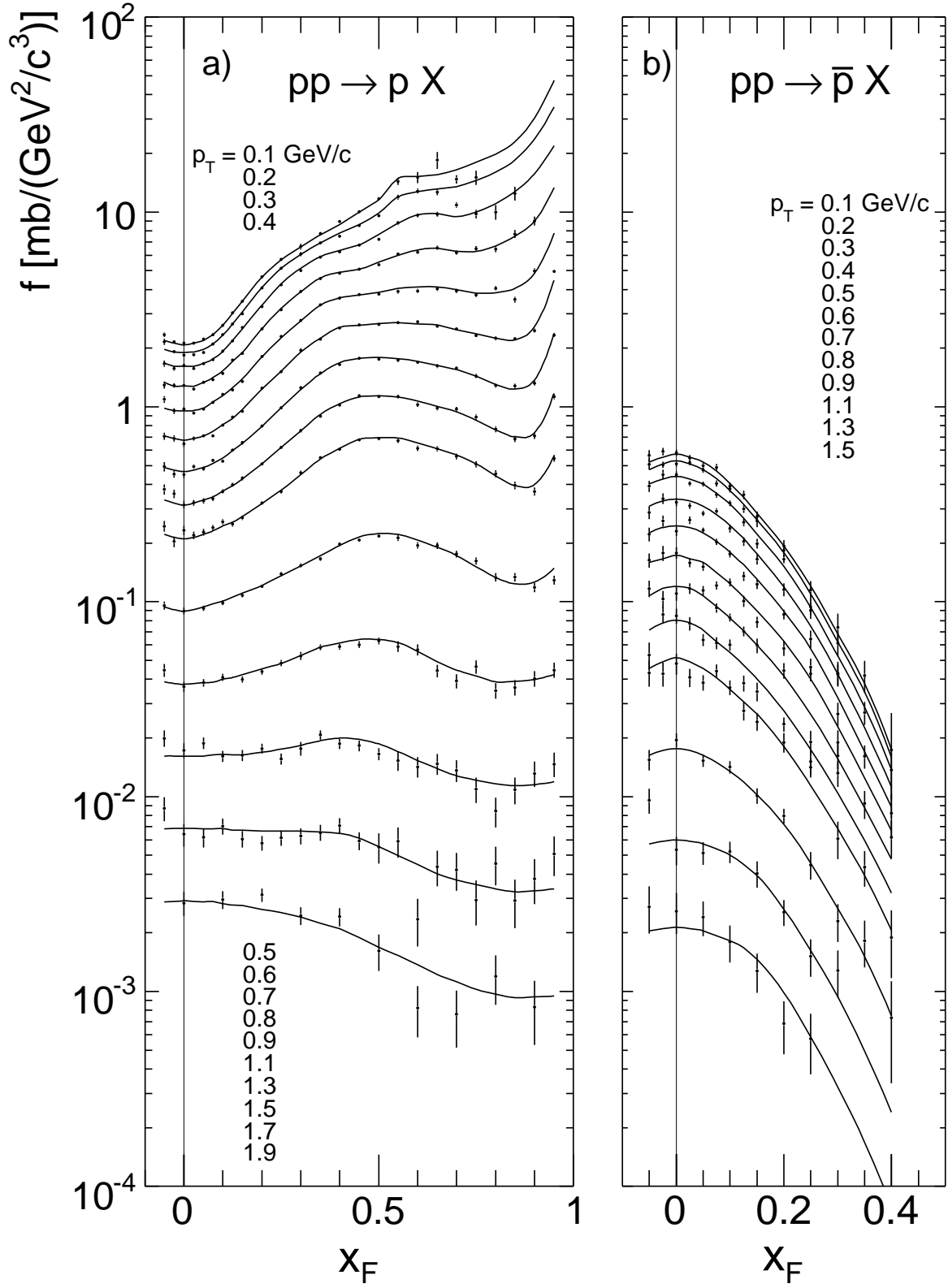}
  	\caption{Double differential invariant cross section $f(x_F,p_T)$ [mb/(GeV$^2$/c$^3$)] as a function 
  	              of $x_F$ at fixed $p_T$ for a) protons
                  and b) anti-protons produced in p+p collisions at 158~GeV/c beam momentum}
  	\label{fig:xfdist}
  \end{center}
\end{figure}

The shape of the invariant cross section as functions of $p_T$
and $x_F$ is shown in Figs.~\ref{fig:ptdist} and \ref{fig:xfdist}. These Figures include the data
interpolation discussed above. In order to clearly demonstrate
the shape evolution and to avoid the overlap of the interpolated
curves and of the error bars, subsequent $p_T$ distributions have been
multiplied by factors of 0.5 (Fig.~\ref{fig:ptdist}).

%
%
\subsection{$\overline{\textrm{p}}$/p ratios}
\vspace{3mm}
\label{sec:rat}

\begin{figure}[b!]
  \begin{center}
  	\includegraphics[width=14.5cm]{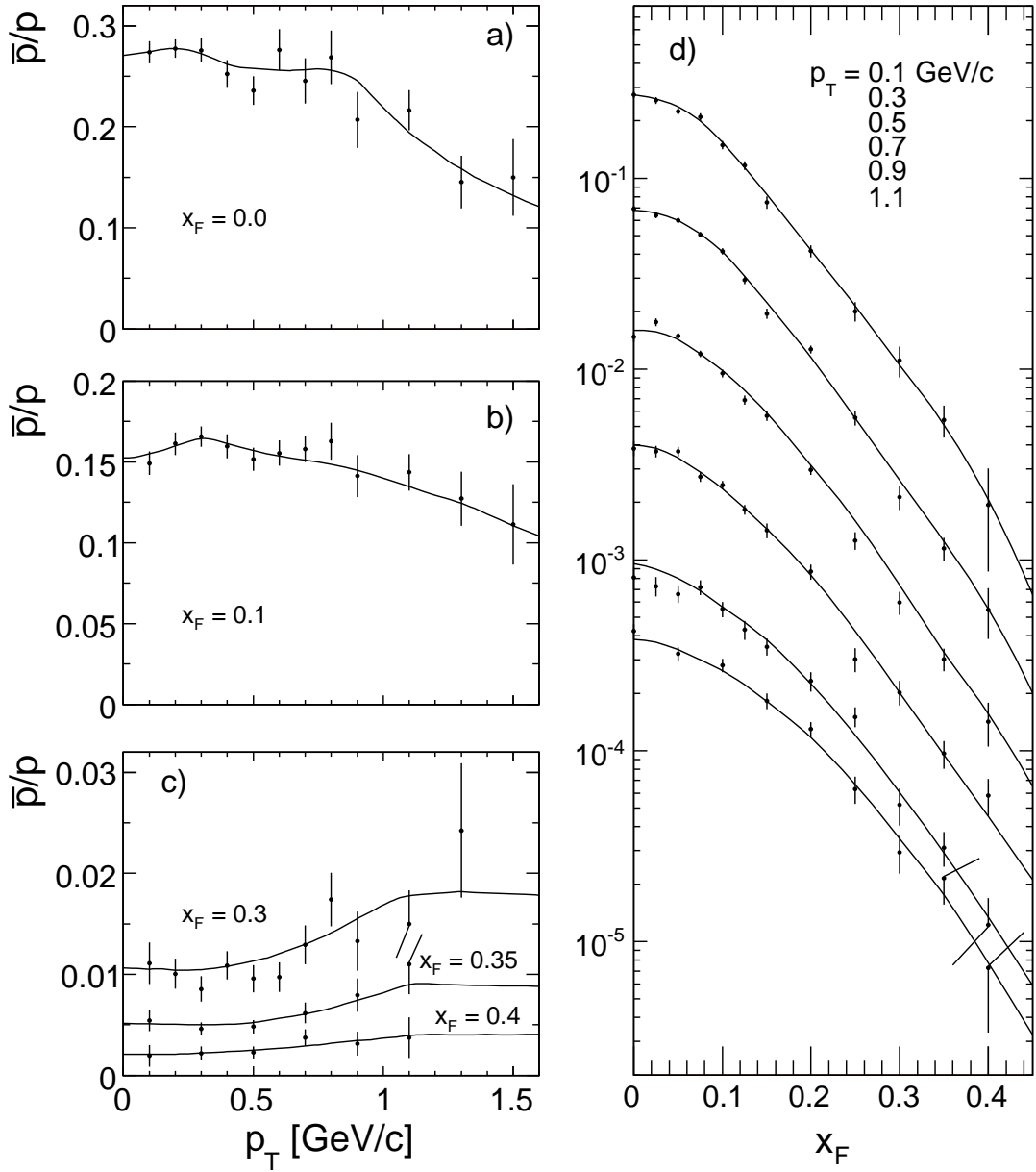}
  	\caption{Ratio of invariant cross section for anti-protons and protons ($\overline{\textrm{p}}$/p) 
  	              as a function of a), b) and c) $p_T$ at fixed $x_F$ and d) $x_F$ at fixed $p_T$. The data in
  	              panel d) were successively divided by 4 for better separation}
  	\label{fig:ratdist}
  \end{center}
\end{figure}

The phase space distributions of protons and anti-protons are
rather similar in transverse momentum, Fig.~\ref{fig:ptdist}, but they
show important differences in longitudinal momentum, Fig.~\ref{fig:xfdist}.
Here the invariant proton cross sections increase with $x_F$
whereas the anti-protons fall off steeply with $x_F$ similar
to mesonic production \cite{bib:pp_paper}. It is therefore interesting to
scrutinize the $\overline{\textrm{p}}$/p ratios quantitatively in both co-ordinates.
This is presented in Fig.~\ref{fig:ratdist} which shows this ratio as a function
of $p_T$ for fixed $x_F$ (left panels) and as a function of $x_F$ for
fixed $p_T$ (right panel). In all plots the results from the
two-dimensional interpolation discussed above are shown as
lines through the data points.

Several features emerge from this comparison. The $\overline{\textrm{p}}$/p
ratio falls with increasing $p_T$ at $x_F \leq$~0.1 and increases with
$p_T$ at $x_F >$~0.15. The ratio between $p_T$~=~0.1 and $p_T$~=~1.5~GeV/c is
about 2 at low $x_F$ and about 0.5 at high $x_F$. This means that the $p_T$ distribution of the anti-protons
flattens out with increasing $x_F$ until it becomes significantly
broader than the one for protons at $x_F >$~0.3.

The $\overline{\textrm{p}}$/p ratio as a function 
of $x_F$ at fixed $p_T$ also shows distinctive trends. Here the steep $x_F$ 
dependence at low $p_T$ (a factor of about 130 between $x_F$~=~0 and $x_F$~=~0.4) 
flattens out at higher $p_T$ (a factor of only 30 over the same $x_F$ range).

\begin{figure}[h]
  \begin{center}
  	\includegraphics[width=15cm]{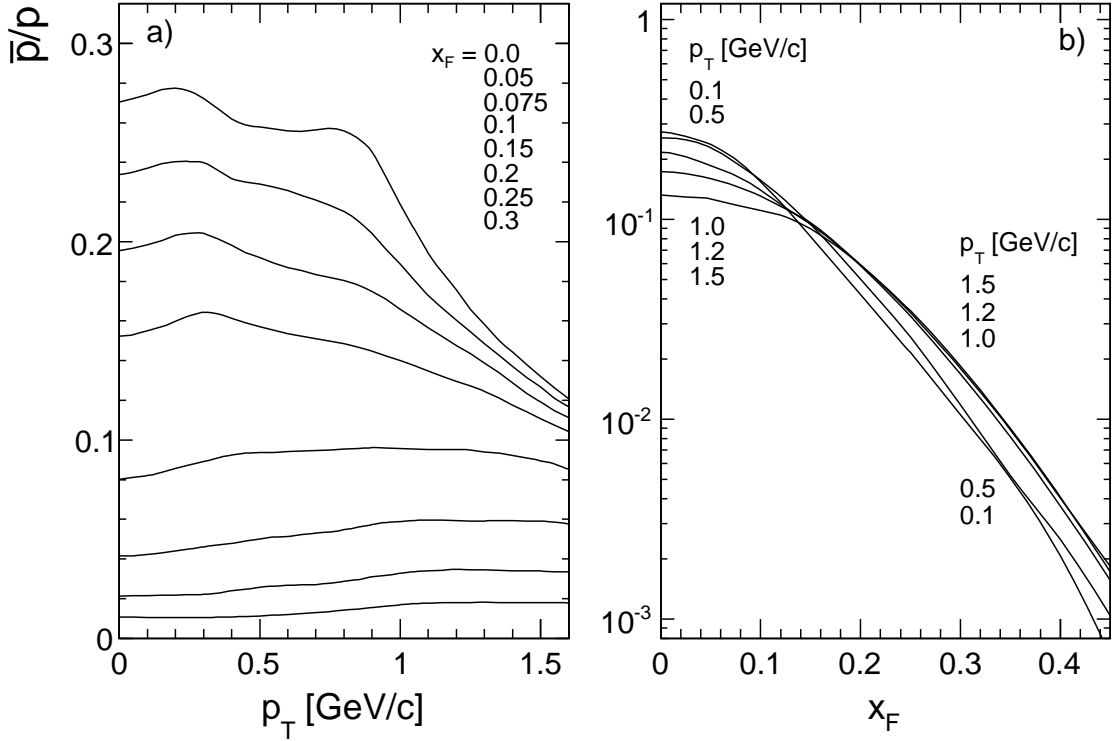}
  	\caption{ Interpolated ratios for anti-protons and protons ($\overline{\textrm{p}}$/p) 
  	              as a function of a) $p_T$ at fixed $x_F$ and b) $x_F$ at fixed $p_T$}
  	\label{fig:ratfit}
  \end{center}
\end{figure}

The situation is clarified by the summary plots of Fig.~\ref{fig:ratfit}
where only the interpolated lines are shown as functions
of $p_T$ and $x_F$, respectively.

%
%
\subsection{Rapidity and transverse mass distributions}
\vspace{3mm}
\label{sec:rap}

\begin{figure}[t!]
  \begin{center}
  	\includegraphics[width=15.8cm]{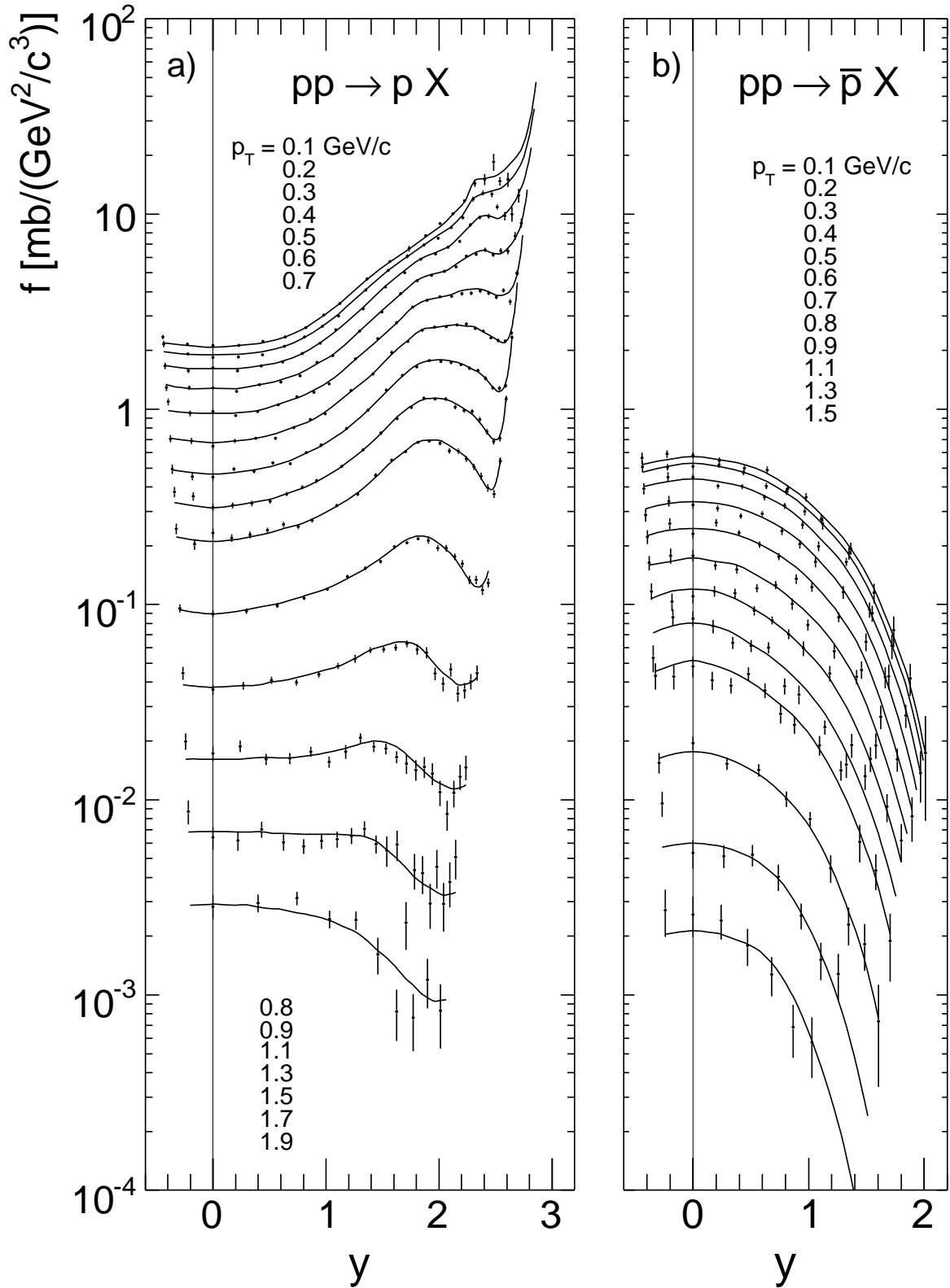}
  	\caption{Double differential invariant cross section $f(x_F,p_T)$ [mb/(GeV$^2$/c$^3$] as a function  
  	              of $y$ at fixed $p_T$ for a) protons
                  and b) anti-protons produced in p+p collisions at 158~GeV/c beam momentum}
  	\label{fig:ydist}
  \end{center}
\end{figure}

As in references \cite{bib:pp_paper,bib:pc_paper} the invariant cross sections are
also presented, for convenience, as a function of rapidity at
fixed $p_T$ in Fig.~\ref{fig:ydist}. Here the absence of a "rapidity plateau" both for protons
(with the exception of the region at $p_T >$~1.5~GeV/c) and for
anti-protons should be noted.

\begin{figure}[t]
  \begin{center}
  	\includegraphics[width=11cm]{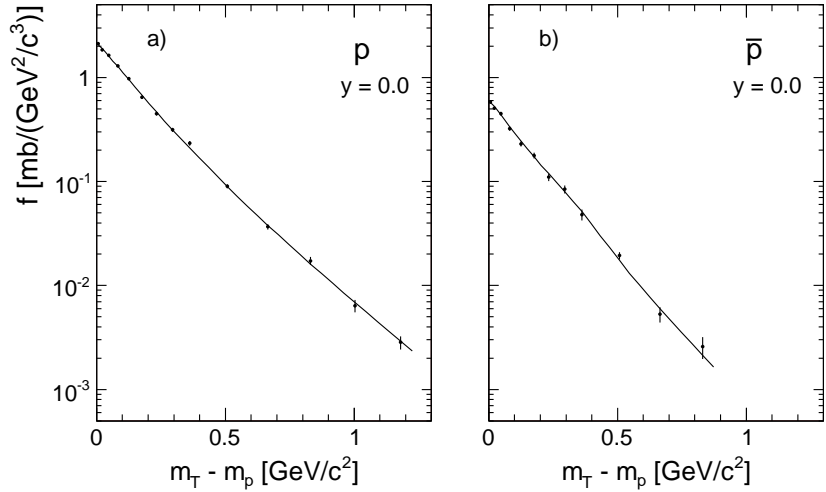}
  	\caption{Invariant cross section as a function of $m_T - m_p$ for a) p and
                   b) $\overline{\textrm{p}}$ produced at $y$~=~0.0}
  	\label{fig:mtdist}
  \end{center}
\end{figure}

\begin{figure}[h]
  \begin{center}
  	\includegraphics[width=11cm]{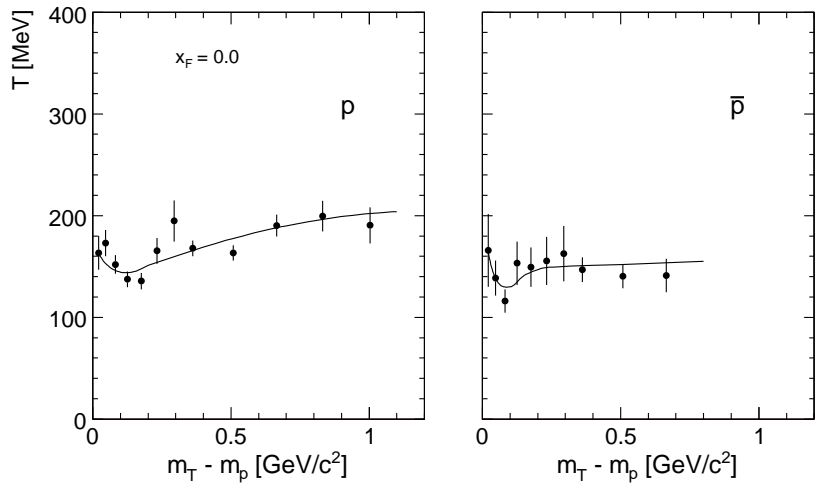}
  	\caption{Local slope of the $m_T$ distribution as a function of $m_T - m_p$
                 for p and $\overline{\textrm{p}}$. The lines corresponds to the data
                 interpolation, Sect.~\ref{sec:interp}}
  	\label{fig:mtslope}
  \end{center}
\end{figure}

Transverse mass distributions, with $m_T$~=~$\sqrt{m_p^2 + p_T^2}$,
are shown in Fig.~\ref{fig:mtdist} for $x_F$~=~$y$~=~0.
In accordance with the $\overline{\textrm{p}}$/p ratios discussed above,
a systematic difference between p and $\overline{\textrm{p}}$ is visible.
The proton distribution is clearly not compatible with simple exponential shape,
whereas the anti-proton distribution happens to be close to exponential up
to the experimental limit of $m_T-m_p$~=~0.8 GeV/c$^2$.This is quantified
by the dependence of the local logarithmic inverse slopes of  $m_T-m_p$ 
given in Fig.~\ref{fig:mtslope}. Here the slope defined by three successive data points
has been used. In Fig.~\ref{fig:mtslope} also the inverse slopes obtained from the data
interpolation, Sect.~\ref{sec:interp}, are shown.

%
%
\subsection{Baryon to pion ratios}
\vspace{3mm}
\label{sec:p2pi}

The NA49 data on charged pions \cite{bib:pp_paper} offer a phase space coverage
which is comparable in completeness, density and statistical 
accuracy to the results on baryons presented here. It is therefore 
indicated to compare the respective cross sections. This is
done in the following section by inspecting the corresponding
ratios of invariant inclusive cross sections as functions of
$x_F$ and $p_T$. 

\begin{figure}[h]
  \begin{center}
  	\includegraphics[width=13.5cm]{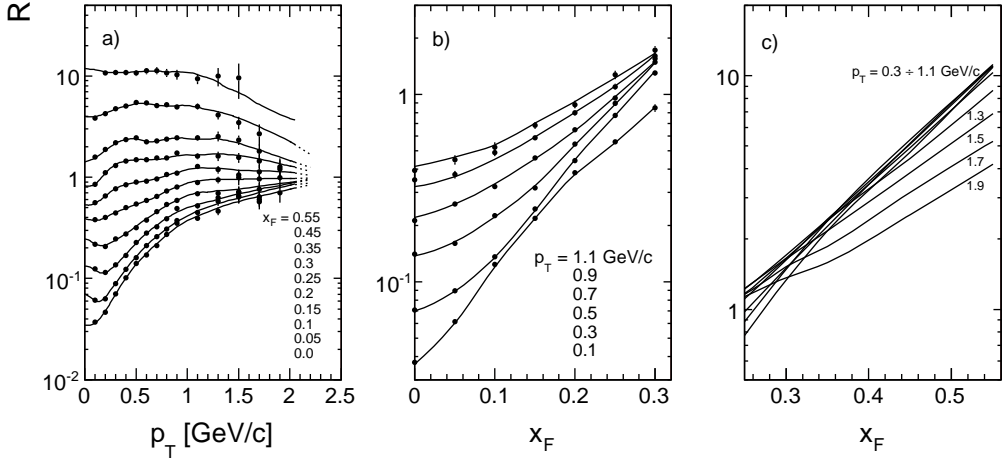}
  	\caption{$R=f_p/\langle f_{\pi} \rangle$: a) as a function of $p_T$ at fixed $x_F$, b) and c) as a 
                  function of $x_F$ at fixed $p_T$. The full lines represent ratios of interpolated cross sections. 
                  Due to the close similarity of the $p_T$ distributions in the range 0.3~$<p_T<$~1.1~GeV/c
                  only the interpolated lines are shown in panel c)}
  	\label{fig:p2pi_comp}
  \end{center}
\end{figure}

For protons, the ratio $R=f_p/\langle f_{\pi} \rangle$, where $\langle f_{\pi} \rangle$ indicates the mean pion
cross section $0.5\cdot(f_{\pi^+}+f_{\pi^-})$, is presented in Fig.~\ref{fig:p2pi_comp}a as
a function of $p_T$ at fixed $x_F$ and in Fig.~\ref{fig:p2pi_comp}b as a function of $x_F$
for fixed $p_T$. For each data sample the corresponding interpolated
cross section ratios are superimposed as full lines.

The $p_T$ dependence, Fig.~\ref{fig:p2pi_comp}a, reveals structure at low $p_T$ which has been
shown in \cite{bib:pp_paper} to result from resonance decay, together with a strong 
increase of the ratio by almost three orders of magnitude between
$x_F$~=~0 and $x_F$~=~0.5. This increase is progressively reduced with increasing
$p_T$ to less than an order of magnitude at $p_T \sim$~2~GeV/c. In fact $R$
approaches unity in the high $p_T$ region for all $x_F$ values shown, and
the extrapolation of the data interpolation (full lines) beyond the 
measured $p_T$ range indicates a convergence point at $R \sim$~1 for $p_T\sim$~2.5~GeV/c.
This is again an indication of resonance decay. A study of the pion
cross sections resulting from the decay of an ensemble of 13 known
resonances \cite{bib:site,bib:andrzej} has indeed shown that the inclusive pion yields are 
saturated in the range 1.5~$< p_T <$~3~GeV/c, at SPS energy, by two-body
resonance decays. The high $p_T$ pions originate either from high mass 
resonances or from the high mass Breit-Wigner tails of lower mass
states. In both cases the available momentum $q$ in the resonance cms
becomes high enough so that the dependence on the mass $m$ of the decay
particle induced by the energy term

\begin{equation}
E_{\textrm{cms}} = \sqrt{q^2+m^2}
\end{equation}
in the Lorentz-transformation from the resonance cms to the experimental
system becomes small. This means, always considering two-body decays, that
the yield dependence on the kinematical variables
$x_F$ and $p_T$ should become similar for pions and
protons and therefore their ratio should tend
to be stable against these variables. The actual
limiting value of p/$\langle \pi \rangle$ depends however on the
details of the isospin structure of the baryonic
and mesonic resonances contributing to the proton
and pion production in this section of phase space \cite{bib:site,bib:andrzej}.

The $x_F$ dependence at fixed $p_T$, Fig.~\ref{fig:p2pi_comp}b, shows again the strong 
increase of $R$ with $x_F$ in the low $p_T$ region, with a progressive
tendency to flatten out with increasing $p_T$. This results, at $p_T$
up to about 1~GeV/c, in a convergence point at $x_F \sim$~0.5 where $R$
becomes practically $p_T$ independent predicting the equality of 
mean $p_T$ for pions and protons in this $x_F$ region shown in Sect.~\ref{sec:ptint_dist},
Fig.~\ref{fig:ratint}. At $p_T >$~1.1~GeV/c
and $x_F >$~0.3, see Fig.~\ref{fig:p2pi_comp}c, the $p_T$ distribution of protons
becomes steeper than the one for pions. The ratio $R$ thus approaches
unity from above, whereas at $x_F <$~0.3, Fig.~\ref{fig:p2pi_comp}b, the opposite
trend is visible as discussed above.

\begin{figure}[h]
  \begin{center}
  	\includegraphics[width=11.5cm]{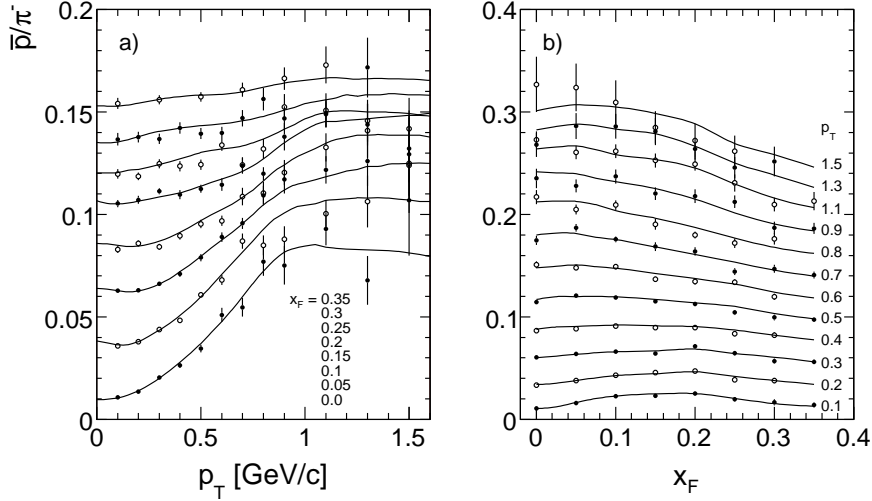}
  	\caption{$\overline{\textrm{p}}$/$\pi^-$ as a function of a) $p_T$ at fixed $x_F$, b) $x_F$ at 
  	             fixed $p_T$. The full lines represent data interpolation. Data points and 
                 interpolated lines of subsequent $x_F$ ($p_T$) values are shifted
                 upwards by 0.02 in $\overline{\textrm{p}}$/$\pi^-$}
  	\label{fig:a2pi_comp}
  \end{center}
\end{figure}

Concerning the relation of anti-protons to pions it is indicated to rather
study the $\overline{\textrm{p}}$/$\pi^-$ ratios. This is due to the similar isotriplet structure of
both the baryon-pair and the pion production \cite{bib:fischer}, see also Sect.~\ref{sec:neut_na49}. 
The $\overline{\textrm{p}}$/$\pi^-$ ratios are shown in Fig.~\ref{fig:a2pi_comp} both as a function 
of $p_T$ at fixed $x_F$ and as a function of $x_F$ at fixed $p_T$.

Due to the larger error bars for $\overline{\textrm{p}}$ production together with the smaller
range of variation, data points and interpolated lines of subsequent $x_F$ (Fig.~\ref{fig:a2pi_comp}a)
and $p_T$ (Fig.~\ref{fig:a2pi_comp}b) values are shifted upwards by 0.02 in $\overline{\textrm{p}}$/$\pi^-$. 
The full lines correspond again to the two-dimensional interpolation of the invariant cross sections.

Similar to what has been shown for p/$\langle \pi \rangle$, the $\overline{\textrm{p}}$/$\pi^-$ ratios increase 
strongly with $p_T$ at low $x_F$ by about one order of magnitude, Fig.~\ref{fig:a2pi_comp}a. And 
similarly, this increase reduces for larger $x_F$ to a factor of only $\sim$~2 at 
the limit of the measurements at $x_F$~=~0.35. In contrast there is a general 
flattening of the $p_T$ dependence for $p_T$ beyond about 1.2~GeV/c.

As far as the $x_F$ dependence is concerned, Fig.~\ref{fig:a2pi_comp}b, the strong increase 
observed for p/$\langle \pi \rangle$ with $x_F$ is inverted to a general modest decrease 
which amounts to a factor of about four between $x_F$~=~0 and 0.35 at the 
highest $p_T$ values. At $p_T$ below 0.4~GeV/c however the ratios show a distinct 
maximum at $x_F \sim$~0.2 and little if any difference comparing the values at 
$x_F$~=~0 and 0.35.

\begin{figure}[h]
  \begin{center}
  	\includegraphics[width=12cm]{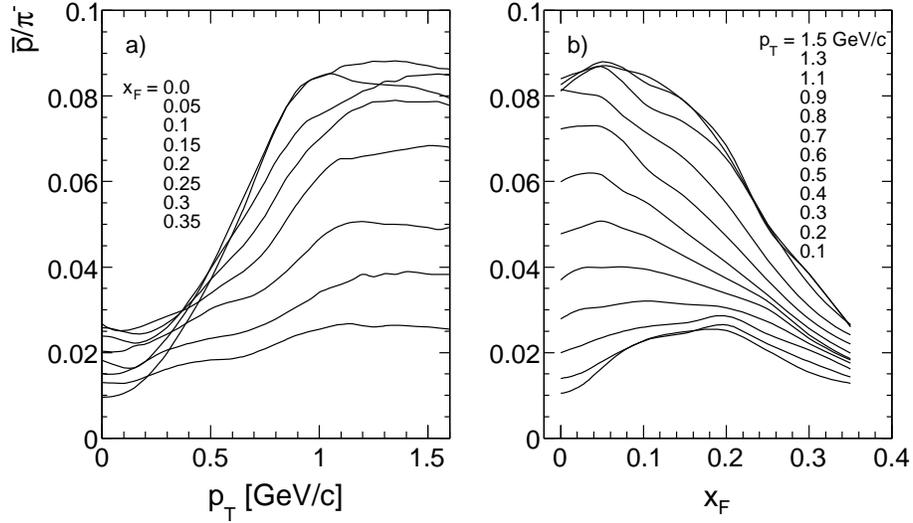}
  	\caption{$\overline{\textrm{p}}$/$\pi^-$ from data interpolation as a function of 
                  a)  $p_T$ at fixed $x_F$, b) $x_F$ at fixed $p_T$}
  	\label{fig:a2pi_inter}
  \end{center}
\end{figure}

In order to bring out the trends described above more clearly, the 
ratios of the interpolated cross sections are shown, without 
scale shift, separately in Figs.~\ref{fig:a2pi_inter}a and \ref{fig:a2pi_inter}b.

Here again, it is worth to note the flattening of the $p_T$ dependence 
above 1.2~GeV/c and the convergence of the ratios for the higher $x_F$ range 
where the mean $p_T$ for $\overline{\textrm{p}}$ and pions becomes comparable.

%
%
\section{Comparison to Fermilab data}
\vspace{3mm}
\label{sec:comp}

%
%
\subsection{The feed-down problem}
\vspace{3mm}
\label{sec:comp_fd}

Before proceeding to a detailed comparison with the data
sets specified in Sect.~\ref{sec:exp_sit}, the general problem of
baryon feed-down from weak decays of strange hyperons
has to be discussed. In the case of the NA49 data a
feed-down correction has been performed (Sect.~\ref{sec:feed}).
It amounts to typically 5-20\% of the measured baryon
yields, with specific $x_F$ and $p_T$ dependences. This is
only a fraction of the total hyperon decay contribution
as the TPC tracking system of the NA49 detector has 
a resolution of the primary vertex position sufficient to 
reject a major part of the decay baryons. This is
not a priori true for the reference data. As most of
the corresponding experiments date from the 1970's to
the early 1980's, micro-vertex detection or precision
tracking was not available and therefore a large fraction
if not all of the decay baryons contributed to the measured
cross sections. What counts here is the distance of the
first tracking elements from the primary vertex
in relation to the typical hyperon decay length.

For the CERN ISR collider it may be stated that,
given the different detector layouts for the $x_F$ and
$p_T$ ranges covered, and given the large dimension of 
the interaction diamond, practically all baryonic decay products
are included in the published data. A correction for this
feed-down has not been attempted by any of the quoted
experiments.       

For fixed-target experiments the situation is somewhat more
complicated as the range of lab momenta covered shows a
much larger variation. Measurements in the target hemisphere
with lab momenta comparable to the range at colliders are
definitely prone to feed-down contamination. But even
in the forward direction with momenta in the range of several
tens of GeV/c, in many cases the first active detector elements
are many meters away from the primary vertex, not to mention
the general absence of precision tracking. A precise
simulation of trajectories through the detectors and the
aperture-defining collimators would be mandatory to come
to a quantitative determination of the feed-down contributions.

\begin{figure}[h]
  \begin{center}
  	\includegraphics[width=12cm]{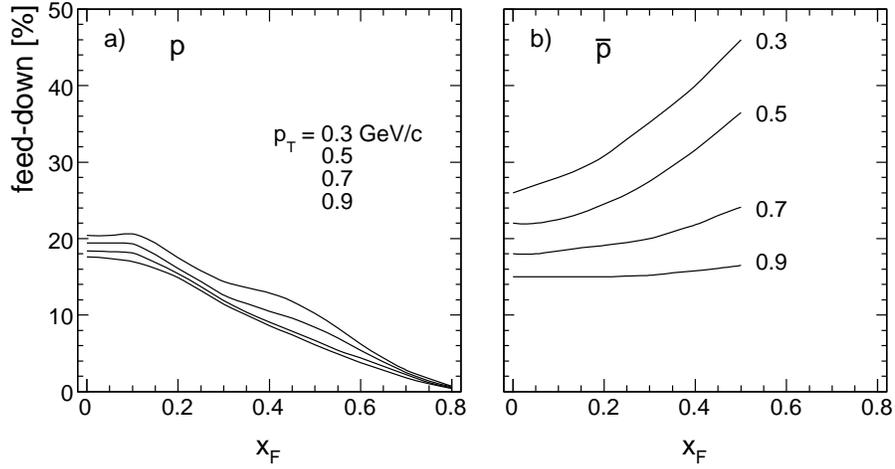}
  	\caption{Total feed-down for a) protons and b) anti-protons as function of $x_F$ for different $p_T$ values}
  	\label{fig:tot_fd}
  \end{center}
\end{figure}

A feeling for the size of the corresponding corrections may
be obtained from Fig.~\ref{fig:tot_fd} where the total yield of decay products
is given in percent of the direct baryon cross section for
protons and anti-protons at $\sqrt{s}$~=~17.2~GeV/c.

Whereas this fraction tends to decrease below the 10\% level
at $x_F >$~0.4 for protons, it stays constant or even increases with
$x_F$ for anti-protons, with a sizeable $p_T$ dependence.
The data comparisons carried on below will therefore attempt
to address this problem on a case-to-case basis. 

%
%
\subsection{The Brenner et al. data, \cite{bib:brenner}}
\vspace{3mm}
\label{sec:comp_bren}

\begin{figure}[b]
  \begin{center}
  	\includegraphics[width=13cm]{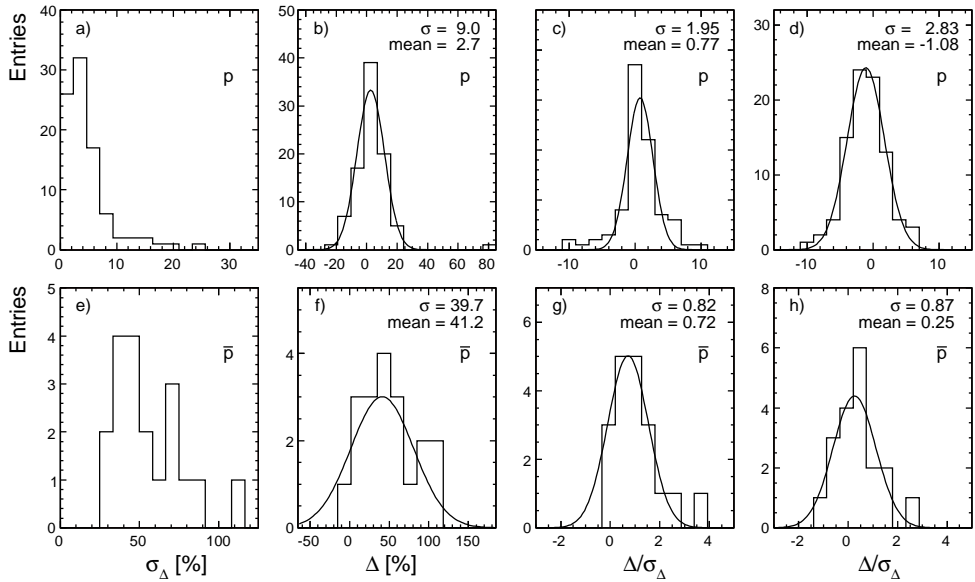}
  	\caption{Statistical analysis of the difference between the measurements of  \cite{bib:brenner} 
  	               and NA49 for protons (upper four panels) and anti-protons (lower four panels): 
  	               a) and e) error of the difference of the measurements; b) and f) difference of 
  	               the measurements; c) and g) difference divided by the error; d) and h) difference 
  	               divided by the error after feed-down correction of data from \cite{bib:brenner}}
  	\label{fig:bren_stat}
  \end{center}
\end{figure}

This experiment offers 90 overlapping data points for protons
and 19 points for anti-protons at the two beam momenta of 100
and 175~GeV/c. If the statistical errors of the proton sample
are typically on the 1-10\% level, the ones for anti-protons
are considerably larger and vary between 20 and 50\%. The
situation is quantified in Fig.~\ref{fig:bren_stat} which shows the distributions
of the statistical errors, the differences to the interpolated
NA49 data and the differences normalized to the statistical
errors for protons and anti-protons, with and without feed-down
correction of \cite{bib:brenner}. This latter distribution should be 
centered at zero with variance unity if the two measurements are 
compatible on an absolute scale.

Evidently the feed-down correction helps to reduce the almost
50\% average difference for anti-protons, but over-corrects for 
protons. It should however be realized that the mean differences
are for protons on the $\pm$4\% level which signals good agreement if 
compared to the quoted absolute normalization errors. This result
verifies the excellent agreement found in \cite{bib:pp_paper} for pions.

\begin{figure}[h]
  \begin{center}
  	\includegraphics[width=13.cm]{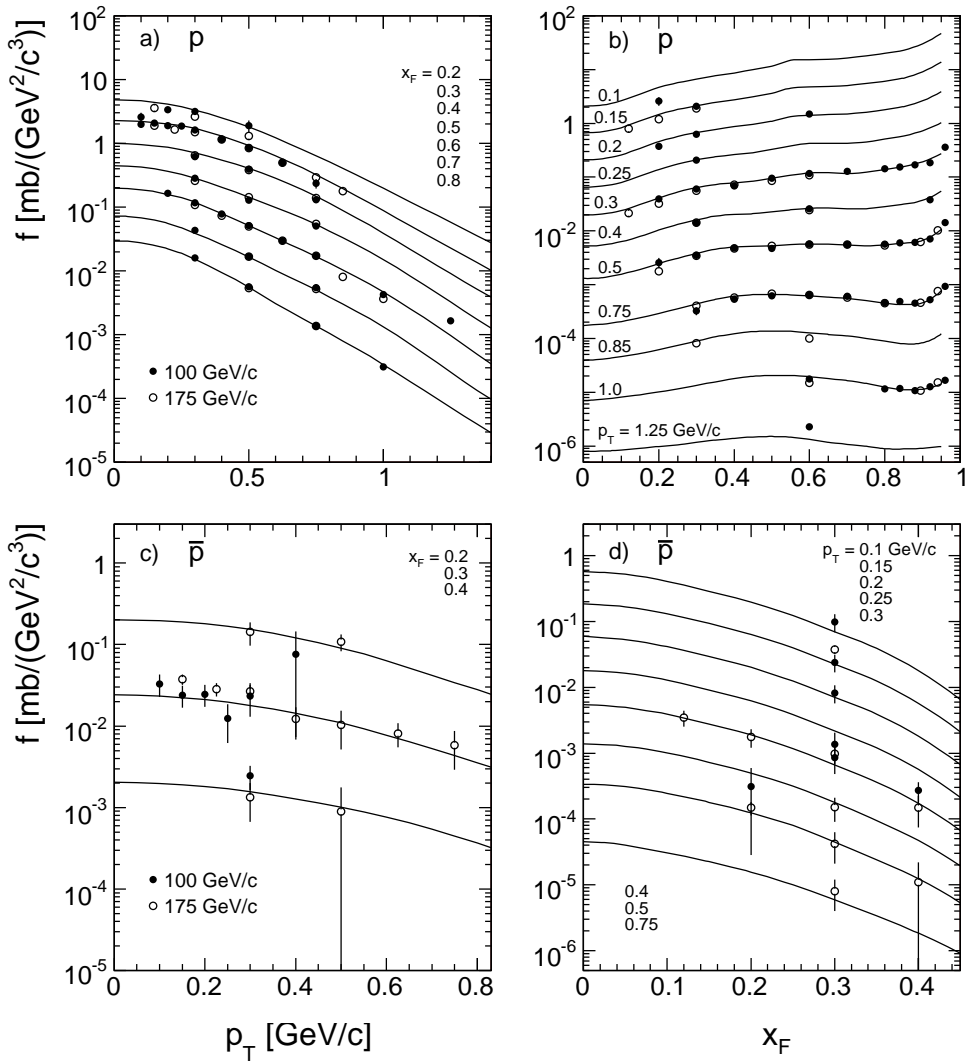}
  	\caption{Comparison of invariant cross section between NA49 (lines) and measurements from 	
  	                \cite{bib:brenner} at 100 (full circles) and 175~GeV/c (open circles) for protons as a function 
  	                of a) $p_T$ at fixed $x_F$ and b) $x_F$ at fixed $p_T$, and for anti-protons as a function of 
  	                c) $p_T$ at fixed $x_F$ and d) $x_F$ at fixed $p_T$. The data were successively divided by
  	                3 for better separation}
  	\label{fig:bren_comp}
  \end{center}
\end{figure}

The distribution of the comparison data over phase space may be
judged from Fig.~\ref{fig:bren_comp} where the $x_F$ and $p_T$ distributions of the
data points from \cite{bib:brenner} are given against the interpolated
NA49 data.

%
%
\subsection{The Johnson et al. data, \cite{bib:john}}
\vspace{3mm}
\label{sec:comp_john}

From this experiment 54 and 26 data points for protons and 
anti-protons, respectively, may be used for comparison. The
data were obtained at 100, 200 and 400~GeV/c beam momentum.
As in the Brenner experiment, there is a large difference
between the statistical errors of protons (2-6\%) and anti-protons (10-30\%).

\begin{figure}[h]
  \begin{center}
  	\includegraphics[width=13cm]{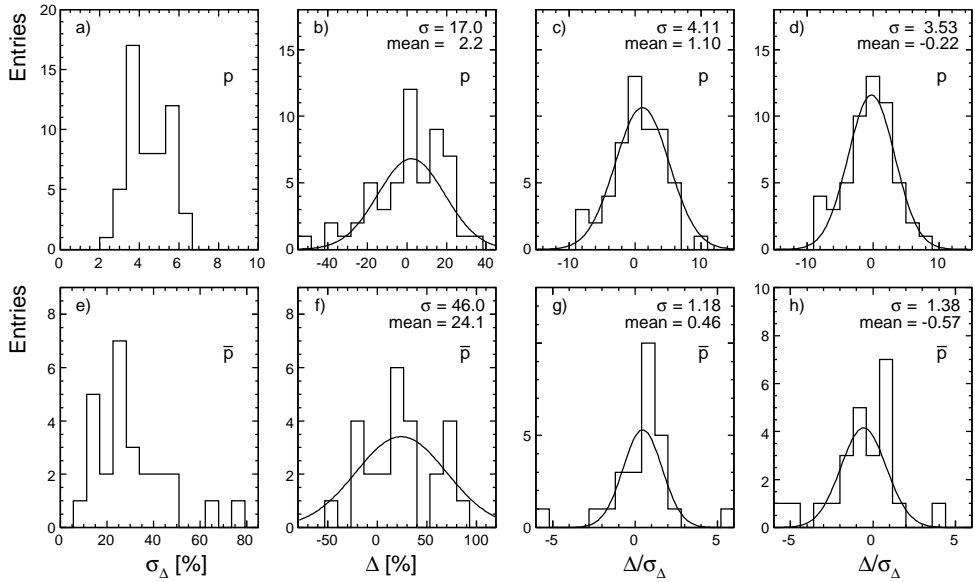}
  	\caption{Statistical analysis of the difference between the measurements of  \cite{bib:john} 
  	               and NA49 for protons (upper four panels) and anti-protons (lower four panels): 
  	               a) and e) error of the difference of the measurements; b) and f) difference of 
  	               the measurements; c) and g) difference divided by the error; d) and h) difference 
  	               divided by the error after feed-down correction of data from \cite{bib:john}}
  	\label{fig:john_stat}
  \end{center}
\end{figure}

As the measurements were done in the backward hemisphere up
to maximum lab momenta of 2.3~GeV/c and as the aperture
defining first magnet is placed at about 7 decay lengths
for the maximum contributing hyperon momentum, a major
fraction of the feed-down baryons must be expected to be
contained in the data sample. This is visible in Fig.~\ref{fig:john_stat}
where again the distributions of the statistical error, 
of the difference and the relative difference to the NA49 data
with and without feed-down correction are presented.

Particularly for protons an improvement of
the experimental differences is visible with feed-down
correction, with mean deviations on the few percent level.
The large rms values of the relative differences are, however,
noteworthy. As was already the case for the pion comparison \cite{bib:pp_paper},
this speaks for additional fluctuations beyond those from 
counting statistics proper in this experiment. Why the mean
relative deviations are below one standard deviation for the 
baryons and about 3 standard deviations for pions \cite{bib:pp_paper} remains however
an open question.

\begin{figure}[h]
  \begin{center}
  	\includegraphics[width=13.cm]{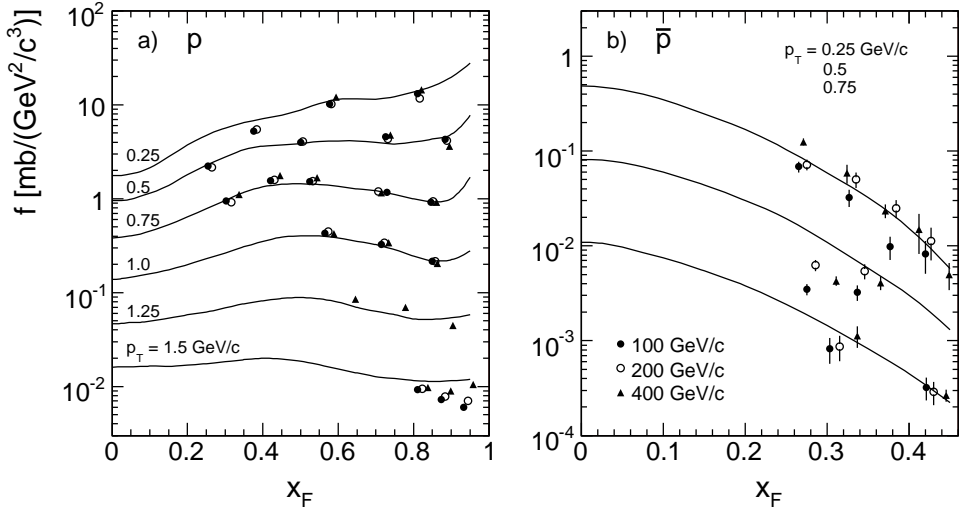}
  	\caption{Comparison of invariant cross section between NA49 (lines) and measurements from 	
  	                \cite{bib:john} at 100 (full circles), 200 (open circles) and 400~GeV/c (full triangles) 
  	                as a function of $x_F$ at fixed $p_T$ for a) protons and b) anti-protons. The anti-proton data 
  	                were successively divided by 3 for better separation}
  	\label{fig:john_comp}
  \end{center}
\end{figure}

The phase space distribution of the Johnson data, compared to the NA49
data interpolation, is shown in Fig.~\ref{fig:john_comp}.

%
%
\subsection{The Antreasyan et al. data, \cite{bib:cronin}}
\vspace{3mm}
\label{sec:comp_cron}

This so-called "Cronin" experiment represents the only measurement
near $x_F$~=~0 in the SPS energy range. As it is overlapping with the
lower ISR energy range there is a long standing problem 
with an unresolved discrepancy of the proton yields by about 
a factor of 1.3--1.4 and of the anti-proton yields by a factor of 2,
whereas there is reasonable agreement of the pion cross sections \cite{bib:alper}.
The experiment which was aiming at high $p_T$ production contributes
just a couple of cross sections at the $p_T$ values of 0.77 and 1.54~GeV/c 
in the NA49 $p_T$ range.

A first problem is connected with the fact that the spectrometer
was set to a constant lab angle of 77~mrad at all energies and
for all particle masses. This results in an $\sqrt{s}$ and $p_T$
dependent offset in $x_F$ which introduces non-negligible variations
of the cross sections. This is quantified in Table~\ref{tab:cron_comp} which gives
the corresponding deviations in $x_F$ and of proton and anti-proton cross
sections $\Delta f$, referred to $x_F$~=~0.

\begin{table}[h]
\footnotesize
\begin{center}
\begin{tabular}{ccr@{}lr@{}lr@{}l}
\hline
 \multirow{2}{8mm}{$p_T$}
&$p_{\textrm{beam}}$&\multicolumn{2}{c}{200}&\multicolumn{2}{c}{300}&\multicolumn{2}{c}{400}\\        
  &    $\sqrt{s}$        & \multicolumn{2}{c}{19.3}& \multicolumn{2}{c}{23.7}& \multicolumn{2}{c}{27.3} \\
  \hline 
 \multirow{5}{8mm}{0.77}                                            
 &      $x_F$                                                    &  -0.0&28      &   -0.0&45     &  -0.0&53    \\ 
 & $\Delta f_{\overline{\textrm{p}}}$ [\%] &    -3&.5        &    -8&.9        &  -12&.0      \\        
  &  $\Delta f_{\textrm{p}}$ [\%]                 &     1&.2        &     3&.1        &     4&.2    \\        
  & $R_{\overline{\textrm{p}}}$   &  0.713&$\pm$0.081  & 0.972&$\pm$0.097 & 0.956&$\pm$0.101  \\  
  &   $R_{\textrm{p}}$                   &  0.726&$\pm$0.084 & 0.797&$\pm$0.082 &  0.760&$\pm$0.081 \\ 
     \hline 
 \multirow{5}{8mm}{1.54}                                             
  &     $x_F$                                                     &     0.0&13     &    -0.0&20    &   -0.0&37    \\
  & $\Delta f_{\overline{\textrm{p}}}$ [\%] &    -0&.3        &    -0&.8        &   -2&.2      \\        
  &   $\Delta f_{\textrm{p}}$ [\%]                 &     --&           &       --&     &   --&   \\        
  & $R_{\overline{\textrm{p}}}$  & 0.756&$\pm$0.058 & 1.230&$\pm$0.059 & 1.540&$\pm$0.059  \\  
  &   $R_{\textrm{p}}$                    & 0.728&$\pm$0.044 & 0.824&$\pm$0.044 & 0.809&$\pm$0.074  \\ \hline
 \end{tabular}
\end{center}
\caption{Offset in $x_F$ and difference $\Delta f$ in the cross section due to this offset at different $\sqrt{s}$ and
               $p_T$. The cross section ratio $R$ between the data from \cite{bib:cronin} and NA49.}
\label{tab:cron_comp}
\end{table}

A second problem is also here connected to feed-down. As the first,
aperture-defining collimators of the spectrometer are about 18~m
downstream of the target, a good fraction of the feed-down baryons
may enter into the acceptance. Following Fig.~\ref{fig:tot_fd} this may well
give downward corrections of up to 18\% for protons and 13 to 16\%
for anti-protons in the given $p_T$ range.

\begin{figure}[h]
  \begin{center}
  	\includegraphics[width=8.5cm]{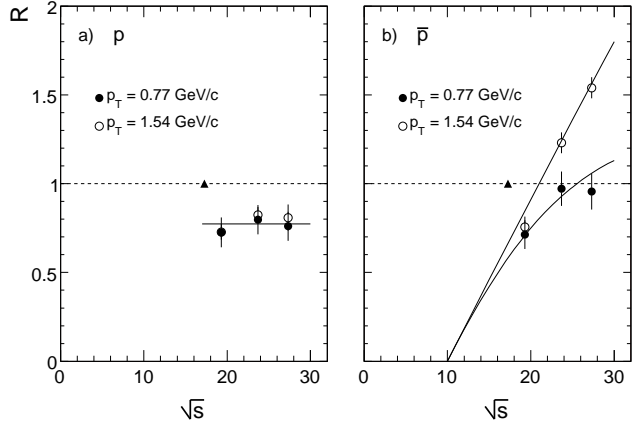}
  	\caption{The cross section ratios $R$ between the data from \cite{bib:cronin} and NA49  
  	                as a function of $\sqrt{s}$ for two values of $p_T$ for a) protons and b) anti-protons. 
  	                In both of the panels the NA49 point is indicated with full triangle }
  	\label{fig:cron_comp}
  \end{center}
\end{figure}

The cross section ratios $R$ between the data from \cite{bib:cronin} and NA49 are
also presented in Table~\ref{tab:cron_comp}. They are shown in Fig.~\ref{fig:cron_comp}a for protons
and Fig.~\ref{fig:cron_comp}b for anti-protons as a function of $\sqrt{s}$. For protons
there is to first order an $s$-independent offset at 0.75, whereas
for anti-protons the expected strong $s$-dependence of anti-baryon
production is evident. In this case however, when extrapolating
this dependence to $\sqrt{s}$~=~17.2~GeV, there is a discrepancy of about 
a factor of two as compared to the NA49 data. Similar discrepancies
have been mentioned above with respect to the ISR data.

\begin{figure}[h]
  \begin{center}
  	\includegraphics[width=8.5cm]{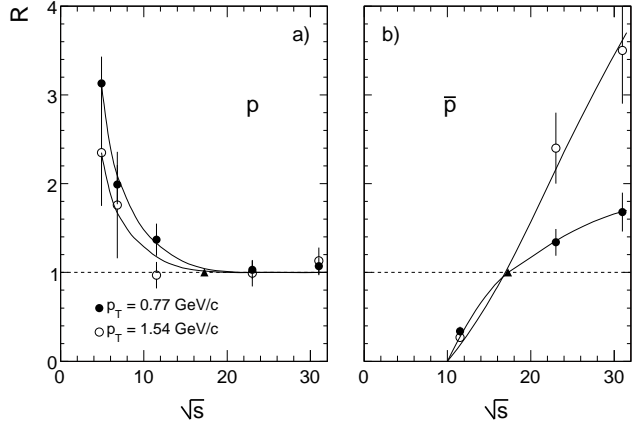}
  	\caption{The cross section ratios $R$ between the data from \cite{bib:blobel,bib:abramov,bib:alper,bib:guettler} 
  	               and NA49 as a function of $\sqrt{s}$ for two values of $p_T$ for a) protons and b) anti-protons. 
  	               In both of the panels the NA49 point is indicated with full triangle }
  	\label{fig:rat_comp}
  \end{center}
\end{figure}

In order to clarify this experimental situation one may
take reference to data at lower $\sqrt{s}$ and at ISR energies.
In Table~\ref{tab:isr_ps_comp} the cross section ratios between the PS experiment
of \cite{bib:blobel} at $\sqrt{s}$~=~4.9 and 6.8~GeV, the Serpukhov
experiment of \cite{bib:abramov} at $\sqrt{s}$~=~11.5~GeV and the
ISR measurements \cite{bib:alper,bib:guettler} at $\sqrt{s}$~=~23 and 31~GeV, and NA49 are 
given.  These data ratios are presented in Fig.~\ref{fig:rat_comp} as a function of $\sqrt{s}$.

\begin{table}[h]
\footnotesize
\begin{center}
\begin{tabular}{ccl@{$\pm$}rl@{$\pm$}rl@{$\pm$}rl@{$\pm$}rl@{$\pm$}r}
\hline
   $p_T$/$\sqrt{s}$ &   &  \multicolumn{2}{c}{4.9}     &\multicolumn{2}{c}{6.8}                      
                         &   \multicolumn{2}{c}{11.5} &  \multicolumn{2}{c}{23.0}   &\multicolumn{2}{c}{31.0} \\  
    \hline
  0.77   & $R_{\overline{\textrm{p}}}$   &    \multicolumn{2}{c}{}   & \multicolumn{2}{c}{}  &     
                                                            0.338 & 0.05 &   1.34 & 0.15 & 1.68  & 0.22   \\
            & $R_{\textrm{p}}$          & 3.13 & 0.30 & 1.99 & 0.22  & 1.37  & 0.18  & 1.02 & 0.10  & 1.07 & 0.10   \\ 
  1.54   & $R_{\overline{\textrm{p}}}$     &    \multicolumn{2}{c}{}   & \multicolumn{2}{c}{}  &    
                                                           0.270 & 0.05 & 2.40 & 0.40  & 3.50 & 0.60    \\
            & $R_{\textrm{p}}$            & 2.35 & 0.60 & 1.76 & 0.60 & 0.970 & 0.15 & 0.992 & 0.15 & 1.13 & 0.15 \\
            \hline
 \end{tabular}
\end{center}
\caption{The cross section ratios $R$ between the data from \cite{bib:blobel,bib:abramov,bib:alper} and NA49}
\label{tab:isr_ps_comp}
\end{table}

In Fig.~\ref{fig:rat_comp}a the very strong decrease of the central invariant proton
cross section up to SPS energies is evident. This decrease is compensated
by the comparatively strong increase of pair produced protons from
Serpukhov through SPS up to ISR energies which produces an effective
flattening of the $s$-dependence between $\sqrt{s}$~=~17.2 and 31~GeV
followed by a steady increase at higher energies. As explained in 
detail in \cite{bib:fischer} the proper subtraction of the yield of pair-produced 
protons results in a continued decrease of the net proton yield 
to about zero at the highest ISR energies.

As shown in Fig.~\ref{fig:rat_comp}b the increase of the anti-proton cross sections
from threshold through Serpukhov and SPS to ISR energies gives a 
consistent picture in the comparison of the experiments quoted
in Table~\ref{tab:isr_ps_comp}. The difference in the $s$-dependence between the lower
$p_T$ range at 0.77~GeV/c and the $p_T$ of 1.54~GeV/c should be noted.
It is evident also in the Cronin data, Fig.~\ref{fig:cron_comp}b.

In conclusion it appears that the data from \cite{bib:cronin} seem to be low
for baryons in comparison to all other available data, by $\sim$~25\% for protons
and $\sim$~50\% for anti-protons.

%
%
\section{ Comparison to ISR and RHIC data}
\vspace{3mm}
\label{sec:comp_isr}

As shown in Fig.~\ref{fig:cov} the ISR data on baryons cover the $x_F$/$p_T$ plane
quite extensively with a series of different spectrometer experiments
in the range of $\sqrt{s}$ from 23 to 63~GeV. The present paper will
limit the detailed comparison to the forward region at $x_F >$~0.1,
with the exception of the preceding chapter where a few points
at $x_F$~=~0 were included in order to clarify the experimental
situation. The reason for this limitation lies in the rapid
evolution of both the proton and anti-proton yields at central
rapidity and in the difficulties of defining "net" protons as
the difference between proton and pair-produced proton cross
sections. Here, the use of data from the isospin-reflected reaction
n+p $\rightarrow$ p, $\overline{\textrm{p}}$ is mandatory in order to fully understand the
isospin structure of baryon pair production \cite{bib:fischer}. The central area
will therefore be treated in a subsequent publication including
the neutron beam data available to NA49.

The main interest in regarding the forward ISR region of baryon 
production lies in a detailed study of $s$-dependence both of the
proton and anti-proton cross sections, especially in relation
to scaling concepts and to the question of form stability of
the $p_T$ and $x_F$ distributions. Two collaborations 
\cite{bib:alb_prot1,bib:alb_prot2,bib:alb_prot3,bib:alb_prot4,bib:alb_aprot1,bib:alb_aprot2,bib:capi}
have contributed data in forward direction, with more than 
1200 data points for protons and a comparatively rather limited
set of only about 100 points for anti-protons.

It should be remarked that all ISR data are corrected by us for baryon
feed-down from hyperon decay as described in Sect.~\ref{sec:comp_fd}.  

%
%
\subsection{Proton data \cite{bib:alb_prot2,bib:alb_prot3,bib:alb_prot4} from ISR}
\vspace{3mm}
\label{sec:prot_alb}

\begin{figure}[h]
  \begin{center}
  	\includegraphics[width=7.5cm]{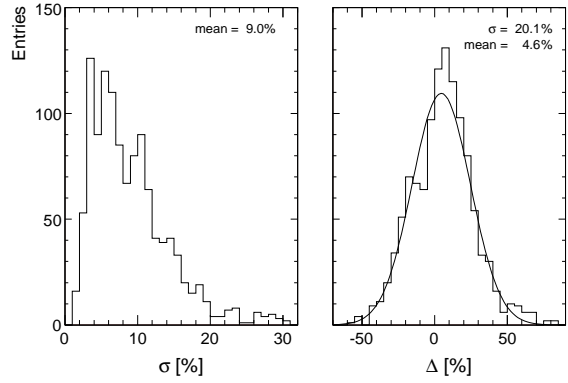}
  	\caption{Statistical analysis of the difference of the ISR measurements 
  	               \cite{bib:alb_prot1,bib:alb_prot2,bib:alb_prot3,bib:alb_prot4,bib:alb_aprot1} with 
  	               respect to NA49: a) error of the difference and b) difference of the measurements}
  	\label{fig:isr_stat}
  \end{center}
\end{figure}

The rich data set of 
\cite{bib:alb_prot2,bib:alb_prot3,bib:alb_prot4}, if compared directly and as a whole
to the NA49 data, reveals a discouragingly wide distribution of 
differences, Fig.~\ref{fig:isr_stat}, with an rms of twice the mean statistical error 
and a full width at base of more than $\pm$50\%. It will be demonstrated
below that this may be understood as the combination of two effects,
namely an apparent normalization uncertainty of about 10\% rms and
a very sizeable shape change of the $x_F$ distributions in the region
$x_F >$~0.7 which introduces systematic deviations of up to 30\%. In order to
bring this out clearly the comparison will be conducted in several
steps.

\begin{figure}[b]
  \begin{center}
  	\includegraphics[width=8.5cm]{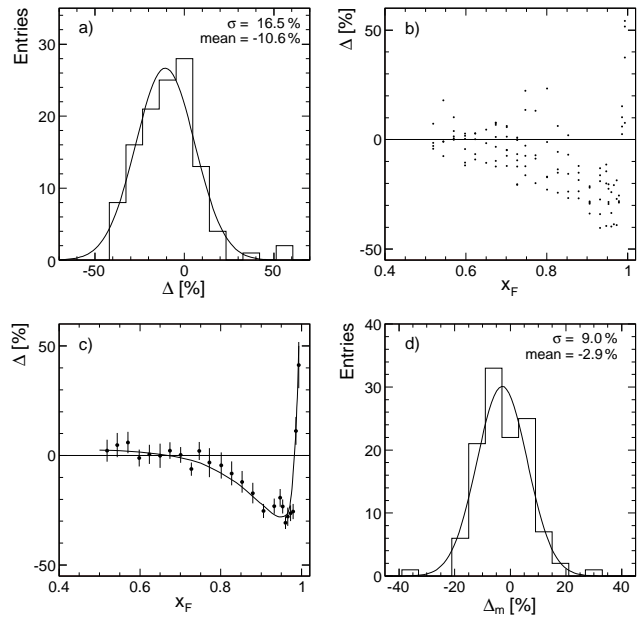}
  	\caption{Comparison between \cite{bib:alb_prot2} and NA49: a) distribution of differences,
  	               b) scatter plot of differences vs. $x_F$, c) mean values of differences over $p_T$ as a
  	               function of $x_F$, d) distribution of point-by-point difference to the mean value $\Delta_m$}
  	\label{fig:isr31}
  \end{center}
\end{figure}
   
A first step regards the data \cite{bib:alb_prot2} at $\sqrt{s}$~=~31 GeV (118 points),
covering a range from 0.5 to 1 in $x_F$ and from 0.47 to
1.08~GeV/c in $p_T$. The necessary feed-down correction to these
data is relatively small, ranging from 8\% at the lowest $x_F$ to
zero for $x_F >$~0.85.
The overall distribution of differences against NA49 is shown
in Fig.~\ref{fig:isr31}a where again the large width and a considerable offset
are evident. When however plotting the differences for each of the
25 available $x_F$ values separately, Fig.~\ref{fig:isr31}b, a sizeable depletion
of the ISR data above $x_F$~=~0.7 becomes visible, followed by a
rapid increase towards the diffractive peak at $x_F >$~0.97. The mean
values over $p_T$ at each $x_F$, Fig.~\ref{fig:isr31}c, indicate this trend with
good precision. When plotting the point-by-point differences to
this curve, Fig.~\ref{fig:isr31}d, the rms width is reduced to the expected
mean value of the statistical errors.

It should be pointed out that in the region below $x_F$~=~0.7 the mean 
difference is flat and close to zero with an offset of about +2.5\% 
with respect to the NA49 data. This is a first indication of
approximate scaling.

In a second step the data \cite{bib:alb_prot3} are compared to NA49. This data
set (134 points) covers a wide $p_T$ range from 0.17 to about 2~GeV/c 
with $x_F$ ranging from 0.3 to 0.7. These data are therefore below the 
region of depletion discussed above. The $\sqrt{s}$ ranges from 31 to
53~GeV.

\begin{figure}[h]
  \begin{center}
  	\includegraphics[width=8cm]{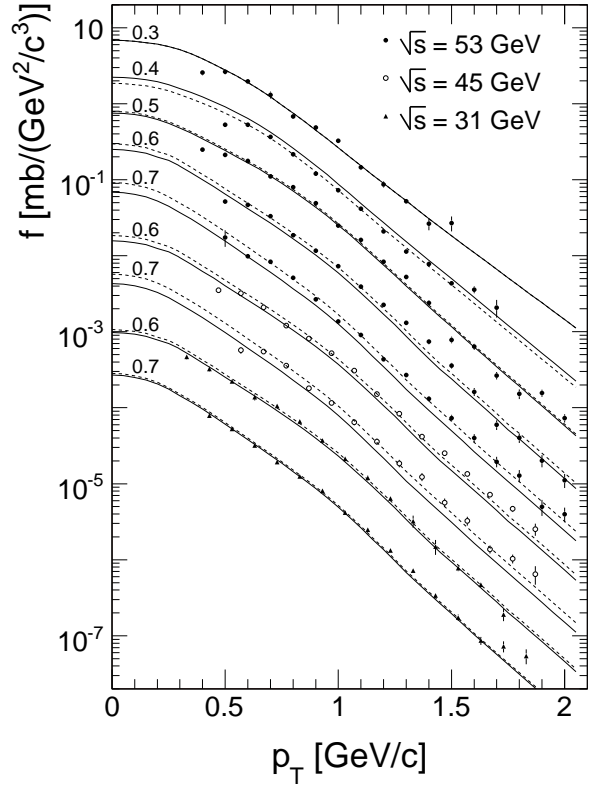}
  	\caption{Comparison of $p_T$ distribution at fixed $x_F$ values (indicated in the plot) from
  	               \cite{bib:alb_prot3} to interpolated NA49 data (full lines) and to interpolated NA49 data
  	               corrected with factors from Table~\ref{tab:offset} (dashed lines)}
  	\label{fig:isr_ptdist}
  \end{center}
\end{figure}

A first look at the 9 available $p_T$ distributions at the different
$\sqrt{s}$ and $x_F$ values as compared to the interpolated NA49 data,
Fig.~\ref{fig:isr_ptdist}, shows good agreement as far as the shape over the full
range of $p_T$ is concerned. 

There are however noticeable offsets with respect to NA49
which may be described by multiplicative factors as shown in
Table~\ref{tab:offset}.

\begin{table}[h]
\small
\begin{center}
\begin{tabular}{cccc}
\hline
$x_F$/$\sqrt{s}$  &   31   &    45    &   53     \\  \hline
             0.3            &          &            &  1.00   \\
             0.4            &          &            &  0.83   \\
             0.5            &          &            &  1.05   \\
             0.6            &  1.09 &  1.18   &  1.18   \\
             0.7            &  1.05 &  1.33   &  1.33   \\
\hline
 \end{tabular}
\end{center}
\caption{Offset factors with respect to NA49}
\label{tab:offset}
\end{table}
 
\begin{figure}[h]
  \begin{center}
  	\includegraphics[width=15cm]{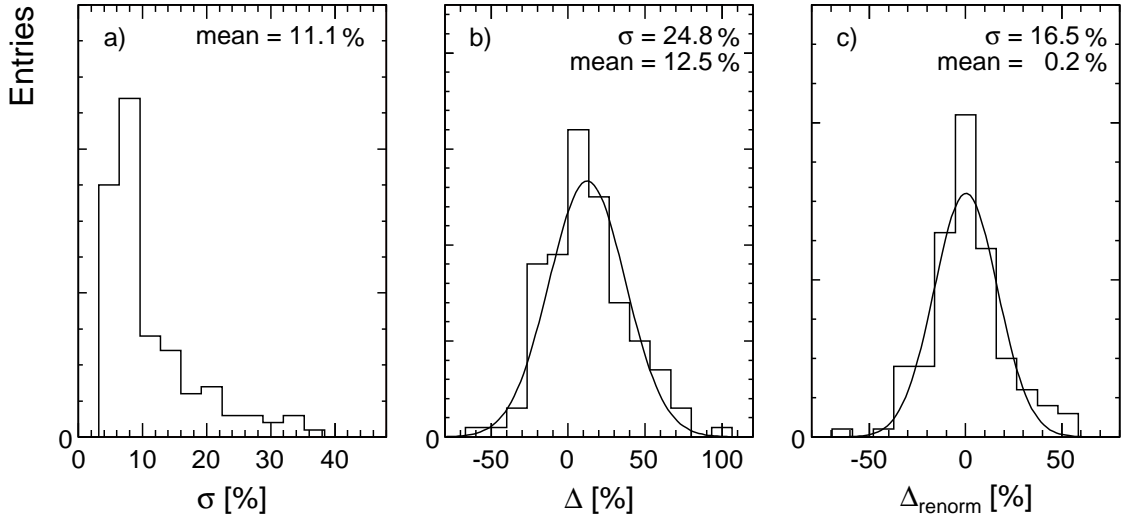}
  	\caption{Statistical analysis of the difference of the ISR measurement 
  	               \cite{bib:alb_prot3} with 
  	               respect to NA49: a) error of the difference, b) difference of the measurements and
  	               c) difference of the measurements after renormalization using the factors
                    of Table~\ref{tab:offset}}
  	\label{fig:isr1_diff}
  \end{center}
\end{figure}

Applying these factors to the ISR data the distribution of
differences to NA49 becomes centered at zero with a variance
which corresponds to the mean of the given statistical errors,
Fig.~\ref{fig:isr1_diff}.

The mean of the offset factors, including the first data set 
discussed above, amounts to 1.10. This might indicate a general
increase of the proton cross sections at ISR energies compared
to the SPS by this amount. The sizeable fluctuation of the offset 
with both $x_F$ and $\sqrt{s}$ shown in Table~\ref{tab:offset} indicates however at
least an additional normalization problem.

This problem can be quantified in a third step by comparing the
large data set \cite{bib:alb_prot4} with about 1000 data points spread over
9 different values of $\sqrt{s}$ from 23 to 62~GeV, with $p_T$ and
$x_F$ ranges of 0.3--1.7~GeV/c and 0.64--0.96, respectively. It should
be mentioned that this experiment did not have particle 
identification so that in the lower $x_F$ range a correction for
$\pi^+$ and $K^+$ had to be applied (see Sect.~\ref{sec:forw_rat}).

\begin{figure}[h]
  \begin{center}
  	\includegraphics[width=12cm]{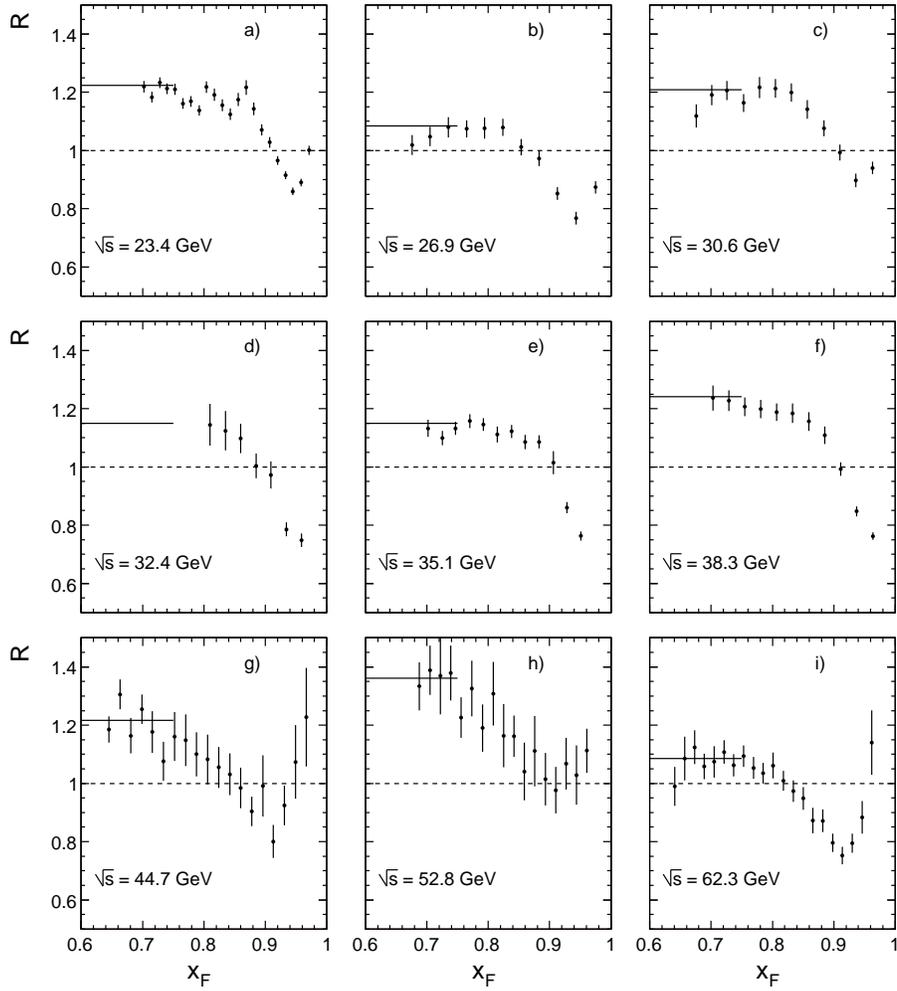}
  	\caption{The $p_T$ averaged ratio $R$ of results from \cite{bib:alb_prot4} to NA49 as a function of $x_F$
  	               for 9 different values of $\sqrt{s}$. The full lines denote the normalization factors shown in 
  	               Fig.~\ref{fig:normfac}}
  	\label{fig:isr_mean}
  \end{center}
\end{figure}

A first impression of the evolution of the invariant cross section
in the region above $x_F$~=~0.65 may be obtained from Fig.~\ref{fig:isr_mean} which
shows the $p_T$ averaged deviations from the NA49 data as a function
of $x_F$ for the nine $\sqrt{s}$ values. Although the depletion at
$x_F >$~0.8 is generally similar to the one shown at $\sqrt{s}$~=~31~GeV
(Fig.~\ref{fig:isr31}c) rather important overall deviations from unity in
the flat region below $x_F$~=~0.7 are visible, similar to the ones
given in Table~\ref{tab:offset} for $\sqrt{s}$~=~45 GeV. 

\begin{figure}[b]
  \begin{center}
  	\rotatebox{270}{\includegraphics[width=6cm]{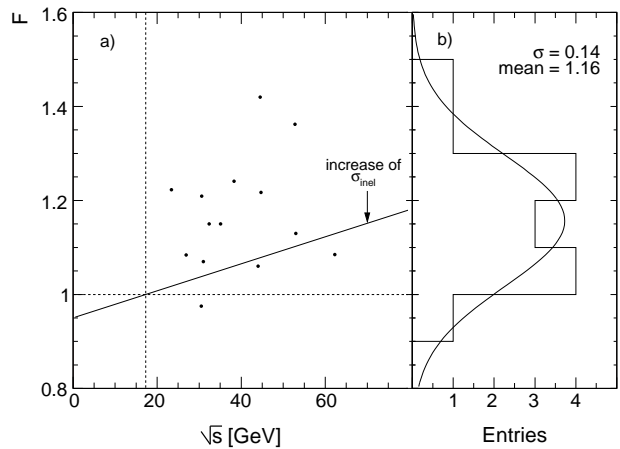}}
  	\caption{a) Normalization factors as a function of $\sqrt{s}$, b) distribution of normalization factors}
  	\label{fig:normfac}
  \end{center}
\end{figure}

Tentatively normalizing this lower $x_F$ region to the NA49 data 
one obtains the normalization factors given in Fig.~\ref{fig:normfac} as a
function of $\sqrt{s}$,including also the ones from Table~\ref{tab:offset}. The
projection of this distribution on the vertical axis shows
a wide spread with an rms of about 14\% and a mean of 1.16.

The variance is in agreement with the normalization uncertainty
given by the experiment. The offset might indicate a general
increase of the invariant cross section over the ISR energy
range by about this amount. This will be discussed in more detail
below.
 
\begin{figure}[t]
  \begin{center}
  	\includegraphics[width=16cm]{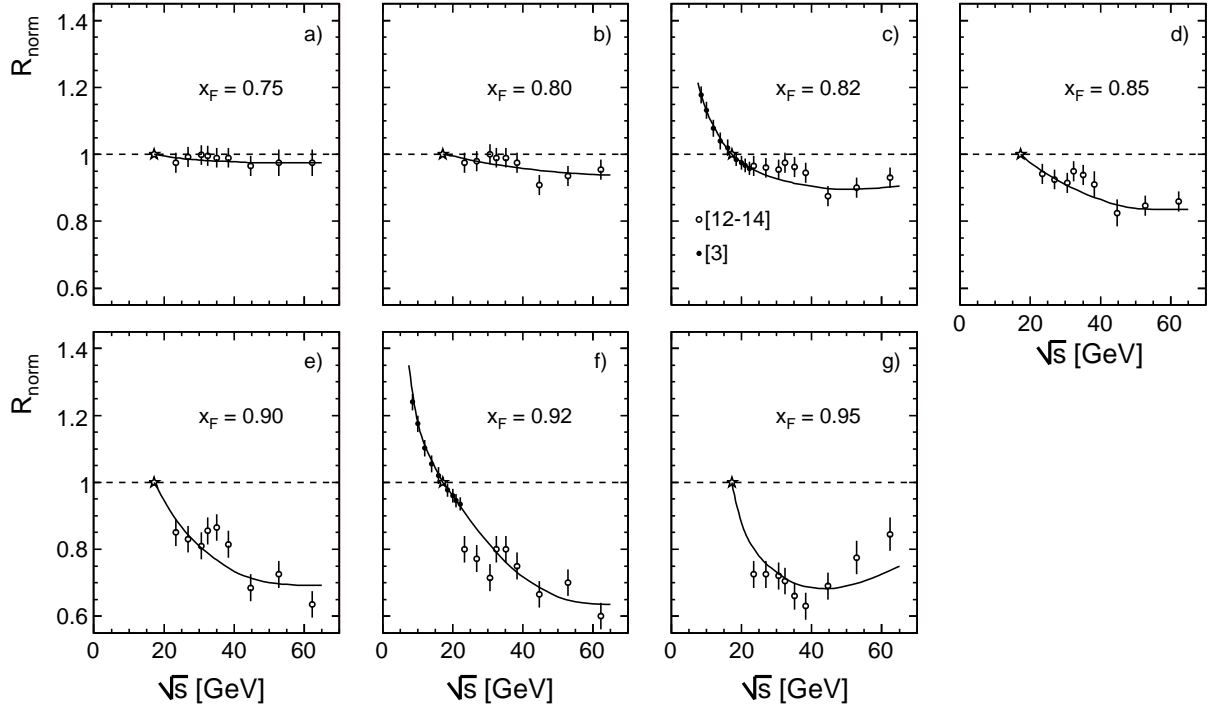}
  	\caption{Normalized mean ratios $R_{\textrm{norm}}$ as a function of $\sqrt{s}$ at fixed $x_F$. The star
  	              indicates the NA49 point }
  	\label{fig:isr_sdist}
  \end{center}
\end{figure}

As visible from Fig.~\ref{fig:isr_mean} the depletion at high $x_F$ develops in a
characteristic fashion as a function of $\sqrt{s}$. In order to
bring this evolution out more clearly the ISR data are normalized
to NA49 using the low-$x_F$ correction factors of Fig.~\ref{fig:normfac} and the mean
ratios plotted as a function of $\sqrt{s}$ at fixed $x_F$ in Fig.~\ref{fig:isr_sdist}.

In this Figure, the data from \cite{bib:sannes} have been included.
Their slope against $\sqrt{s}$ has been used in Sect.~\ref{sec:high_xf} to
correct the high-$x_F$ Fermilab data for $s$-dependence. As is visible
from Fig.~\ref{fig:isr_sdist} a consistent, smooth decrease of the invariant
cross sections from $\sqrt{s}$~=~8 to $\sqrt{s}$~=~63 GeV/c is experimentally
established. It continues to lower $\sqrt{s}$ with the data from
\cite{bib:blobel} not shown here. Seen as a function of $x_F$
this decrease starts at $x_F \sim$~0.75 with a few percent depletion
and reaches its maximum at $x_F \sim$~0.90--0.95 with an almost 40\% effect.

\begin{figure}[h]
  \begin{center}
  	\includegraphics[width=13cm]{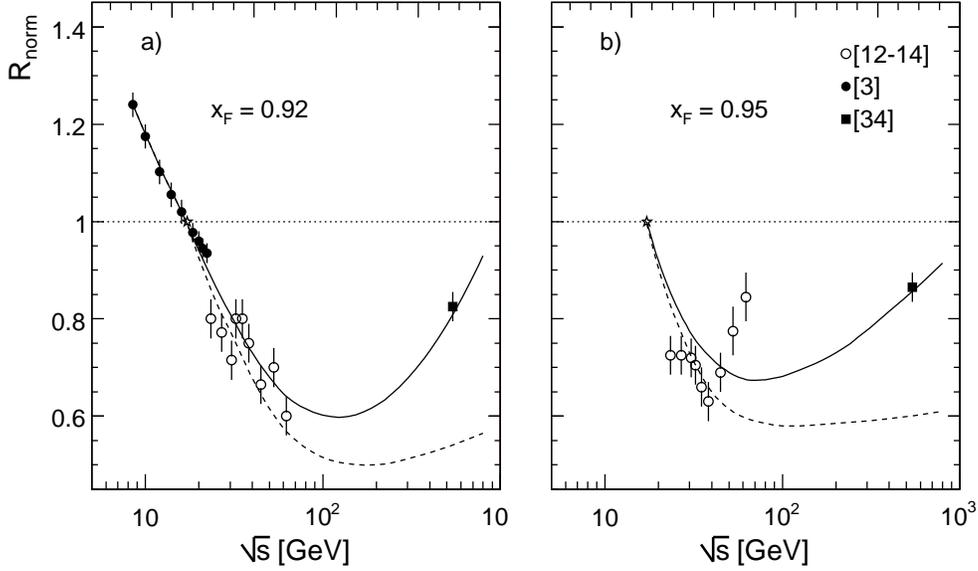}
  	\caption{Normalized mean ratio $R_{\textrm{norm}}$ as a function of $\sqrt{s}$ at fixed $x_F$ including the 
  	               UA4 measurement. The star indicates the NA49 point, full lines: invariant cross sections and
                   dashed lines: proton density per inelastic event}
  	\label{fig:isr_ua4}
  \end{center}
\end{figure}

In this context it is of course interesting to look at the 
higher $\sqrt{s}$ range of the p+$\overline{\textrm{p}}$ colliders. Only one data
set from the UA4 experiment \cite{bib:ua4} is available here which covers
the $x_F$ range from 0.92 to 1 with four $p_T$ values between 0.74 and
1.07~GeV/c. Applying the same method described above by averaging
over $p_T$ and normalizing to the NA49 data, the $\sqrt{s}$ dependence
shown in Fig.~\ref{fig:isr_ua4} is obtained.

Although the compatibility of the UA4 data with ISR results has
been noted in \cite{bib:ua4} the strong $s$-dependence from lower energies
implies a minimum of the invariant cross section at about
RHIC energy and a subsequent rise towards p+$\overline{\textrm{p}}$ collider energy,
Fig.~\ref{fig:isr_ua4}. This raises another question concerning baryon number
conservation. As the total inelastic cross section rises by
13\% at the highest ISR energy and by 48\% at $\sqrt{s}$~=~540~GeV as
compared to SPS energies, the proton density at high $x_F$ will
decrease faster than the invariant cross section with increasing
$s$. This is indicated by the dashed line in Fig.~\ref{fig:isr_ua4} which shows
the evolution of proton density rather than invariant cross 
section. In this case a flattening of the $s$-dependence up to
collider energy is not excluded. As at the same time the central
net proton density decreases at the higher ISR
energy range \cite{bib:fischer} the eventual scaling of the invariant cross 
section in the intermediate $x_F$ range has to be questioned.
Unless the whole decrease of proton density at low and high $x_F$ plus 
the increase of the inelastic cross section is absorbed into
increased neutron or heavy flavour (mostly strangeness) production, 
there should be problems with baryon number conservation. In this  
sense the mean increase by about 20\% of the invariant proton cross 
section which is visible in the average cross section ratios 
of Fig.~\ref{fig:normfac} might be real. In fact the percentage rise
of the inelastic cross section over the ISR energy range is also
indicated in this Figure. Although of course the large systematic
uncertainties in the ISR data do not allow for a definite 
statement, an upwards scaling violation of the invariant proton
cross section of 10-20\% over the ISR region cannot be excluded
at this stage. The interesting intermediate $\sqrt{s}$ region at RHIC 
energy is only covered in the interval 0.1~$< x_F <$~0.3 by the BRAHMS 
experiment \cite{bib:brahms} in two rapidity windows, see Sect.~\ref{sec:brahms}. 
However, recent data from deep inelastic leptoproduction at HERA 
\cite{bib:prot_hera,bib:neut_hera} help to fill the gap in $x_F$ up to the kinematic
limit at $\sqrt{s} \sim$~130~GeV, see Sect.~\ref{sec:hera}.

%
%
%
%
\subsection{Proton data \cite{bib:capi} from ISR}
\vspace{3mm}
\label{sec:prot_capi}

The data of Capiluppi et al. \cite{bib:capi} offer an additional set of proton
cross sections with 184 points at four ISR energies and $p_T$ and $x_F$
ranges of 0.16--1.38~GeV/c and 0.05--0.6 respectively. This coverage
has some overlap with the data \cite{bib:alb_prot2,bib:alb_prot3,bib:alb_prot4} discussed above.

\begin{figure}[h]
  \begin{center}
  	\rotatebox{270}{\includegraphics[width=5cm]{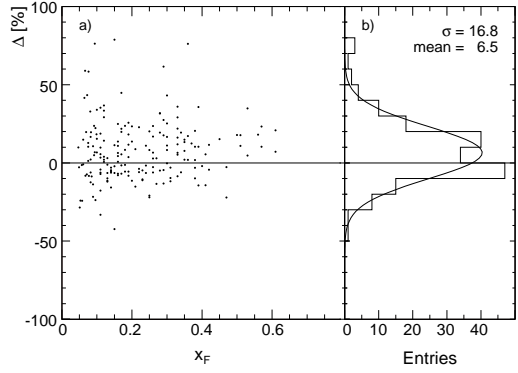}}
  	\caption{Comparison of the ISR measurement \cite{bib:capi} to 
  	               the NA49 results: a) difference $\Delta$ as a function of $x_F$ and b) distribution of the differences}
  	\label{fig:capi_prot}
  \end{center}
\end{figure}

Plotting again, after feed-down correction, the point-by-point 
differences to the NA49 data, Fig.~\ref{fig:capi_prot}, a picture similar to Fig.~\ref{fig:isr_stat}
emerges with an average offset of +6\% and an rms of 17\%. 

\begin{figure}[h]
  \begin{center}
  	\includegraphics[width=6cm]{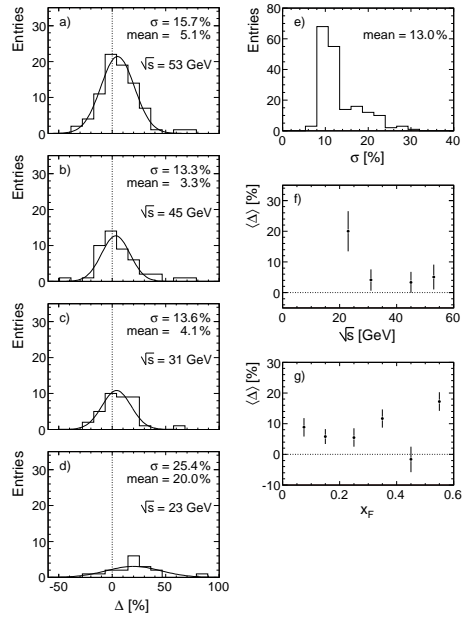}
  	\caption{Distributions of the differences $\Delta$ for different energies: a) $\sqrt{s}$~=~53~GeV,
  	               b) $\sqrt{s}$~=~45~GeV, c) $\sqrt{s}$~=~31~GeV, d) $\sqrt{s}$~=~23~GeV; 
  	               e) Distribution of the errors of \cite{bib:capi}; Mean difference as a function of f) $\sqrt{s}$
  	               and g) $x_F$}
  	\label{fig:capi_diff}
  \end{center}
\end{figure}

Given a mean statistical error of the data  \cite{bib:capi} of 13\%, Fig.~\ref{fig:capi_diff},
this variance indicates again additional normalization and/or $x_F$ and 
$\sqrt{s}$ dependences which are however much smaller than the ones found 
in the forward data of \cite{bib:alb_prot2,bib:alb_prot3,bib:alb_prot4}.
The distributions of the differences $\Delta$ with respect
to the NA49 data plotted separately for the four $\sqrt{s}$ values,
Fig.~\ref{fig:capi_diff}, indicate only a small if any s-dependence.  
There is also, within the statistical uncertainties, no discernible
$x_F$ dependence as shown by the mean differences as a function of $x_F$
in Fig.~\ref{fig:capi_diff}.

It should however be mentioned that in this $x_F$ region 
there are two counteracting phenomena to be taken into account.
Firstly there is the decrease of central net proton density with
increasing $\sqrt{s}$ in the approach to baryon transparency \cite{bib:fischer}.
Secondly there is the strong increase of pair produced protons
with $\sqrt{s}$, see Sect.~\ref{sec:aprot_alb} below. Both phenomena extend
over the region of $x_F$ studied here. A detailed discussion has
to take into account, as already mentioned in Sect.~\ref{sec:comp_cron} above,
the isospin structure of baryon pair production. This will be
elaborated in a subsequent publication. The observed overall
offset of about +6\% indicates again a possible upwards scaling
violation of the invariant cross section in the ISR energy range
on the 10\% level.

%
%
\subsection{Anti-proton data \cite{bib:alb_aprot1,bib:alb_aprot2} from ISR}
\vspace{3mm}
\label{sec:aprot_alb}

\begin{figure}[h]
  \begin{center}
  	\includegraphics[width=8.cm]{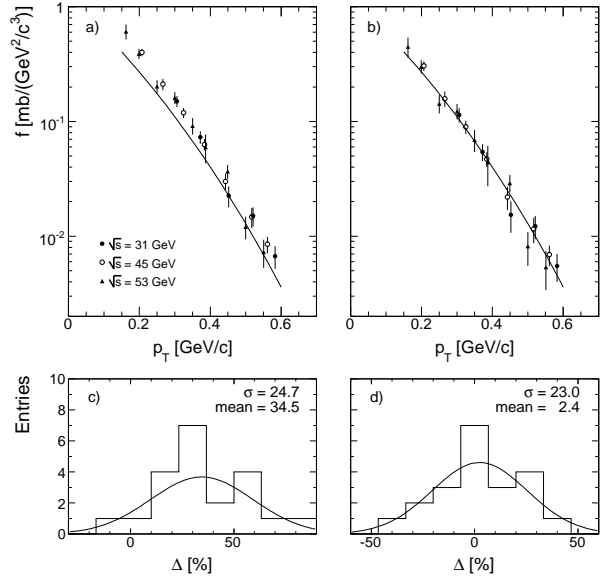}
  	\caption{Comparison between data from \cite{bib:alb_aprot2} (points) and NA49 results (lines) 
  	               as function of $p_T$: a) without feed-down correction of data \cite{bib:alb_aprot2}
  	               and b) with feed-down correction. Distribution of the differences: c) without feed-down correction
  	               and d) with feed-down correction}
  	\label{fig:isr_aprot}
  \end{center}
\end{figure}

The data of Albrow et al. \cite{bib:alb_aprot2} have been obtained at fixed angle
and for three ISR energies of 31, 45 and 53~GeV. They cover a range 
of 0.12 to 0.6 in $x_F$ and 0.16 to 0.8~GeV/c in $p_T$. The comparison
to the NA49 data is shown in Fig.~\ref{fig:isr_aprot}a without and \ref{fig:isr_aprot}b 
with feed-down subtraction. 

\begin{figure}[h]
  \begin{center}
  	\includegraphics[width=8.cm]{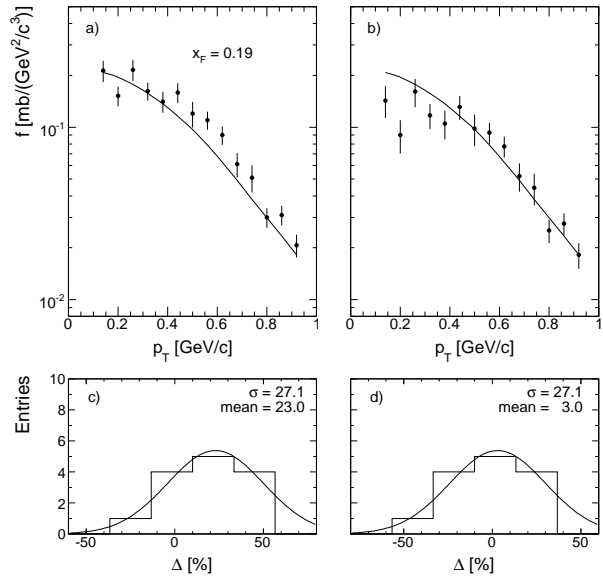}
  	\caption{Comparison between data from \cite{bib:alb_aprot1} (points) and NA49 results (lines) 
  	               as function of $p_T$: a) without feed-down correction of data \cite{bib:alb_aprot1}
  	               and b) with feed-down correction. Distribution of the differences: c) without feed-down correction
  	               and d) with feed-down correction}
  	\label{fig:isr_aprot30}
  \end{center}
\end{figure}

The difference distributions of Fig.~\ref{fig:isr_aprot} show an offset of 34\% for
the non-subtracted case which reduces to 2\% applying the feed-down
correction. The variance of the distributions is again somewhat
larger than the mean statistical error of 17\% necessitating an
additional fluctuation of the normalization of about 13\% rms
which complies with the estimated margin. There is no discernible
$s$-dependence in the ISR data itself, and no $s$-dependence up from
SPS energy after feed-down subtraction. This somewhat surprising
result is verified by the second measurement \cite{bib:alb_aprot1} which provides
14 data points at fixed $x_F$~=~0.19 and $p_T$ ranging from 0.14 to 0.92~GeV/c 
and $\sqrt{s}$~=~53~GeV/c, Fig.~\ref{fig:isr_aprot30}. 

Again there is an offset of +23\% without feed-down correction
which reduces to +3\% after subtraction. In this case the rms
fluctuation of the differences is about a factor of 1.8 above
the given statistical errors.

Taken at face value these results would establish a perfect scaling
of the anti-proton cross sections from $\sqrt{s}$~=~17 to $\sqrt{s}$~=~53 GeV
in the overlapping $x_F$ range between 0.1 and 0.4.   

%
%
\subsection{Anti-proton data \cite{bib:capi} from ISR}
\vspace{3mm}
\label{sec:aprot_capi}

\begin{figure}[h]
  \begin{center}
  	\rotatebox{270}{\includegraphics[width=8cm]{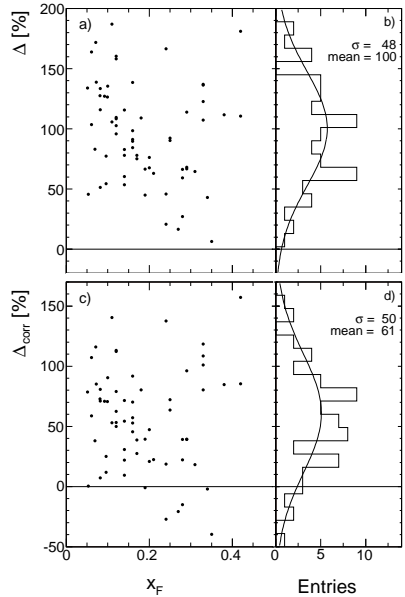}}
  	\caption{Comparison of the ISR measurement \cite{bib:capi} to the NA49 results without feed-down correction
  	              of \cite{bib:capi}: a) difference $\Delta$ as a function of $x_F$ and b) distribution of the differences,
  	              and with feed-down correction: c) difference $\Delta_{\textrm{corr}}$ as a function of $x_F$ and 
  	              d) distribution of the differences}
  	\label{fig:capi_aprot}
  \end{center}
\end{figure}

The anti-proton data from Capiluppi et al. \cite{bib:capi} cover, for the
four ISR energies 23, 31, 45 and 53~GeV, the $x_F$ range from 0.05 to 
0.42 and the $p_T$ range from 0.18 to 1.29~GeV/c. Hence there is
almost complete overlap with the data \cite{bib:alb_aprot1,bib:alb_aprot2}. Contrary to
\cite{bib:alb_aprot1,bib:alb_aprot2} however, the data comparison with NA49 shows a large
positive offset, see Fig.~\ref{fig:capi_aprot}, with means of +100\% without 
and +60\% with feed-down subtraction.

\begin{figure}[h]
  \begin{center}
  	\includegraphics[width=6cm]{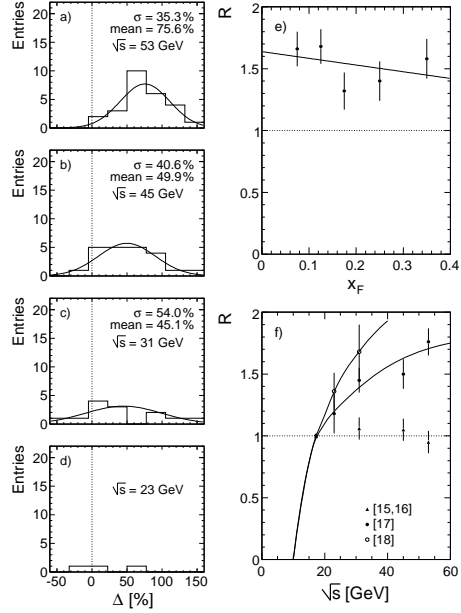}
  	\caption{Distributions of the differences $\Delta$ for different energies: a) $\sqrt{s}$~=~53~GeV,
  	               b) $\sqrt{s}$~=~45~GeV, c) $\sqrt{s}$~=~31~GeV, d) $\sqrt{s}$~=~23~GeV; 
  	                Mean difference as a function of e) $x_F$ and f) $\sqrt{s}$}
  	\label{fig:capi_diff_aprot}
  \end{center}
\end{figure}

When plotting the difference distributions separately for the 
different $\sqrt{s}$ values, Fig.~\ref{fig:capi_diff_aprot}a-d, a clear $s$-dependence becomes
evident, with mean values varying from +19\% at $\sqrt{s}$~=~23~GeV to
+74\% at $\sqrt{s}$~=~53~GeV. A small $x_F$ dependence cannot be excluded
as shown in Fig.~\ref{fig:capi_diff_aprot}e. In Fig.~\ref{fig:capi_diff_aprot}f the different $s$-dependences
treated in this paper, \cite{bib:guettler} at $x_F$~=~0 and $p_T$~=~0.77 GeV/c and 
\cite{bib:alb_aprot2} overlapping with \cite{bib:capi} at $\langle x_F \rangle $~=~0.19  
and $\langle p_T \rangle$~=~0.56~GeV/c are shown for comparison. Given the 
apparent strong $s$-dependence of the central anti-proton yields \cite{bib:alper,bib:guettler}, 
see Fig.~\ref{fig:rat_comp}, and the eventual decrease with $x_F$, Fig.~\ref{fig:capi_diff_aprot}, 
the results from \cite{bib:alper,bib:guettler} and \cite{bib:capi} may be
regarded as compatible within the sizeable systematic errors. The results from
Albrow et al. \cite{bib:alb_aprot1,bib:alb_aprot2} can however not be reconciled with the
observed dependences. This discrepancy remains unexplained,
especially in view of the fact that the proton and pion \cite{bib:pp_paper} yields
from the same experiment do not show deviations of comparable
magnitude. 
   
%
%
\subsection{Proton and anti-proton data  \cite{bib:brahms} from RHIC}
\vspace{3mm}
\label{sec:brahms}

\begin{figure}[b]
  \begin{center}
  	\includegraphics[width=13.5cm]{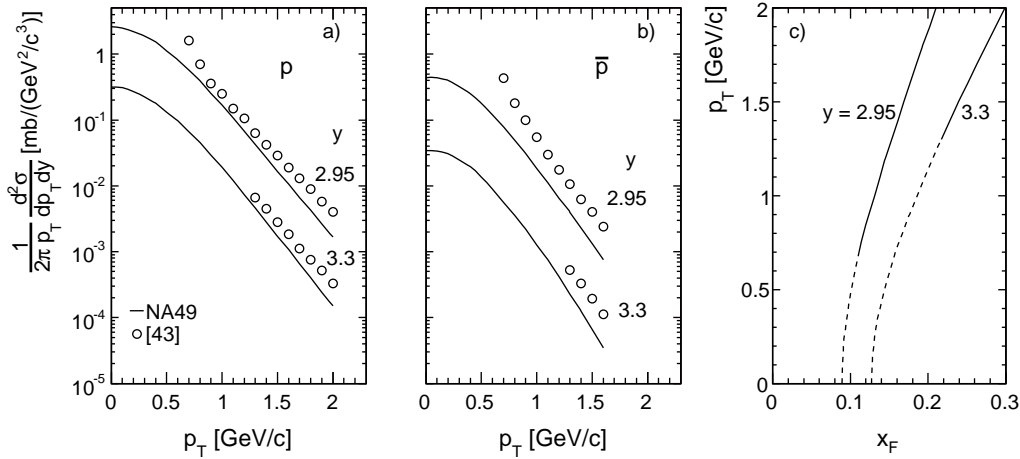}
  	\caption{Comparison of data from \cite{bib:brahms} with NA49 results as a function of $p_T$ at two
  	                rapidity values for a) protons and b) anti-protons. The measurements at $y$~=~3.3 are multiplied
  	                by 0.1 for better separation. Panel c) values of $x_F$ and $p_T$ corresponding to the two
                    rapidity windows of the BRAHMS experiment}
  	\label{fig:brahms_comp}
  \end{center}
\end{figure}

As explained above in the beginning of Sect.~\ref{sec:comp_isr}, the present
paper limits itself to the discussion and comparison of data
in the range $x_F >$~0.1. In view of the discussion of $s$-dependence
in Sects.~\ref{sec:prot_alb} and \ref{sec:prot_capi} it is of particular interest to include
data from RHIC into the comparison. The BRAHMS collaboration
has recently presented baryonic data  \cite{bib:brahms} from p+p collisions
at $\sqrt{s}$~=~200~GeV at the forward rapidities of 2.95 and 3.3
and at transverse momenta larger than 0.7 and 1.3~GeV/c, respectively. 
Viewed in the scaling variable $x_F$, Fig.~\ref{fig:brahms_comp}c, this
corresponds to a range from 0.1 to 0.3 which offers
considerable overlap with the NA49 experiment and the
ISR data of \cite{bib:capi}.
        
The invariant $p_T$ distributions of protons and anti-protons for the 
two rapidities are presented in 
Fig.~\ref{fig:brahms_comp}a,b together with the NA49 data interpolated to
the corresponding ($x_F$,$p_T$) values.

\begin{figure}[h]
  \begin{center}
  	\includegraphics[width=15cm]{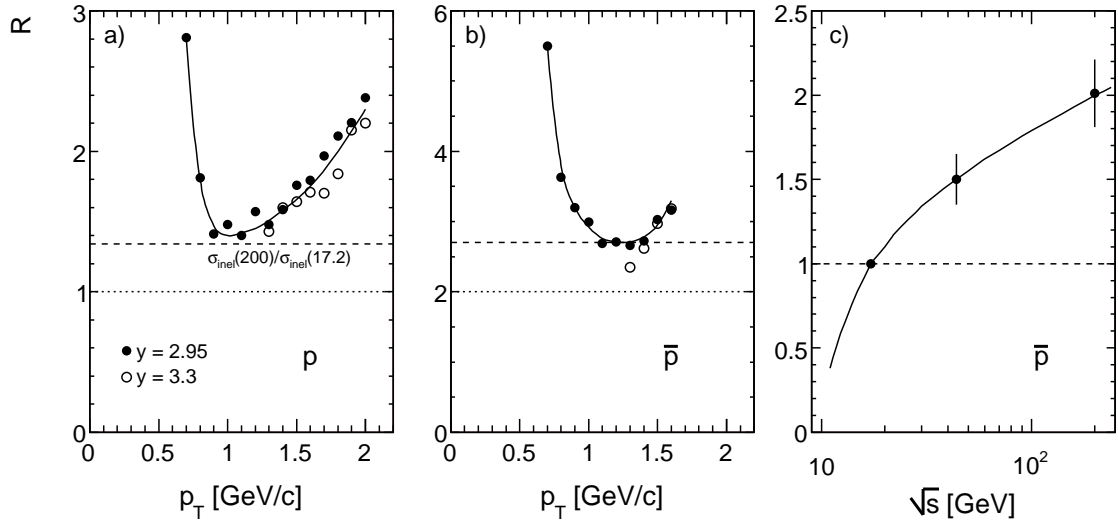}
  	\caption{Ratio $R$ between BRAHMS and NA49 cross sections as a function of $p_T$ at two
  	                rapidity values for a) protons and b) anti-protons; c) $R$ as function of $\sqrt{s}$ 
  	                including point from ISR \cite{bib:capi}.
  	                The ratio of inelastic cross sections $\sigma_{\textrm{inel}}(200)/\sigma_{\textrm{inel}}(17.2)$ is
  	                indicated in panel a) with a dashed line}
  	\label{fig:brahms_rat}
  \end{center}
\end{figure}

Several features are noteworthy in this comparison:

\begin{itemize}
\item the BRAHMS data for protons are very close for the two rapidity
         windows in the common $p_T$ range from 1.3 to about 1.6~GeV/c, see also Fig.~\ref{fig:brahms_rat}.
\item the same is true for the NA49 data. In the range $p_T >$~1.6 GeV/c
          the cross sections at the higher rapidity are depleted by
          similar amounts in both experiments.
\item at $p_T <$~0.9~GeV/c the BRAHMS data diverge sharply upwards from the
          NA49 distribution. 
\item a similar pattern emerges for the anti-proton data although
          the comparison is here limited to $p_T <$~1.6~GeV/c due to the
          counting statistics of NA49. There is however a general 
          depletion of the cross sections in passing from
          2.95 to 3.3 units of rapidity. 
\end{itemize}

This situation is quantified by the cross section ratios plotted
in Fig.~\ref{fig:brahms_rat}. The proton ratios, 
Fig.~\ref{fig:brahms_rat}a, decrease sharply from 2.8 at the
lower limit of the BRAHMS acceptance to values of about 1.4
in the range 0.9~$< p_T <$~1.3~GeV/c. This ratio is close to the
ratio of inelastic cross sections $\sigma_{\textrm{inel}}(200)/\sigma_{\textrm{inel}}(17.2)$
indicated as a line at 1.34 in Fig.~\ref{fig:brahms_rat}a. Tentatively attributing 
the low-$p_T$ divergence to an edge effect of the BRAHMS acceptance 
one may argue that in the region below $p_T \sim$~1~GeV/c the invariant cross
sections are scaled up by just the ratio of the inelastic cross
sections, see also the argumentation in Sect.~\ref{sec:prot_alb} concerning
$s$-dependence. This would mean that the proton densities in
the $x_F$ range considered here are $s$-independent with the exception
of the high-$p_T$ region above about 1~GeV/c where a substantial increase of 
$R$ is visible. Compare also the
discussion of the HERA data at $\sqrt{s}$~=~130~GeV in Sect.~\ref{sec:hera}
at low $p_T <$~0.6 GeV/c. 

For the anti-proton ratios, Fig.~\ref{fig:brahms_rat}b, a qualitatively similar
picture emerges, with the important exception that there is a
general increase of the invariant yields beyond the ratio of
the inelastic cross sections. The divergence at $p_T <$~0.9~GeV/c
is quantitatively the same as the one observed for protons, indicating 
problems at the lower edge of the BRAHMS acceptance also for anti-protons.
As already observed for protons the ratios for $y$~=~2.95 and $y$~=~3.3 
are quite similar and they tend to be constant at 1.0~$< p_T <$~1.4~GeV/c
followed by an increase towards higher $p_T$. For anti-protons
however, the flat part of the ratio corresponds to a value of 2.7.
Repeating the argument for protons by taking into account the 
increase of the total inelastic cross section, an effective 
increase by a factor of 2 of the anti-proton density is resulting. 
Combining this increase with the one observed at the mean $\sqrt{s}$ 
of the ISR data \cite{bib:capi}, see Fig.~\ref{fig:capi_diff_aprot}e, in this $x_F$ range, the $s$-dependence 
shown in Fig.~\ref{fig:brahms_rat}c may be extracted. 

In conclusion and within the $x_F$ range of 0.1 to 0.3 the scaling of proton densities 
rather than inclusive cross sections may be established from SPS through 
ISR up to RHIC energies. For anti-protons, a smooth  increase by about a 
factor of two is seen over the same $\sqrt{s}$ interval. For both particle 
types the yields increase towards  higher $p_T$ reaching for protons a 
factor of about 1.7 at 2~GeV/c as compared to SPS energy.
This increase should be confronted with the apparent $s$-independence of 
the shape of the proton $p_T$ dependences up to $\sqrt{s}$~=~53~GeV in 
this $p_T$ range as demonstrated in Sect.~\ref{sec:prot_alb}, Fig.~\ref{fig:isr_ptdist}. 
Taken at face value this would mean that there is a strong evolution of 
the transverse momentum dependence between ISR and RHIC energies.
Some basic differences between the ISR and RHIC experiments
have, however, to be taken into account in this respect. If the ISR experiments 
were triggering on typically more than 90\% of the total inelastic cross section, this is
not true for the RHIC situation. The BRAHMS experiment for instance triggers on only 
70\% of the inelastic cross section with a trigger device which spans angles between
0.6 and 4.4 degrees with respect to the beams. In addition, a coincidence between 
both rapidity hemispheres is requested. This means that single as well as double 
diffractive events are excluded from the trigger. If this in itself might not
introduce grave biases at least for proton production in the forward BRAHMS acceptance, 
see Sect.~\ref{sec:s4}, it is the apparent azimuthal asymmetry of the beam-beam 
trigger system on the spectrometer side which might cause systematic effects.
By pointing away from the spectrometer acceptance in the medium to high $p_T$ 
region it will tend to increase the measured high $p_T$ yield from simple energy-momentum 
conservation arguments. A strong azimuthal correlation between forward 
hadrons has indeed been observed in p+p interactions at the ISR \cite{bib:negra} 
with trigger particles at 1~$< p_T <$~4~GeV/c \cite{bib:cottrell} in the
$x_F$/$p_T$ wedge of the BRAHMS trigger. This correlation increases strongly with 
$p_T$ of both the trigger particle and the observed hadrons in the opposite azimuthal 
hemisphere. It is trivially explained by resonance decay governing the $p_T$ region in 
question \cite{bib:site,bib:andrzej}. In addition, comparing the forward pion 
yields measured by BRAHMS to the NA49 results \cite{bib:pp_paper} an increase 
of a factor of five is found at $p_T$~=~2~GeV/c and $y$~=~2.95, again in contrast 
to results at ISR energies. Also this effect is expected to follow from resonance 
production and decay. If extracting corrections for this trigger bias from microscopic 
hadronization models it must be ensured that production and decay of high mass states are properly
contained in these models, see also the discussion in \cite{bib:site,bib:andrzej}.  

%
%
\section{Integrated data}
\vspace{3mm}
\label{sec:ptint}

%
%
\subsection{$\bf p_T$ integrated distributions}
\vspace{3mm}
\label{sec:ptint_dist}

The $p_T$ integrated non-invariant and invariant baryonic yields 
are defined by:

\begin{align}
\label{eq:int}
  dn/dx_F &= \pi/\sigma_{\textrm{inel}} \cdot \sqrt{s}/2 \cdot 
             \int{f/E \cdot dp_T^2} \nonumber \\
  F &= \int{f \cdot dp_T^2}  \\
  dn/dy &= \pi/\sigma_{\textrm{inel}} \cdot \int{f \cdot dp_T^2} \nonumber  
\end{align} 
with $f= E \cdot d^3\sigma/dp^3$, the invariant double differential cross
section. The integrations are performed numerically using the
two-dimensional data interpolation (Sect.~\ref{sec:interp}). Table~\ref{tab:integr}
gives the numerical values and the first and second moments
of the $p_T$ distributions, as functions of $x_F$ and rapidity.

\vspace{3mm}
\begin{table}[h]
\scriptsize
\renewcommand{\tabcolsep}{0.10pc} 
\renewcommand{\arraystretch}{1.05} 
\begin{center}
\begin{tabular}{|l|cc|cc|cc|cc||cc|cc|cc|cc||c|c|c|}
\hline
 & \multicolumn{8}{|c||}{p} &\multicolumn{8}{|c||}{$\overline{\textrm{p}}$} & & p & $\overline{\textrm{p}}$ \\
\hline
 $x_F$& $F$& $\Delta$& $dn/dx_F$& $\Delta$& $\langle p_T \rangle$& $\Delta$
      & $\langle p_T^2 \rangle$ & $\Delta$
      & $F$& $\Delta$& $dn/dx_F$& $\Delta$& $\langle p_T \rangle$& $\Delta$
      & $\langle p_T^2 \rangle $& $\Delta$
      & $y$& $dn/dy$ & $dn/dy$ \\ \hline
 0.0  & 0.7413 & 0.21 & 0.5749  & 0.21 & 0.5165 & 0.08 & 0.3601 & 0.16 &
            0.1874 & 0.44 & 0.1477  & 0.42 & 0.4880 & 0.17 & 0.3156 & 0.31 &
 0.0   & 0.07364 & 0.01869 \\ 
 0.025  & 0.7494 & 0.16 & 0.5696  & 0.16 & 0.5187 & 0.09 & 0.3629 & 0.16 &
                0.1823 & 0.36 & 0.1407  & 0.35 & 0.4897 & 0.18 & 0.3176 & 0.34 &
  0.1   & 0.07412 & 0.01860 \\
 0.05  & 0.7746 & 0.14 & 0.5576  & 0.14 & 0.5212 & 0.06 & 0.3658 & 0.13 &
              0.1708 & 0.32 & 0.1247  & 0.31 & 0.4924 & 0.13 & 0.3216 & 0.27 &
  0.2   & 0.07477 & 0.01815 \\
 0.075   & 0.8169 & 0.14 & 0.5439  & 0.14 & 0.5226 & 0.05 & 0.3671 & 0.10 &
                 0.1532 & 0.34 & 0.1031  & 0.33 & 0.4972 & 0.13 & 0.3286 & 0.23 &
  0.3   & 0.07551 & 0.01759 \\
  0.1  & 0.8802 & 0.13 & 0.5351   & 0.13 & 0.5214 & 0.06 & 0.3655 & 0.12 &
             0.1348 & 0.34 & 0.08245 & 0.34 & 0.5038 & 0.16 & 0.3378 & 0.31 &
  0.4   & 0.07718 & 0.01681 \\
 0.125  & 0.9630 & 0.13 & 0.5321   & 0.13 & 0.5151 & 0.06 & 0.3585 & 0.09 &
                0.1155 & 0.45 & 0.06394 & 0.45 & 0.5109 & 0.18 & 0.3478 & 0.31 &
  0.5   & 0.07943 & 0.01587 \\
  0.15  & 1.0741  & 0.13 & 0.5388    & 0.13 & 0.5099 & 0.06 & 0.3510 & 0.11 &
               0.09723 & 0.50 & 0.04872 & 0.50 & 0.5185 & 0.24 & 0.3581 & 0.42 &
  0.6   & 0.08226 & 0.01479 \\
  0.2   & 1.3620   & 0.11 & 0.5682    & 0.11 & 0.4980 & 0.05 & 0.3341 & 0.09 &
              0.06671 & 0.52 & 0.02772  & 0.52 & 0.5252 & 0.24 & 0.3665 & 0.46 &
  0.7   & 0.08558 & 0.01360 \\
  0.25  & 1.6853   & 0.14 & 0.5944   & 0.14 & 0.4923 & 0.06 & 0.3242 & 0.10 &
               0.04198 & 0.77 & 0.01475 & 0.77 & 0.5296 & 0.33 & 0.3710 & 0.62 &
  0.8   & 0.09024 & 0.01237 \\
  0.3  & 2.0307   & 0.16 &  0.6165     & 0.16 & 0.4930 & 0.06 & 0.3216 & 0.11 &
             0.02401 & 1.08 &  0.007262 & 1.08 & 0.5361 & 0.43 & 0.3789 & 0.76 &
  0.9   & 0.09627 & 0.01103 \\
  0.35 & 2.3807   & 0.08 &  0.6323     & 0.08 & 0.4953 & 0.04 & 0.3220 & 0.07 &
              0.01318 & 1.13 &  0.003491 & 1.13 & 0.5394 & 0.49 & 0.3826 & 0.82 &
  1.0   & 0.10463 & 0.009639 \\      
  0.4   & 2.6341     & 0.10 &  0.6205     & 0.10 & 0.4978 & 0.04 & 0.3248 & 0.07 &
             0.006648 & 2.13 &  0.001562 & 2.14 & 0.5499 & 0.84 & 0.3911 & 1.53 &
  1.1   & 0.11465 & 0.008296 \\      
  0.45 & 2.8083 & 0.10 &  0.5938  & 0.10 & 0.4952 & 0.05 & 0.3237 & 0.08 &
          &&  &&  &&  &&  
  1.2   & 0.12713 & 0.007015 \\
  0.5   & 3.0140 & 0.14 &  0.5778 & 0.14 & 0.4830 & 0.07 & 0.3108 & 0.11 &
          &&  &&  &&  &&  
  1.3   & 0.14188 & 0.005733 \\
  0.55  & 3.2814 & 0.21 &  0.5740 & 0.21 & 0.4616 & 0.11 & 0.2891 & 0.18 &
           &&  &&  &&  &&  
  1.4   & 0.15901 & 0.004543 \\
  0.6   & 3.3827 & 0.25 &  0.5458 & 0.25 & 0.4498 & 0.16 & 0.2746 & 0.22 &
          &&  &&  &&  &&  
   1.5  & 0.17881 & 0.003424 \\
  0.65   &  3.3668 & 0.29 &  0.5032 & 0.29 & 0.4413 & 0.16 & 0.2645 & 0.25 &
           &&  &&  &&  &&  
   1.6  & 0.20063 & 0.002444 \\
  0.7   &  3.2902 & 0.36 &  0.4577 & 0.36 & 0.4326 & 0.17 & 0.2559 & 0.27 &
           &&  &&  &&  &&  
   1.7   & 0.22404 & 0.001646 \\ 
  0.75  &  3.3055 & 0.45 &  0.4301 & 0.45 & 0.4168 & 0.20 & 0.2402 & 0.32 &
           &&  &&  &&  &&  
  1.8   & 0.24574 & 0.001052 \\
  0.8   &  3.4796 & 0.54 &  0.4252 & 0.54 & 0.3978 & 0.19 & 0.2195 & 0.33 &
           &&  &&  &&  &&  
  1.9    & 0.25980 & 0.000615 \\ 
  0.85  &  3.7868 & 0.51 &  0.4362 & 0.51 & 0.3826 & 0.16 & 0.2032 & 0.29 &
           &&  &&  &&  &&  
  2.0    & 0.26832 & 0.000298 \\ 
  0.9    &  4.5527 & 0.53 &  0.4877 & 0.53 & 0.3663 & 0.19 & 0.1875 & 0.33 &
           &&  &&  &&  &&  
  2.1    & 0.27770 & 0.000110 \\ 
  0.95  &  6.8665 & 0.50 &  0.7056 & 0.50 & 0.3674 & 0.18 & 0.1859 & 0.31 &
           &&  &&  &&  &&  
  2.2    & 0.29182 & 0.000028 \\ 
           & &&  &&  &&  &&  
           &&  &&  &&  &&  
  2.3    & 0.30972 & 0.000005 \\ 
           & &&  &&  &&  &&  
           &&  &&  &&  &&  
  2.4    & 0.31161 & 0.0000007\\ 
           & &&  &&  &&  &&  
           &&  &&  &&  &&  
  2.5    & 0.30474 & \\ 
           & &&  &&  &&  &&  
           &&  &&  &&  &&  
  2.6    & 0.33347 & \\ 
           & &&  &&  &&  &&  
           &&  &&  &&  &&  
  2.7    & 0.41145 & \\ 
           & &&  &&  &&  &&  
           &&  &&  &&  &&  
  2.8    & 0.51284 & \\ 
           & &&  &&  &&  &&  
           &&  &&  &&  &&  
  2.9    & 0.26117 & \\  \hline
\end{tabular}
\end{center}
\caption{$p_T$ integrated invariant cross section $F$ [mb$\cdot$c],
         density distribution $dn/dx_F$, mean transverse momentum $\langle p_T \rangle $
         [GeV/c], mean transverse momentum squared $\langle p_T^2 \rangle $ 
         [(GeV/c)$^2$] as a function of $x_F$, as well as density distribution 
         $dn/dy$ as a function of $y$ for p and $\overline{\textrm{p}}$. The relative statistical 
         uncertainty $\Delta$ for each quantity is given in \% }
\label{tab:integr}
\end{table}

\begin{figure}[h]
  \begin{center}
  	\includegraphics[width=13cm]{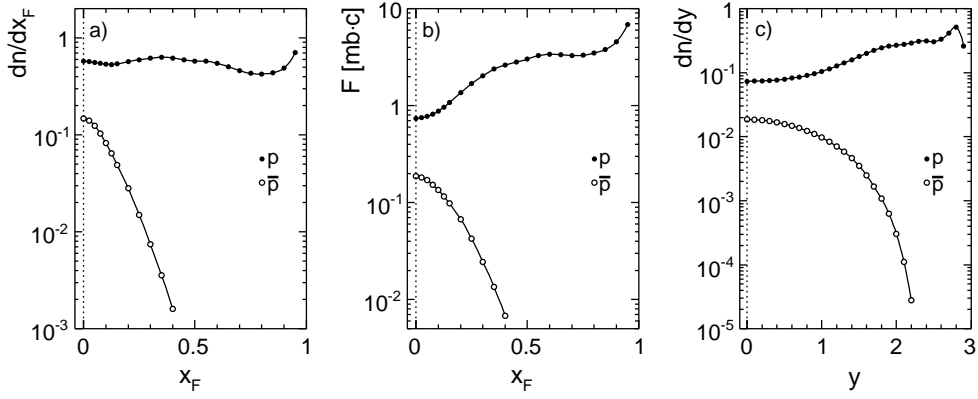}
  	\caption{Integrated distributions of p and $\overline{\textrm{p}}$ produced in p+p 
                   interactions at 158~GeV/c:
                  a) density distribution $dn/dx_F$ as a function of $x_F$;
                  b) invariant cross section $F$ as a function of $x_F$;
                  c) density distribution $dn/dy$ as a function of $y$}
  	\label{fig:ptint}
  \end{center}
\end{figure}

\begin{figure}[h]
  \begin{center}
  	\includegraphics[width=13cm]{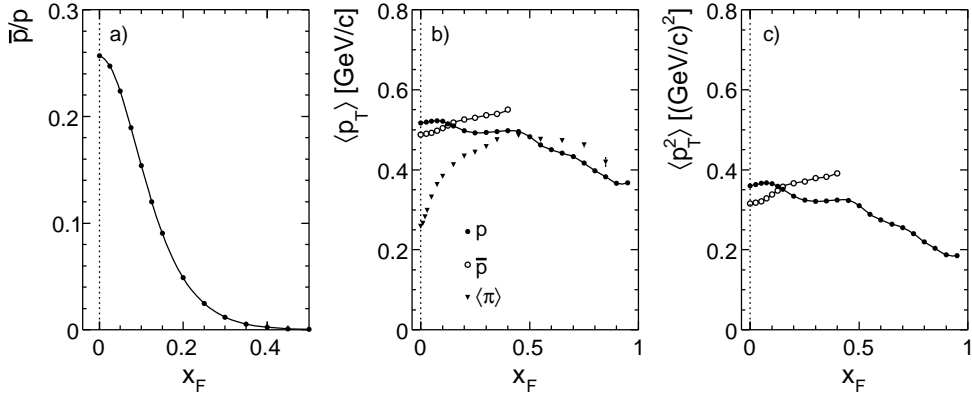}
  	\caption{a) $\overline{\textrm{p}}$/p ratio, b) mean $p_T$, and c) mean $p_T^2$ as a function 
                   of $x_F$ for p and  $\overline{\textrm{p}}$ produced in p+p interactions at 158~GeV/c.
                   In panel b) the mean $p_T$ of $\langle \pi \rangle$ in also shown}
  	\label{fig:ratint}
  \end{center}
\end{figure}

The corresponding distributions are shown in Figs.~\ref{fig:ptint} and \ref{fig:ratint} for
protons and anti-protons. The statistical errors of the integrated quantities are below the
percent level with the exception of the anti-proton yields above
$x_F$~=~0.2 due to the limited size of the total data sample of 
4.8~Mevents. This also sets a limit to the exploration of the
interesting evolution of the mean transverse momentum of the
anti-protons, Fig.\ref{fig:ratint}b, which rises from $x_F$~=~0 to increase above
the values for protons at $x_F >$~0.2. The similar behaviour of the
mean pion transverse momentum \cite{bib:pp_paper} with a cross-over at $x_F$~=~0.5 
is also indicated in this Figure. The sizeable $\langle p_T \rangle$ of about 0.5~GeV/c 
for all particle species at $x_F \sim$~0.5 remains a challenge to most 
current hadronization models. 

%
%
\subsection{Comparison to other data}
\vspace{3mm}
\label{sec:ptint_comp}

\begin{figure}[h]
  \begin{center}
  	\includegraphics[width=12cm]{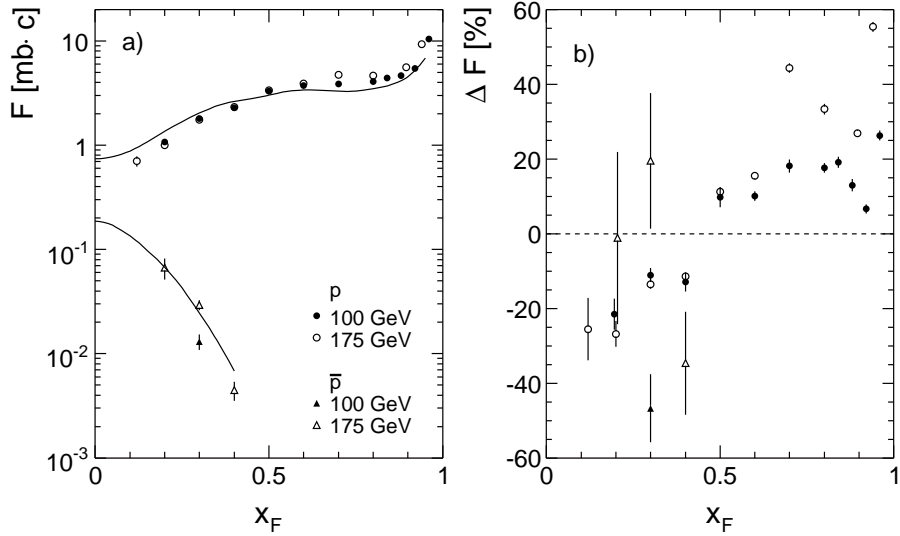}
  	\caption{a) Comparison of $p_T$ integrated invariant cross section $F$ as a function of 
  	               $x_F$ for p and $\overline{\textrm{p}}$ measured by \cite{bib:brenner} to 
  	               NA49 results (represented as lines); 
                   b) Deviation of the measurements of \cite{bib:brenner} from the NA49 results in percent}
  	\label{fig:bren_int_comp}
  \end{center}
\end{figure}

Sufficient $p_T$ coverage is needed to come to a bias-free evaluation
of the integrated quantities defined above. The danger of using
straight-forward analytic descriptions of limited data sets is
illustrated in the comparison to the integrated yields of the
Brenner et al. data \cite{bib:brenner}. As shown in Fig.~\ref{fig:bren_int_comp} large and systematic
deviations are resulting using data which are compatible on the
few percent level for the measured double differential cross 
sections, see Sect.~\ref{sec:comp_bren}. 

Here the apparent under-estimation of the related systematic
uncertainties visible in the given error bars, Fig.~\ref{fig:bren_int_comp}b, is 
especially noteworthy. The systematic trend as a function of
$x_F$ happens to be opposite but equal in size to the one observed
for pions \cite{bib:pp_paper}.

\begin{figure}[t]
  \begin{center}
  	\includegraphics[width=12cm]{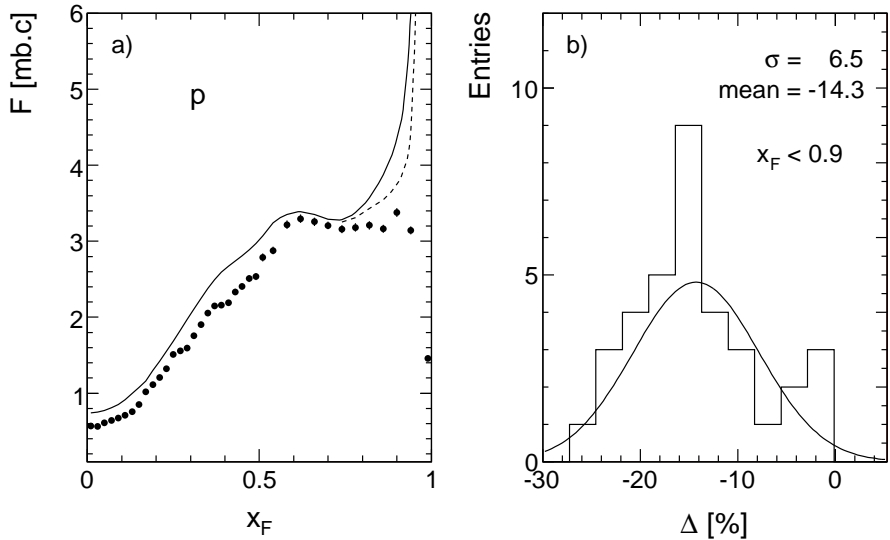}
  	\caption{a) Comparison of $p_T$ integrated invariant cross section $F$ as a function of 
                  $x_F$ for p measured by \cite{bib:ehs} to NA49 results (represented as lines); 
                   b) Distribution of the differences between measurements 
                   of \cite{bib:ehs} and NA49 in percent in the region of $x_F <$~0.9}
  	\label{fig:ehs_fprot_comp}
  \end{center}
\end{figure}

In comparison, the EHS experiment at the CERN SPS \cite{bib:ehs} using
a 400~GeV/c proton beam offers the necessary phase space coverage 
although this collaboration did not publish double differential 
data. The invariant integrated data presented in Fig.~\ref{fig:ehs_fprot_comp} show 
indeed a reasonable overall agreement as a function of $x_F$, with a 
few noticeable exceptions. For protons, Fig.~\ref{fig:ehs_fprot_comp}, there is strong 
disagreement above $x_F$~=~0.9. In fact the EHS data show no indication 
at all of the presence of a diffractive peak. Even correcting 
the NA49 data for the $s$-dependent depletion in this area following
Sect.~\ref{sec:comp_isr} and also shown in Fig.~\ref{fig:ehs_fprot_comp} with a dashed line, 
this discrepancy remains present.

Evidently the trigger efficiency of only 77\% of the total inelastic
cross section (compared to 89\% for the NA49 experiment) leads
to uncorrected losses in the diffraction region of protons.
In addition, correlated trigger bias corrections similar but
sizeably bigger than in the NA49 case, see Sect.~\ref{sec:s4} and \cite{bib:pp_paper},
have to be expected. This might explain part of the systematic
downward shift of the invariant density by about 14\% in the
$x_F$ region below 0.9, Fig.~\ref{fig:ehs_fprot_comp}b, which in view of the discussion of 
$s$-dependence in Sect.~\ref{sec:comp_isr} is in contradiction to the accumulated 
ISR data. 

\begin{figure}[h]
  \begin{center}
  	\includegraphics[width=12cm]{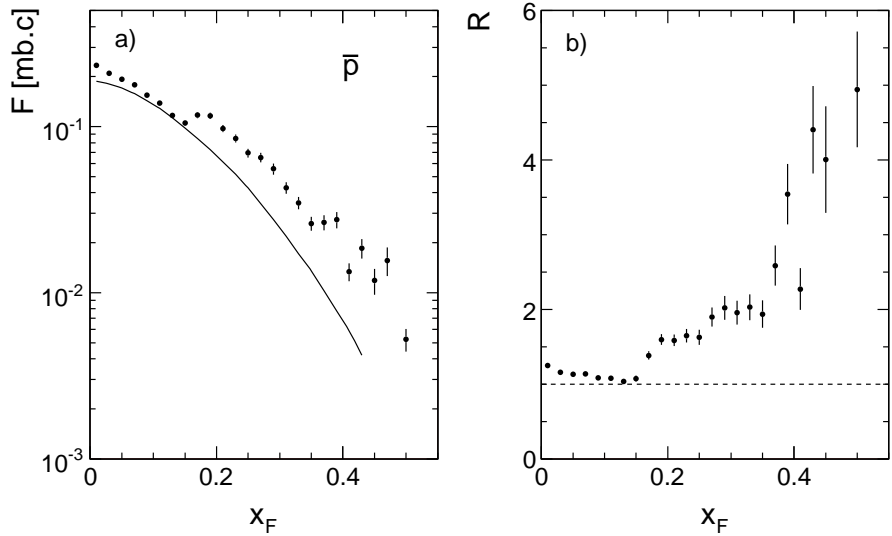}
  	\caption{a) Comparison of $p_T$ integrated invariant cross section $F$ as a function of 
                  $x_F$ for $\overline{\textrm{p}}$ measured by \cite{bib:ehs} to 
                   NA49 results (represented as lines); 
                   b) Ratio $R$ as a function of $x_F$ between measurements of \cite{bib:ehs} and NA49 }
  	\label{fig:ehs_faprot_comp}
  \end{center}
\end{figure}

For anti-protons, Fig.~\ref{fig:ehs_faprot_comp}, an expected increase with $\sqrt{s}$ is
borne out by an overall upward shift of about 12\%.

\begin{figure}[b]
  \begin{center}
  	\includegraphics[width=15cm]{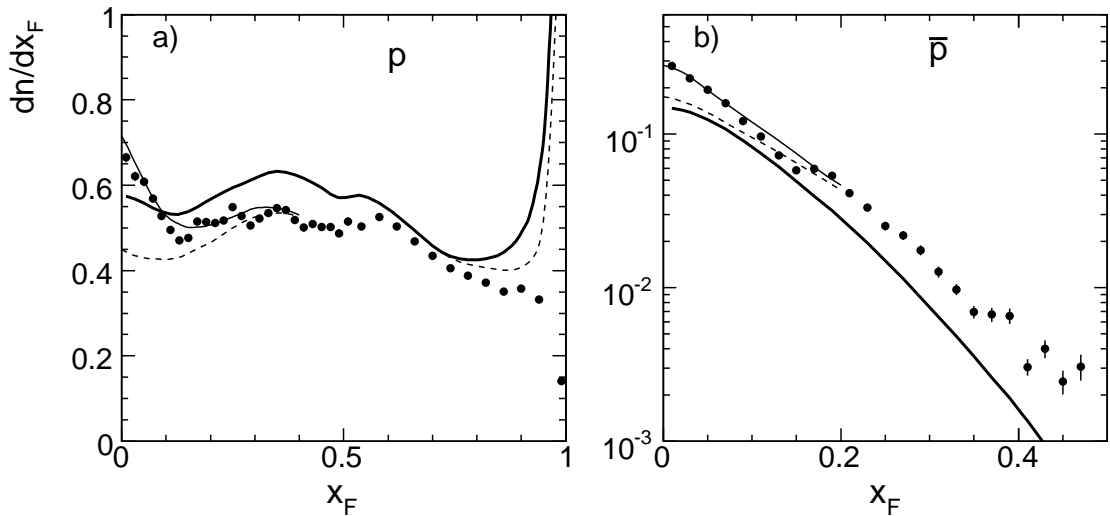}
  	\caption{Comparison of $p_T$ integrated non-invariant density $dn/dx_F$ as a function of 
                  $x_F$ for a) p and b) $\overline{\textrm{p}}$ measured by \cite{bib:ehs} to 
                   NA49 results (represented as thick lines). The difference between the thin and
                   dashed lines shows the influence of the $\sqrt{s}/2E$
                   factor in Eq.~\ref{eq:int} with respect to a scaling invariant
                   cross section }
  	\label{fig:ehs_dndx_comp}
  \end{center}
\end{figure}

\begin{figure}[h]
  \begin{center}
  	\includegraphics[width=15.5cm]{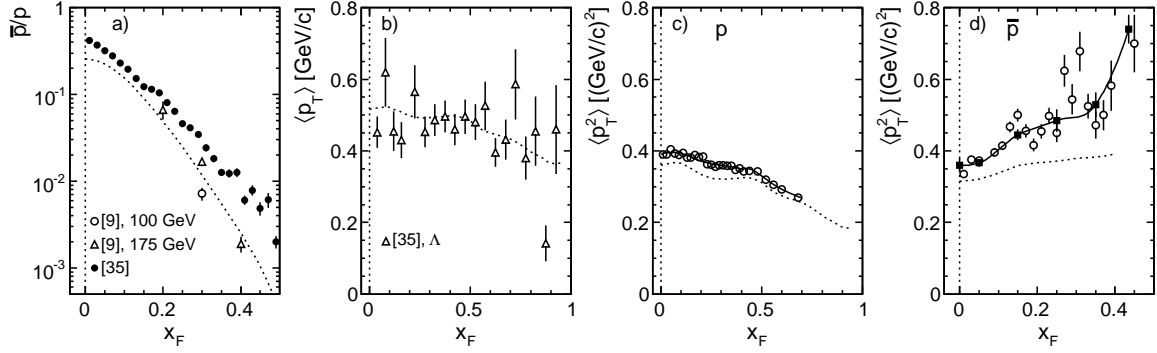}
  	\caption{ Comparison as a function of $x_F$ of 
                   a) $\overline{\textrm{p}}$/p ratio  measured by \cite{bib:ehs} (full circles) and 
                   \cite{bib:brenner} (open symbols) to NA49 (dotted line) , 
                   b) mean $p_T$ of $\Lambda$ measured by \cite{bib:bai} to mean $p_T$ of protons measured
                   by NA49 (dotted line); Comparison of mean $p_T^2$ for 
                   c) p and d) $\overline{\textrm{p}}$ measured by \cite{bib:ehs} to NA49 results (dotted lines)}
  	\label{fig:ehs_rat_comp}
  \end{center}
\end{figure}

There is however a strong local structure at
$x_F$ between 0.1 and 0.2 which is also present
in the proton data (Fig.~\ref{fig:ehs_dndx_comp}a) and which is in
all probability due to apparatus effects. In addition the strong
and apparently divergent increase of the anti-proton
yields for $x_F >$~0.2, Fig.~\ref{fig:ehs_faprot_comp}b, contradicts
the flat $x_F$ dependence of the $\overline{\textrm{p}}$ enhancement
at ISR energies, Fig.~\ref{fig:capi_diff_aprot}e. This effect is probably
connected to the divergence of $\langle p_T^2 \rangle$ in the
same $x_F$ region, Fig.~\ref{fig:ehs_rat_comp}d. It is also to be
compared to the erratic behaviour of the pion cross 
sections from this experiment in the same $x_F$ region  \cite{bib:pp_paper}.
                 
The non-invariant density distributions $dn/dx_F$ for protons and
anti-protons are shown in Fig.~\ref{fig:ehs_dndx_comp}. They demonstrate the strong $s$
dependence introduced by the factor $\sqrt{s}/E$ in Eq.~\ref{eq:int}
above. Only at $x_F >$~0.2 this factor reduces to the simple
multiplicative term $1/x_F$.

The increase of particle density at $x_F \sim$~0 is practically equal to
the increase of $\sqrt{s}$. This means that for an $s$ independent invariant
cross section at low $x_F$ the total proton density will diverge
with $s$ in this region, thus creating a problem with baryon number
conservation \cite{bib:fischer}.

\begin{figure}[h]
  \begin{center}
  	\includegraphics[width=12cm]{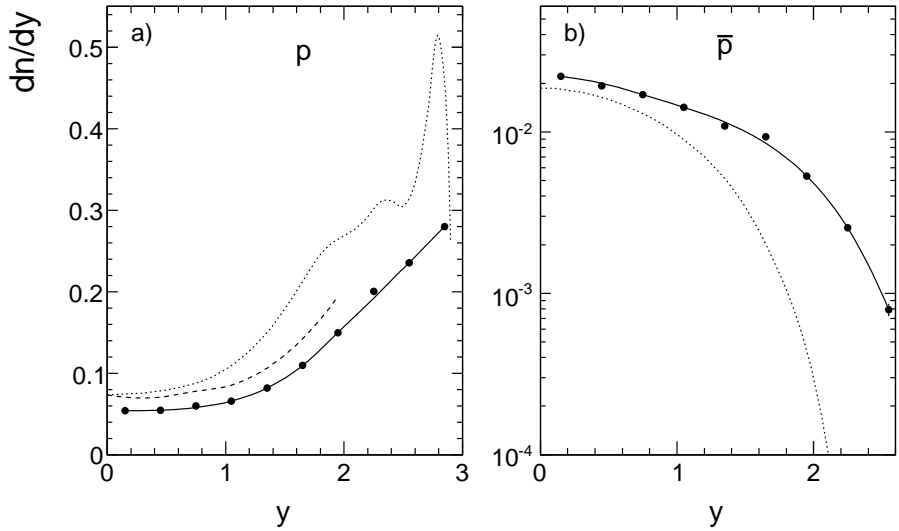}
  	\caption{Comparison of $p_T$ integrated density $dn/dy$ as a function of 
                  $y$ for a) p and b) $\overline{\textrm{p}}$ measured by \cite{bib:ehs} to 
                   NA49 results (dotted lines) }
  	\label{fig:ehs_dndy_comp}
  \end{center}
\end{figure}

The rapidity distributions $dn/dy$ of \cite{bib:ehs}  are presented for protons 
and anti-protons in Fig.~\ref{fig:ehs_dndy_comp}.

Here the extension of the $y$ scale with increasing $\sqrt{s}$ should be noted,
which is visualized in the EHS data re-normalized to NA49 at $x_F$~=~0
also shown in Fig.~\ref{fig:ehs_dndy_comp} (dashed line). The shape comparison of hadronic rapidity 
distributions at different $\sqrt{s}$ hence suffers non-negligible 
systematic effects which are to be carefully taken into account. 

Finally a comparison of the $p_T$ integrated $\overline{\textrm{p}}$/p ratio and of the
first and second moment of the $p_T$ distributions as a function of
$x_F$ is presented in Fig.~\ref{fig:ehs_rat_comp}.

As there is no published $ \langle p_T \rangle$ distribution available, the mean
transverse momentum of Lambdas from EHS \cite{bib:bai} is compared to protons
in Fig.~\ref{fig:ehs_rat_comp}b. As far as $ \langle p_T \rangle$ and $ \langle p_T^2 \rangle $ 
are concerned, the measurements
at the higher $\sqrt{s}$ follow, at increased levels, rather closely
the shape of the NA49 data as a function of $x_F$. This has already
been apparent for pions \cite{bib:pp_paper}.  It remains however to be shown how 
much of the apparent increase has to be imputed to the absence of 
diffraction in the EHS data as opposed to a true $s$-dependence. 
In this context the even smaller fraction of the total inelastic 
cross section generally available for triggering at collider energies 
has to be mentioned. Also here the effects of this trigger bias should
be evaluated before detailed conclusions may be drawn in comparison to
lower energy data.

%
%
\subsection{Total baryonic multiplicities}
\vspace{3mm}
\label{sec:total}

The integration over $x_F$ of the $dn/dx_{F}$ distributions presented in Table~\ref{tab:integr}
results in the following total baryonic yields:

\begin{align}                     
  \langle n_{\textrm{p}}\rangle                                                                     &= 1.1623 \nonumber \\
  \langle n_{\overline{\textrm{p}}}\rangle                                                   &= 0.03860 \\ 
  \langle n_{\overline{\textrm{p}}}\rangle/\langle n_{\textrm{p}}\rangle &= 0.03321 \nonumber
\end{align}         

The statistical errors of these quantities are negligible compared
to the overall systematic uncertainty of about 2--3\% given in 
Table~\ref{tab:syst}.

%
%
\subsection{Availability of the presented data}
\vspace{3mm}
\label{sec:avail}
As in \cite{bib:pp_paper,bib:pc_paper} the tabulated values of NA49 data are available in 
numerical form on the Web Site \cite{bib:site}. In addition, the ($x_F$,$p_T$) distributions following from the
two-dimensional interpolation, Sect.~\ref{sec:interp}, are made available on this site.
%
%
\section{Neutrons}
\vspace{3mm}
\label{sec:neut}

%
%
\subsection{NA49 results}
\vspace{3mm}
\label{sec:neut_na49}

The unfolded $x_F$ distribution of the $p_T$ integrated neutron
yield has been shown in Fig.~\ref{fig:neut_raw}. In this yield there is no
distinction between the different neutral hadronic particles.
The measured cross section is therefore the sum of neutrons, pair
produced neutrons, anti-neutrons and K$^0_L$ particles which are 
experimentally inseparable. As in the proton cross sections 
presented in this paper the contribution of pair produced protons 
has not been subtracted, the neutron yield may be defined
as the total measured neutral hadron yield minus the K$^0_L$
and the anti-neutron contribution. 

\begin{figure}[h]
  \begin{center}
  	\includegraphics[width=10cm]{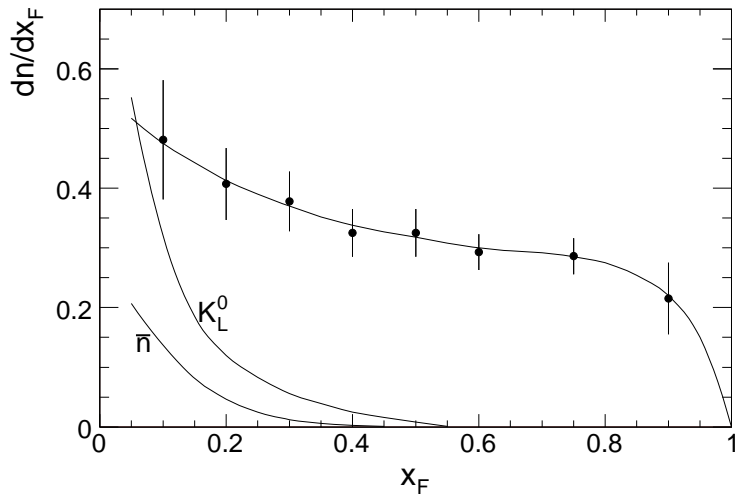}
  	\caption{$p_T$ integrated density distribution $dn/dx_F$ as a function of $x_F$ of neutrons produced in p+p 
                   interactions at 158~GeV/c. The subtracted K$^0_L$ and anti-neutron distributions are also shown}
  	\label{fig:neut_xfdist}
  \end{center}
\end{figure}

The K$^0_L$ cross section can be described, invoking isospin
symmetry, by the average charged kaon yield which is available 
to the NA49 experiment \cite{bib:ka_paper}. The corresponding $p_T$ integrated $x_F$
distribution is shown in Fig.~\ref{fig:neut_xfdist}.

The situation with pair produced neutrons is somewhat more
complicated. In fact it has been shown that baryon pairs
may be described as an isospin $I$~=~1 triplet \cite{bib:fischer} with the structure
given in Table~\ref{tab:iso}.

\begin{table}[h]
	\begin{center}
		\begin{tabular}{cccc}
  \hline 
      $I_3$        &                 -1                            &                     0                        &                    1                           \\ \hline
 \multirow{2}{25mm}{baryon pairs}  
                       &   $\overline{\textrm{p}}$n  &  $\overline{\textrm{p}}$p  &  $\overline{\textrm{n}}$p    \\
                       &                                                &  $\overline{\textrm{n}}$n  &                                                 \\  \hline
 \multirow{2}{25mm}{relative yield}  
                       &                 0.5                          &                    1                        &                 1.5                           \\
                       &                                                &                    1                        &                                                 \\  \hline                       
		\end{tabular}
	\end{center}
 	\caption{Isospin structure and relative yields of baryon pair 
                   production in p+p collisions}
     \label{tab:iso}
\end{table}

In p+p interactions it is reasonable to assume the relative yields
given above which are typical of heavy isovectors with a relatively
large suppression of the $I_3$~=~-1 component with respect to $I_3$~=~+1.
From this table one gets the following predictions:

\begin{align}                     
   \textrm{p}(\textrm{pair produced})/\overline{\textrm{p}}           &= 1.66 \nonumber \\
   \textrm{n}(\textrm{pair produced})/\overline{\textrm{n}}           &= 0.60 \\ 
   \overline{\textrm{n}}/\overline{\textrm{p}}                                  &= 1.66 \nonumber
\end{align}               

The first ratio is consistent with the result obtained by NA49
with a neutron beam \cite{bib:fischer}. In view of this it seems reasonable to
subtract from the total neutral yield 1.66 times the anti-proton
yield in order to obtain a definition of neutron production
compatible with the one for proton production. 

The resulting subtracted neutron $dn/dx_F$ distribution as a function 
of $x_F$ is shown in Fig.~\ref{fig:neut_xfdist} together with the anti-neutron and K$^0_L$
distributions used. Evidently these contributions represent an
important background to be taken into account below $x_F \sim$~0.4.

The numerical values of the neutron yields are presented in Table~\ref{tab:neut}.

\begin{table}[h]
	\begin{center}
		\begin{tabular}{|ccc|}
		\hline
		$x_F$    &   $dn/dx_F$     &   $\Delta$  \\  \hline
		  0.1       &    0.481           &      20.8      \\
		  0.2       &    0.407           &      14.7      \\
		  0.3       &    0.378           &      13.2      \\
		  0.4       &    0.325           &      11.5      \\
		  0.5       &    0.325           &      12.3      \\
		  0.6       &    0.293           &      10.2      \\
		  0.75     &    0.286           &      10.5      \\
		  0.9       &    0.215           &      27.9      \\  \hline
		\end{tabular}
	\end{center}
 	\caption{$p_T$ integrated density distribution $dn/dx_F$ for neutrons. The relative error 
 	               $\Delta$ is given in \%. It is governed 
 	               by the systematic uncertainties quoted in Table~\ref{tab:syst} }
     \label{tab:neut}
\end{table}
%
%
\subsection{Comparison with other experiments}
\vspace{3mm}
\label{sec:neut_comp}

\begin{figure}[b]
  \begin{center}
  	\includegraphics[width=11.5cm]{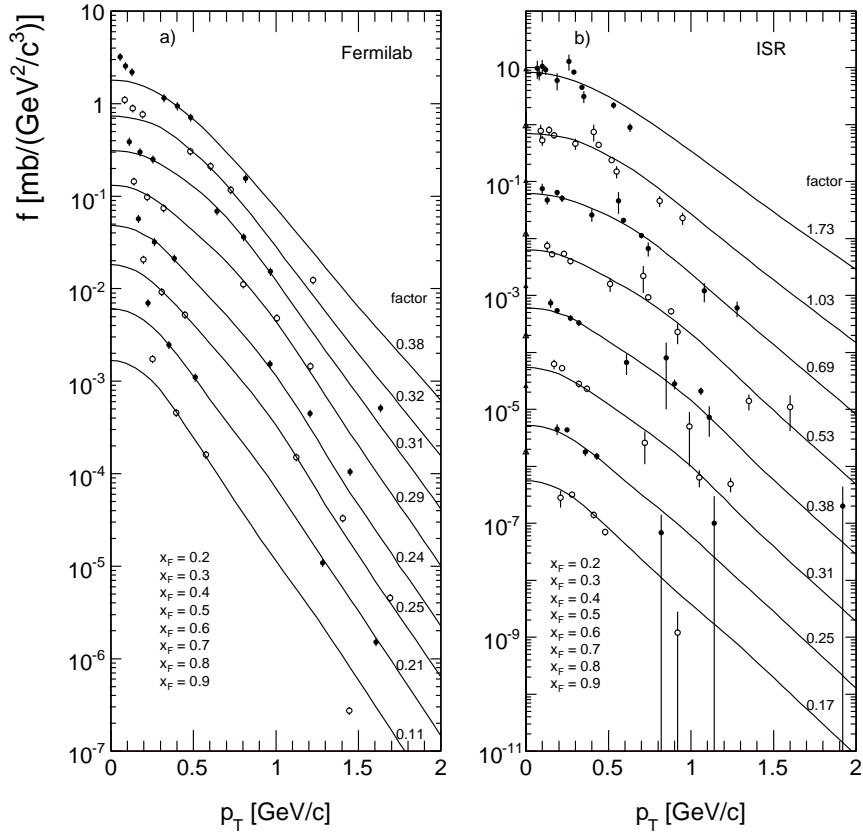}
  	\caption{Neutron $p_T$ distributions at fixed $x_F$ for a) Fermilab \cite{bib:fermilab_neut} 
                   and b) ISR\cite{bib:engl_neut}, superimposed with the interpolated NA49 
  	               proton data (lines) scaled with an appropriate normalization factor (indicated in figure). The data 
  	               were successively divided by 3 for Fermilab distributions and by 10 for ISR distributions}
  	\label{fig:neut_comp_pt}
  \end{center}
\end{figure}

As shown in Sect.~\ref{sec:exp_sit} there are only two available measurements
of neutron production in the SPS/ISR energy range, \cite{bib:fermilab_neut,bib:engl_neut,bib:flau_neut}.
Both experiments have produced double-differential cross sections
measured at a set of fixed angles. The Fermilab data cover lab
angles between 0.7 and 10~mrad, the ISR experiment between 0 and
119~mrad. The corresponding $p_T$ distributions at fixed $x_F$ are shown
in Fig.~\ref{fig:neut_comp_pt} for both cases, superimposed with the NA49 proton data
scaled with an appropriate normalization factor.

\begin{figure}[h]
  \begin{center}
  	\includegraphics[width=12cm]{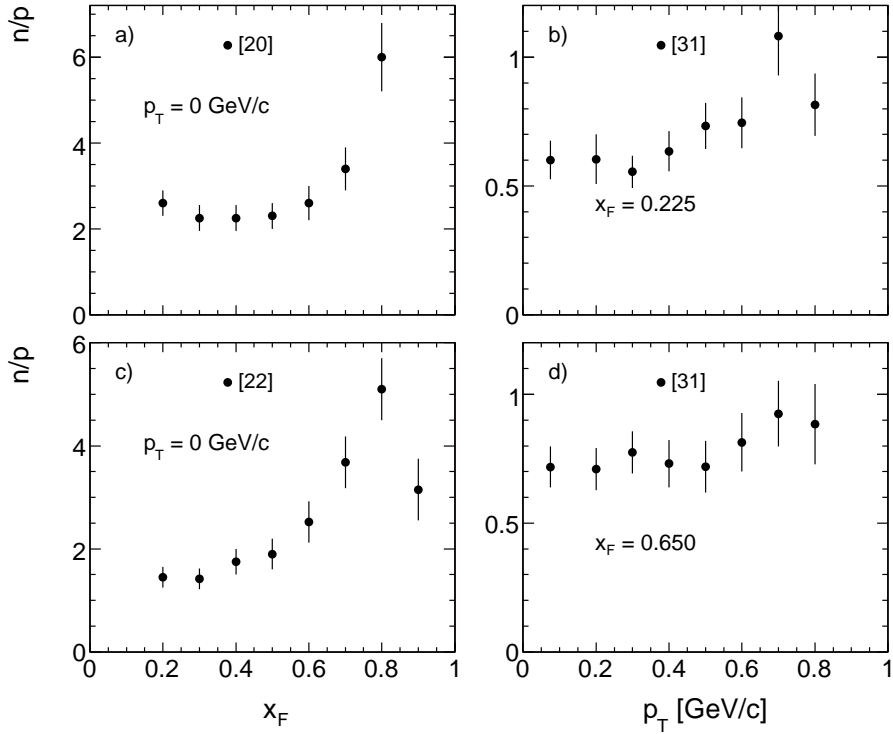}
  	\caption{Neutron to proton cross section ratio: a) \cite{bib:fermilab_neut} and 
  	               c) \cite{bib:flau_neut} at $p_T$~=~0;
                   b) and d) \cite{bib:blobel} at $\sqrt{s}$~=~6.8~GeV/c} 
  	\label{fig:n2p}
  \end{center}
\end{figure}

Evidently the proton transverse momentum distributions provide a
fair description of the neutron data as a function of $p_T$ in the 
range from 0.2 up to 1.7~GeV/c for \cite{bib:fermilab_neut} and from 0.1 to 1.7~GeV/c
for \cite{bib:engl_neut}. However the "zero degree" data from both experiments
with the calorimeter acceptance centered at 0.5~mrad \cite{bib:fermilab_neut} and
0~mrad \cite{bib:flau_neut} (triangles in Fig.~\ref{fig:neut_comp_pt}b) respectively 
exhibit upward deviations which increase 
with $x_F$. This is shown in Fig.~\ref{fig:n2p}a,c by the n/p cross section ratio
at $p_T$~=~0, hand-extrapolated in the case of \cite{bib:fermilab_neut}.

Such an increase is not seen in the n/p ratio of the
lower energy bubble chamber data of Blobel et al. \cite{bib:blobel} plotted 
in Fig.~\ref{fig:n2p}b,d as a function of $p_T$ for two values of $x_F$. Also in the 
forward proton data of NA49 with neutron beam \cite{bib:n_paper} which should, 
by isospin rotation, correspond to neutrons with proton beam, no 
peculiarity at low $p_T$ is visible. On the other hand the effective
$p_T$ window covered by a finite size calorimeter acceptance increases
linearly with $x_F$. It reaches 0.4~GeV/c at $x_F$~=~0.9 for \cite{bib:fermilab_neut}, including
the singular point at $p_T$~=~0 for the lowest angle setting. The proper
evaluation of the bin center and of the binning correction to be
applied can be rather involved in this case. The observed low-$p_T$
enhancement might therefore be assumed to be a detector effect.

\begin{figure}[h]
  \begin{center}
  	\includegraphics[width=12cm]{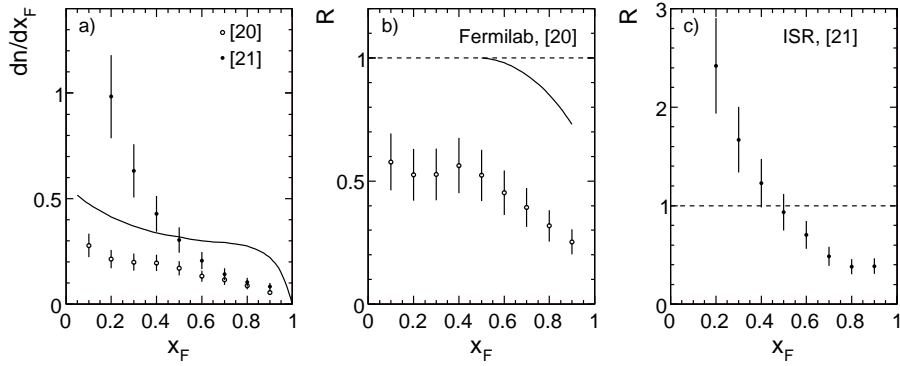}
  	\caption{Comparison as a function of $x_F$ of the NA49 results on neutron density with measurements 
  	               \cite{bib:fermilab_neut,bib:engl_neut}; a) $dN/dx_F$, full line NA49, 
  	                b) ratio $R$ between \cite{bib:fermilab_neut}
  	               and NA49 and c) ratio $R$ between \cite{bib:engl_neut} and NA49. The effect of calorimeter resolution
  	                is shown by the full line in panel b)} 
  	\label{fig:neut_comp_xf}
  \end{center}
\end{figure}

Under this assumption the normalization factors between neutron
and proton $p_T$ distributions, Fig.~\ref{fig:neut_comp_pt}, may be used directly to
determine the $p_T$ integrated neutron yields of \cite{bib:fermilab_neut,bib:engl_neut} from the
NA49 proton yields presented in Table~\ref{tab:integr}. They are compared to the
NA49 neutron measurement in Fig.~\ref{fig:neut_comp_xf}a.

Evidently both measurements deviate strongly from the NA49 results.
These deviations are given as relative factors in Figs.~\ref{fig:neut_comp_xf}b and \ref{fig:neut_comp_xf}c.
A non-trivial pattern emerges. 

For the Fermilab data there seems to be a constant suppression 
of about a factor of 2 up to $x_F \sim$~0.6, followed by a sharp decrease 
towards larger $x_F$. As in \cite{bib:fermilab_neut} no mentioning is made of any calorimeter 
resolution unfolding this decrease is reminiscent of the ratio of 
raw and unfolded data of NA49 also shown in Fig.~\ref{fig:neut_comp_xf}b. For the lower 
$x_F$ range it should be mentioned that a subtraction of K$^0_L$ 
has been performed.

For the ISR data the measured neutron yields are equal to NA49
in a small region between $x_F$~=~0.4 and $x_F$~=~0.5. For lower $x_F$ the
yield ratio increases sharply. As for these data no anti-neutron
and $K^0_L$ correction has been attempted (with the exception of K$^0_L$
subtraction for the 0 degree data), and as the fringe of the 
calorimeter resolution touches $x_F$~=~0 already for the momentum
setting at $x_F$~=~0.2, sizeable contributions from anti-neutrons and
K$^0_L$ must be expected here. In the $x_F$ region above 0.6 again a
sharp drop of the ratio is observed. In this case, however, the
calorimeter resolution has been unfolded at least for the lowest
angle setting.   

The following conclusions may be drawn from the discussion of the
data sets \cite{bib:fermilab_neut,bib:engl_neut,bib:flau_neut}:

\begin{itemize}
 \item The shape of the neutron transverse momentum distributions 
           is well described by the respective proton distributions
           over the full range of $x_F$ measured and for $p_T >$~0.2 GeV/c.
            In the lowest $p_T$ bins containing $p_T$~=~0 both experiments show
            an upward trend with respect to the proton distributions
            which is probably due to apparatus plus binning effects.
  \item The extracted, $p_T$ integrated neutron yields deviate by sizeable 
            factors from the NA49 data. For the Fermilab experiment this 
            difference may be described by a constant factor of $\sim$0.52 
            plus an effect of the non-unfolded calorimeter 
            resolution in the large $x_F$ region.
            For the ISR experiment there are  continuous and large deviations
            over the full $x_F$ scale. At $x_F <$~0.4 the missing K$^0_L$ and $\overline{\textrm{n}}$
            subtraction certainly governs the observed pattern, with neutron
            densities exceeding the measured proton yields already at $x_F$~=~0.3.
            In view of the unfolding procedure claimed in \cite{bib:fermilab_neut,bib:engl_neut} the sharp
            decrease towards high $x_F$ has to remain unexplained.
  \item The use of these data for quantitative yield comparisons is not to be
            recommended.  
\end{itemize}

%
%
\section{Leptoproduction and hadronic factorization}
\vspace{3mm}
\label{sec:hera}

Recent precision data from the ZEUS collaboration at HERA
concerning proton \cite{bib:prot_hera} and neutron \cite{bib:neut_hera} production provide
results at mean energies of about 130~GeV in the 
photon-proton cms. These data allow for a rather detailed
comparison to the p+p interaction in the region above ISR
and up to RHIC energies where little if any experimental
information is available from hadronic reactions.

%
%
\subsection{Proton production}
\vspace{3mm}
\label{sec:prot_hera}

\begin{figure}[h]
  \begin{center}
  	\includegraphics[width=7cm]{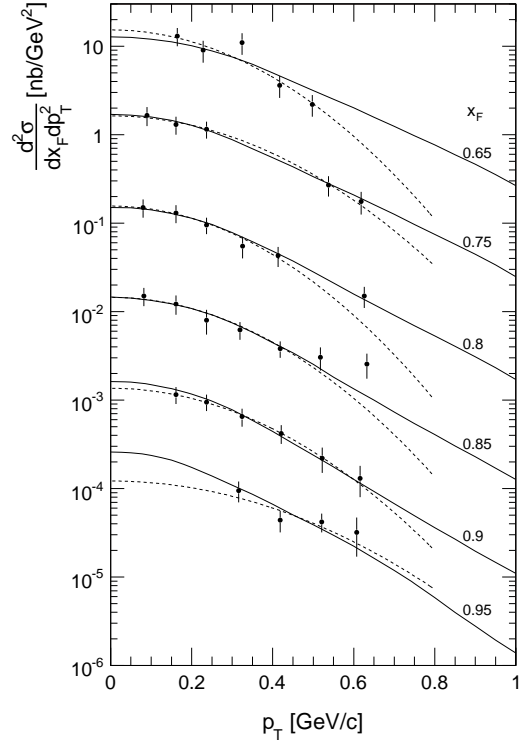}
  	\caption{Comparison of the proton $p_T$ distributions at several $x_F$ values of the NA49 results (full lines) 
  	               with measurements from  \cite{bib:prot_hera}. The data 
  	               were successively divided by 10 for different $x_F$ values for better separation. The dashed lines 
  	               represent the parametrization used in  \cite{bib:prot_hera}} 
  	\label{fig:zeus_prot_pt}
  \end{center}
\end{figure}

The ZEUS proton data \cite{bib:prot_hera} cover ranges from 0.1--0.7~GeV/c in $p_T$
and from 0.6 to 0.99 in $x_F$. Transverse momentum distributions
at 6 values of $x_F$ are compared in shape to the re-normalized
NA49 data in Fig.~\ref{fig:zeus_prot_pt}.

Evidently the HERA data follow the shape of the lower energy
pp data rather precisely within the quoted statistical errors.
This complies with the $s$-independence of the $p_T$ distributions
up to $\sqrt{s}$~=~53~GeV and up to $p_T$~=~1.5~GeV/c in the same $x_F$ 
range, see Fig.~\ref{fig:isr_ptdist}. The Gaussian fits used in \cite{bib:prot_hera} and shown as
dashed lines in Fig.~\ref{fig:zeus_prot_pt} describe the measured cross sections
reasonably well with some exceptions in $x_F$. They
deviate however systematically from the NA49 data already at the highest $p_T$
values available in \cite{bib:prot_hera}. In fact a Gaussian approximation of
the proton transverse momentum distributions is at best only
valid over very restricted regions. This has been discussed
in connection with the low-$p_T$ extrapolation of the hadronic
data in Sect.~\ref{sec:high_xf} and has led to the application of the
two-dimensional interpolation scheme, Sect.~\ref{sec:interp}, which does
not rely on any algebraic parametrization. The extension of
the $p_T$ range of the HERA data up to and beyond the GeV/c region
would of course be very interesting but has to remain on the
wish list for eventual future work on leptoproduction.

\begin{figure}[h]
  \begin{center}
  	\includegraphics[width=15cm]{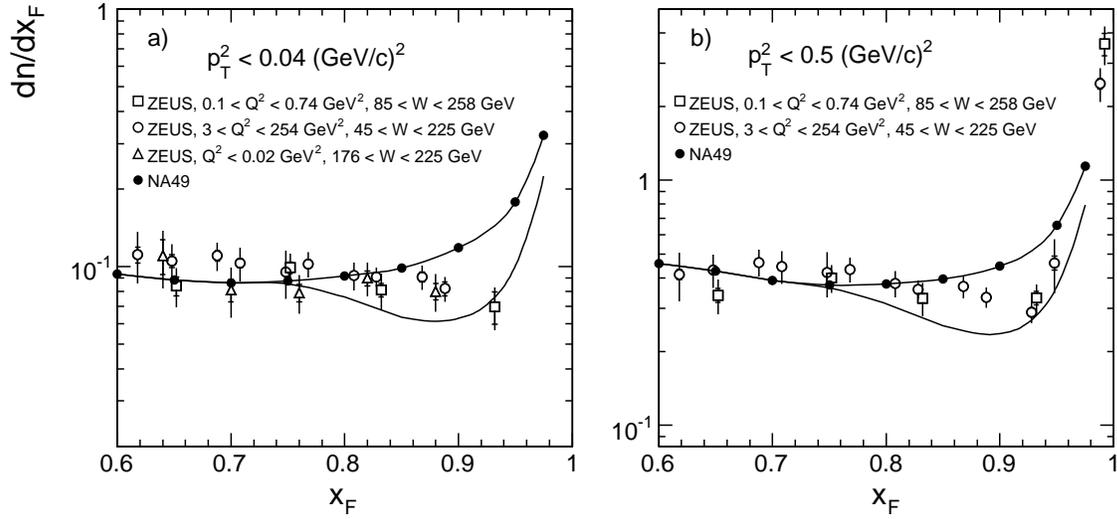}
  	\caption{Comparison of the proton $p_T$ integrated distributions as a function of $x_F$ of the NA49 results with
  	               measurements from  \cite{bib:prot_hera}. The integration is performed in the range of $p_T^2$
  	               less than: a) 0.04 (GeV/c)$^2$ and b) 0.5 (GeV/c)$^2$. The full symbols and lines: NA49} 
  	\label{fig:zeus_prot}
  \end{center}
\end{figure}

A comparison of $p_T$ integrated yields as they are given in
\cite{bib:prot_hera} for the measured ranges of $p_T^2 <$~0.04 and $<$~0.5
(GeV/c)$^2$ to the NA49 data integrated over the same ranges
is presented in Fig.~\ref{fig:zeus_prot}.

\begin{figure}[h]
  \begin{center}
  	\includegraphics[width=8cm]{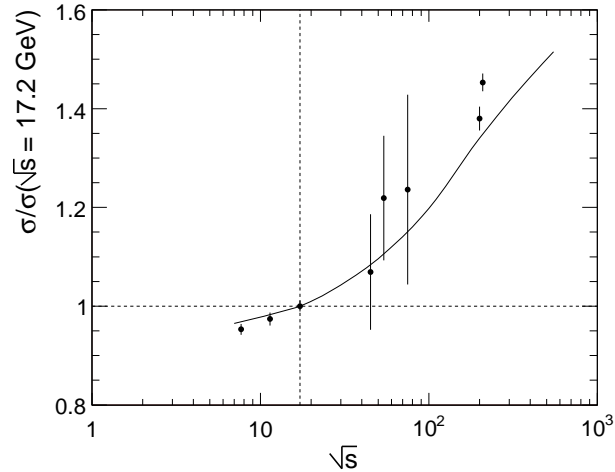}
  	\caption{Total inelastic cross section normalized at $\sqrt{s}$~=~17.2~GeV as a function of 
  	               $\sqrt{s}$ for p+p (line) and $\gamma$+h (circles) interactions} 
  	\label{fig:inel}
  \end{center}
\end{figure}

The good quantitative agreement of the proton densities up to 
$x_F \sim$~0.8 in both $p_T$ windows is noteworthy. This may shed some
light on the question of scaling versus increase of total
inelastic cross section in this energy regime as mentioned in
Sect.~\ref{sec:prot_alb}. As the photonic total cross section rises at 
least as fast as the hadronic one with cms energy, Fig.~\ref{fig:inel},   
a non-scaling of the invariant cross sections as opposed to
particle densities is necessarily implied by baryon number 
conservation.

It is also interesting to regard the high-$x_F$ suppression extracted
in Sect.\ref{sec:comp_isr} from ISR and collider data, Fig.~\ref{fig:isr_ua4}, in connection
with the HERA data. The expected decrease of proton density
above $x_F \sim$~0.7 is indicated by the lower line in Fig.~\ref{fig:zeus_prot}. This
effect will again be discussed in relation to neutrons below.

%
%
\subsection{Neutron production}
\vspace{3mm}
\label{sec:neut_hera}      

\begin{figure}[h]
  \begin{center}
  	\includegraphics[width=7cm]{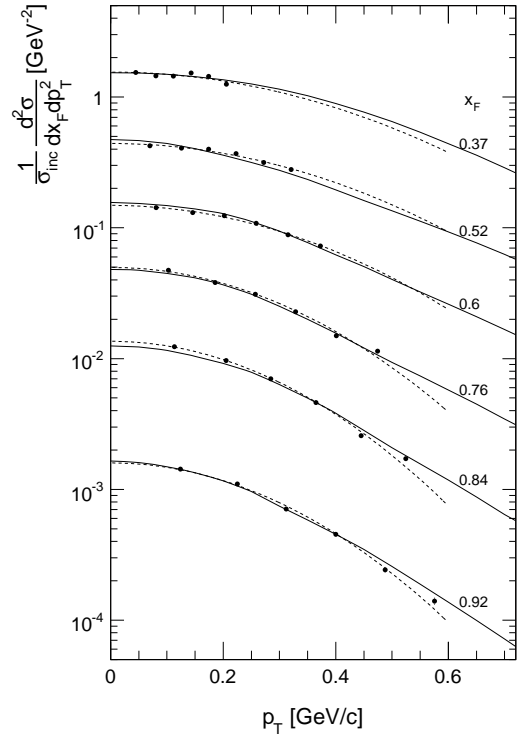}
  	\caption{Comparison of the $p_T$ distributions at different $x_F$ values of protons from NA49 (full lines) 
  	                with neutrons from  \cite{bib:neut_hera}. The data 
  	               were successively divided by 3 for different $x_F$ values for better separation. The dashed lines 
  	               represent the parametrization used in \cite{bib:neut_hera}} 
  	\label{fig:zeus_neut_pt}
  \end{center}
\end{figure}

The ZEUS neutron data \cite{bib:neut_hera} cover ranges from 0.05--0.6~GeV/c in $p_T$
and from 0.26 to 0.97 in $x_F$. As already shown for protons, the
relative shape of the neutron $p_T$ distributions is well described 
by the NA49 proton data in the measured $p_T$ ranges, see Fig.~\ref{fig:zeus_neut_pt}

This shape similarity verifies the result for lower-energy
neutron distributions, Fig.~\ref{fig:neut_comp_pt}, where the comparison reaches
up to $p_T \sim$~1.5~GeV/c. Again the Gaussian parametrization
chosen by \cite{bib:neut_hera} is indicated by dashed lines, and again the
limited applicability of such parametrization is evident
especially if total $p_T$ integrated yields are to be extracted.

\begin{figure}[h]
  \begin{center}
  	\includegraphics[width=11.cm]{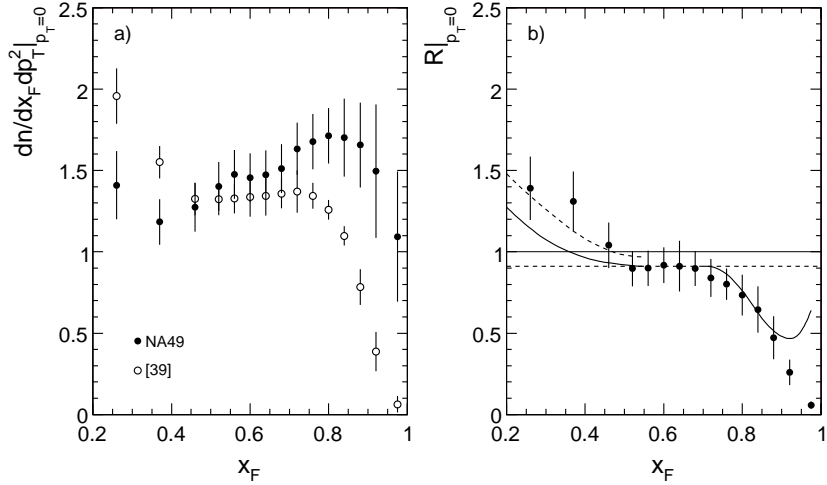}
  	\caption{Comparison of the neutron data from \cite{bib:neut_hera} to
  	               NA49 results as a function of $x_F$ at $p_T$~=~0~GeV/c: a) $dn/dx_Fdp_T^2$,
  	               b) ratio $R$ between \cite{bib:neut_hera} and NA49 results} 
  	\label{fig:zeus2na49}
  \end{center}
\end{figure}

An interesting comparison of the yields $dn/dxdp_T^2$ at $p_T$~=~0  \cite{bib:neut_hera}
with the NA49 data becomes possible under the assumption that the 
$p_T$ distributions of neutrons are identical to the ones of protons 
in p+p interactions. This does not look unreasonable 
in view of the results shown above. With this assumption the 
total measured neutron yields of NA49, Table~\ref{tab:neut} and Fig.~\ref{fig:neut_xfdist}, may 
be converted into $p_T$~=~0 densities using the proton $p_T$ 
distributions shown in Fig.~\ref{fig:zeus_neut_pt}. The resulting absolute densities 
$dn/dxdp_T^2$ are presented in Fig.~\ref{fig:zeus2na49}.
 
This Figure exhibits an interesting pattern. In the region 0.45~$< x_F <$~0.7 
both yields are equal to within about 7-8\%. This difference is 
compatible with the systematic errors of the NA49 data given in 
Table~\ref{tab:syst}. At lower $x_F$ the ZEUS data increase, towards large $x_F$ 
they decrease with respect to the p+p data. This is quantified
in the yield ratio of Fig.~\ref{fig:zeus2na49}b. 

The enhancement of the ZEUS data for $x_F <$~0.5 may be connected to
two effects. A first contribution is given by the production
of K$^0_L$ and anti-neutrons which are experimentally not separable
in the used calorimeter. This contribution appears in the $x_F$ region
in question and has been subtracted from the NA49 data, see Sect.~\ref{sec:neut} 
and Fig.~\ref{fig:neut_xfdist}. At HERA energies the corresponding cross sections must
be expected to increase over their values at $\sqrt{s}$~=~17.2~GeV.
Existing measurements of K$^0_S$ at RHIC and p+$\overline{\textrm{p}}$ collider energies
\cite{bib:rhic} do however not allow for a consistent analysis of this 
situation. An increase by up to a factor of two can nevertheless
not be excluded. Allowing for the same percentage contribution 
to the neutron yield as in the NA49 data, the lower solid line in 
Fig.~\ref{fig:zeus2na49}b is obtained.

A second contribution has to be expected from the feed-down of
neutrons and anti-neutrons from weak decays of strange hyperons.
As the ZEUS calorimeter is placed at a distance of several decay 
lengths of the contributing hyperons, the fraction of decays into 
neutrons defines the principle component. For a quantitative 
elaboration of this effect a detailed simulation of the 
experimental set-up, especially of the aperture limitations, 
is of course mandatory. Adding however the percentage contribution
to the neutron yields as calculated for the NA49 data, Sect.~\ref{sec:feed},
the upper dashed line in Fig.~\ref{fig:zeus2na49}b is obtained. Although this
procedure is of course to be seen as a mere exercise, the
two effects described certainly value a more detailed scrutiny.  

The decrease of the ZEUS data at $x_F >$~0.7 can on the other hand
be connected to the $s$-dependent yield depletion observed for
protons and already invoked in the preceding chapter on proton
production. Indeed there is no reason why neutrons should not
show a similar effect. In fact, due to the absence of a diffractive peak
in the neutron hemisphere, the effect might be enhanced at $x_F >$~0.9.
This is indeed seen in Fig.~\ref{fig:zeus2na49}b. Here the depletion as a function 
of $x_F$ has been evaluated for HERA energy and applied to the NA49
neutron data. This results in the solid line at $x_F >$~0.7 which
describes the rough structure up to $x_F \sim$~0.9. The minimum at $x_F$~=~0.92
and the subsequent increase towards the diffractive proton peak
is of course not to be expected for neutron production.

\begin{figure}[h]
  \begin{center}
  	\includegraphics[width=11.cm]{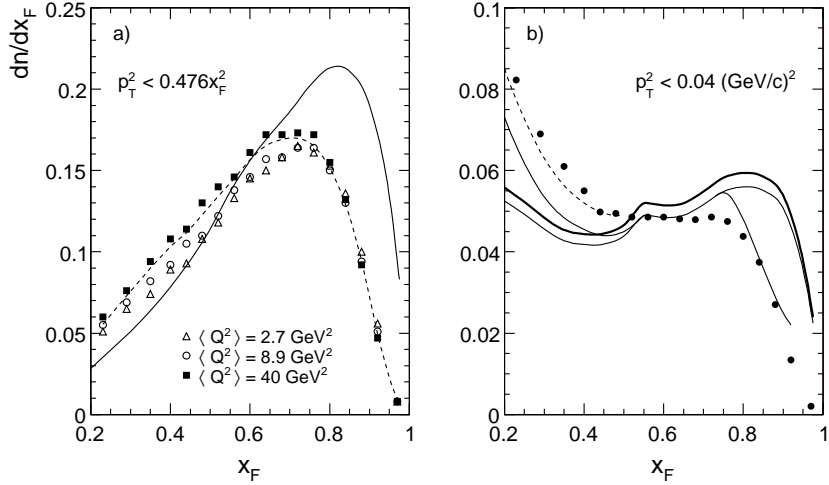}
  	\caption{Comparison of the $p_T$ integrated neutron data from \cite{bib:neut_hera} to
  	               NA49 results (full lines) as a function of $x_F$; a) $p_T$ integration is performed in the range
  	               $p_T <$~0.69$x_F$, b) $p_T$ integration at $p_T <$~0.2~GeV/c} 
  	\label{fig:zeus_neut}
  \end{center}
\end{figure}

As expected from the shape similarity of the transverse momentum 
distributions of neutrons and protons, Fig.~\ref{fig:zeus_neut_pt}, the comparison
of the $p_T$ integrated yields also given in \cite{bib:neut_hera} gives similar
results. Two integrations, one with an $x_F$ dependent $p_T$ window of
$p_T <$~0.69$x_F$ and one with a constant window up to $p_T$~=~0.2~GeV/c
are compared in Fig.~\ref{fig:zeus_neut}a and \ref{fig:zeus_neut}b, respectively. 

In Fig.~\ref{fig:zeus_neut}a the NA49 results are given as solid line, the ZEUS
results for the full DIS sample with $\langle Q^2 \rangle$~=~13~GeV$^2$ as the
dashed line. In addition the ZEUS data points for three
subsamples with different $\langle Q^2 \rangle$ are presented. The pattern of
enhancement below $x_F \sim$~0.5 and depletion above $x_F \sim$~0.7 is very
similar to Fig.~\ref{fig:zeus2na49}a. This is also apparent in Fig.~\ref{fig:zeus_neut}b with
a constant $p_T$ integration window. Here the NA49 results are
given as thick line and compared to the ZEUS data points
corresponding to the full DIS sample. A re-normalization of
the NA49 yield by about 6\% at $x_F$~=~0.6 (see also Fig.~\ref{fig:zeus2na49}) is
indicated as a thin line. The contributions from K$^0_L$ and
$\overline{\textrm{n}}$ production, from hyperon feed-down as well as the high-$x_F$
depletion are referred to this line as in Fig.~\ref{fig:zeus2na49}b.

%
%
\subsection{Hadronic factorization}
\vspace{3mm}
\label{sec:hadr_hera}

The equality, within the experimental errors, of the production 
of forward protons and neutrons in deep inelastic e+p collisions 
to the purely hadronic p+p interaction is reminiscent of
hadronic factorization, that is of the independence of target
fragmentation on the type of hadronic projectile used. This
factorization has been well established with pion, kaon and
baryon beams on a proton target. In this sense the above results
would indicate the virtual photon to act as an $I_3$~=~0 mesonic 
state. The important point here is that the observed factorization 
extends to low $x_F$ values, well into the region of non-diffractive 
hadronic collisions, where it has been shown that neither charge 
nor flavour exchange is present in the hadronic sector, see for 
instance the discussion in \cite{bib:pc_discuss}. The detailed study of 
other particle species also in the region of central rapidity 
and of the long-range correlations (or their absence) with the 
photon hemisphere would be mandatory to further clarify this 
situation.

%
%
\section{Conclusion}
\vspace{3mm}
\label{sec:concl}

New inclusive data on proton, anti-proton and neutron production in
p+p interactions at SPS energy have been presented. These data
represent a continuation of the systematic study of hadronic 
collisions by the NA49 experiment at 158~GeV/c beam momentum.
They offer an unprecedented coverage of the available phase space
with double differential inclusive cross sections featuring 
systematic errors in the few percent range. This allows for a very 
detailed comparison with existing data with the aim at establishing a 
reliable data base up to ISR energies including especially the hitherto 
unclear situation concerning neutron production. In this context several
points are noteworthy:

\begin{itemize}
	\item the consolidation of the wealth of data available in the SPS
             and ISR energy ranges, mostly obtained some 30 years ago,
             has been attempted here with only partially satisfactory
             results, in particular concerning neutrons. 
   \item  the necessity of taking care of baryonic feed-down from strange
             hyperons on a quantitative level has been demonstrated.
   \item  the global independence of transverse momentum distributions
             up to about 1.5~GeV/c on reaction type and $\sqrt{s}$ and their
             equality for protons and neutrons has been shown.
   \item  the $s$-dependence through the ISR energy range and up to HERA
             and p+$\overline{\textrm{p}}$ collider energies has been investigated. In
             particular a specific yield suppression at $x_F >$~0.7 with
             increasing cms energy has been quantified.
  \item   the question of the scaling of baryon yields versus cross sections
             has been addressed in the context of the rapid increase of the
             total inelastic cross sections with interaction energy.
             Scaling of yields rather than cross sections is necessary in
             order not to violate baryon number conservation.
   \item  the comparison to deep inelastic lepton scattering establishes
             hadronic factorization also in this reaction within the 
             experimental uncertainties and the phase space region available.
\end{itemize}

Finally it should be stated that the establishment of a precise base
of single inclusive data on baryon production is only a first step in
an effort to shed some light on the general problem of baryon
number transfer. The transition from the incoming baryonic
target or projectile to the observed final state is, as a part
of the non-perturbative sector of QCD, not understood on the
level of any reliable theory. In hadronic interactions, most
approaches are using ad-hoc assumptions like for instance
the concept of di-quark fragmentation. In electroproduction
baryon production is described in most approaches by the 
scattering of the virtual photon off an exchange pion. There
is no doubt that this situation can only be clarified by
further and more detailed experimental studies which go beyond
the single inclusive level.

One of these experimental openings is the study of resonance
production and decay which is accessible to the NA49 detector
via its good phase space coverage. This widely neglected field
will provide very strong constraints concerning the repartition
of particle species as products of the cascading decay of
heavy resonances, especially concerning the relation between
neutrons and protons as it is given by the isospin structure 
of the initial state. Another field of studies concerns internal 
baryonic correlations. By selecting a leading proton in either 
the target or the projectile hemisphere the forward-backward
correlation of baryon number transfer may be studied, in
particular the feed-over of baryon number from one hemisphere
to the other and its evolution with interaction energy.
The use of neutron projectiles and of non-baryonic, mesonic
beams as they are available in fixed-target operation opens
the possibility of model-independent studies essentially relying 
on baryon number conservation and concepts like isospin symmetry. 

In this context the study of nuclear reactions, in particular
of proton-nucleus and pion-nucleus scattering with controlled centrality,
provides unique access to multiple hadronic interactions. 
The strong dependence of the final state baryon distributions on
the number of projectile subcollisions inside the nucleus,
generally misnamed as "stopping" and as yet not understood on any
theoretical level, offers a further and very strong constraint on
the possible mechanism of baryon number transfer.

\section*{Acknowledgements}
\vspace{3mm}
This work was supported by
the Bundesministerium fuer Forschung und Technologie (06F137), 
the Polish State Committee for Scientific Research (P03B00630),
the Polish Ministry of Science and Higher Education (N N202 078735),
the Hungarian Scientific Research Fund OTKA (T68506),
the Bulgarian National Science Fund (Ph-09/05) and
the EU FP6 HRM Marie Curie Intra-European Fellowship Program.

\vspace{2cm}

\end{document}